\documentclass[journal]{IEEEtran}

\usepackage{amsmath,amsfonts,amssymb}
\usepackage{algorithmic}
\usepackage{algorithm}
\usepackage{array}
\usepackage[caption=false,font=footnotesize]{subfig}
\usepackage{textcomp}
\usepackage{stfloats}
\usepackage[hyphens]{url}
\usepackage{verbatim}
\usepackage{graphicx}
\usepackage{cite}
\usepackage{xcolor}
\usepackage{soul}
\usepackage{orcidlink}

\usepackage{tabularx}
\usepackage{longtable}
\usepackage{booktabs}
\usepackage{multirow}
\usepackage{makecell}

\usepackage{pifont}
\usepackage{ulem}
\usepackage{float}

\usepackage[acronym]{glossaries}
\makeglossaries

\hypersetup{hidelinks}

\newacronym{ml}{ML}{Machine Learning}
\newacronym{ai}{AI}{Artificial Intelligence}
\newacronym{xai}{XAI}{eXplainable AI}
\newacronym{piml}{PIML}{Physics-Informed Machine Learning}
\newacronym{llm}{LLM}{Large Language Model}
\newacronym{kg}{KG}{Knowledge Graph}
\newacronym{rl}{RL}{Reinforcement Learning}
\newacronym{rag}{RAG}{Retrieval-Augmented Generation}

\newacronym{ics}{ICS}{Industrial Control System}
\newacronym{cps}{CPS}{Cyber-Physical System}
\newacronym{cpps}{CPPS}{Cyber-Physical Production System}
\newacronym{icps}{ICPS}{Industrial Cyber-Physical System}
\newacronym{iot}{IoT}{Internet-of-Things}
\newacronym{iiot}{IIoT}{Industrial Internet-of-Things}
\newacronym{iov}{IoV}{Internet-of-Vehicles}
\newacronym{iomt}{IoMT}{Internet-of-Medical-Things}
\newacronym{plc}{PLC}{Programmable Logic Controller}
\newacronym{rtus}{RTUs}{Remote Terminal Units}
\newacronym{hmi}{HMI}{Human-Machine Interface}
\newacronym{scada}{SCADA}{Supervisory Control and Data Acquisition System}
\newacronym{dcs}{DCS}{Distributed Control System}

\newacronym{it}{IT}{Information Technology}
\newacronym{ot}{OT}{Operational Technology}
\newacronym{lan}{LAN}{Local-Area Network}
\newacronym{soc}{SOC}{Security Operations Center}
\newacronym{can}{CAN}{Controller Area Network}

\newacronym{ids}{IDS}{Intrusion Detection System}
\newacronym{dos}{DoS}{Denial-of-Service}
\newacronym{ddos}{DDoS}{Distributed Denial-of-Service}
\newacronym{xss}{XSS}{Cross-Site Scripting}
\newacronym{mitm}{MITM}{Man-in-the-Middle}

\newacronym{shap}{SHAP}{SHapley Additive exPlanations}
\newacronym{lime}{LIME}{Local Interpretable Model-agnostic Explanations}
\newacronym{loco}{LOCO}{Leave-One-Covariate-Out}
\newacronym{cem}{CEM}{Contrastive Explanation Method}
\newacronym{ale}{ALE}{Accumulated Local Effects}
\newacronym{pdp}{PDP}{Partial Dependence Plot}
\newacronym{ice}{ICE}{Individual Conditional Expectancy}
\newacronym{pfi}{PFI}{Permutation Feature Importance}
\newacronym{eli5}{ELI5}{Explain-Like-I-am-5}
\newacronym{trustai}{TRUST}{Transparency Relying Upon Statistical Theory}

\newacronym{svm}{SVM}{Support Vector Machine}
\newacronym{rf}{RF}{Random Forest}
\newacronym{dt}{DT}{Decision Tree}
\newacronym{lr}{LgR}{Logistic Regression}
\newacronym{knn}{kNN}{K-Nearest Neighbors}
\newacronym{nn}{NN}{Neural Network}
\newacronym{cnn}{CNN}{Convolutional Neural Network}
\newacronym{dnn}{DNN}{Deep Neural Networks}
\newacronym{rnns}{RNNs}{Recurrent Neural Networks}
\newacronym{gnn}{GNNs}{Graph Neural Networks}
\newacronym{lstm}{LSTM}{Long Short-Term Memory}
\newacronym{gru}{GRU}{Gated Recurrent Unit}
\newacronym{lgbm}{LGBM}{Light Gradient Boosting Machine}
\newacronym{pca}{PCA}{Principal Component Analysis}
\newacronym{fms}{FMs}{Foundation Models}

\newacronym{cip}{CIP}{Critical Infrastructure Protection}
\newacronym{nerc}{NERC}{North American Electric Reliability Corporation}
\newacronym{nist}{NIST}{National Institute for Standards and Technology}
\newacronym{nis2}{NIS2}{Network and Information Security 2}
\newacronym{eu}{EU}{European Union}
\newacronym{us}{US}{United States}
\newacronym{oecd}{OECD}{Organization for Economic Cooperation and Development}
\newacronym{rmf}{RMF}{Risk Management Framework}

\newacronym{ioc}{IoC}{Indicators of Compromise}
\newacronym{apt}{APT}{Advanced Persistent Threat}
\newacronym{cve}{CVE}{Common Vulnerabilities and Exposures}
\newacronym{xdr}{XDR}{Extended Detection and Response}
\newacronym{soar}{SOAR}{Security Orchestration, Automation and Response}
\newacronym{siem}{SIEM}{Security Information and Event Management}
\newacronym{ir}{IR}{Incident Response}

\begin{document}


\title{
Explainable Artificial Intelligence for Industrial Cybersecurity:
A Review of Methods, Operational Integration, and Research Challenges
}

\author{
Amr~S.~Mohamed\orcidlink{0000-0001-6812-4200},
Charlotte~Fritz\orcidlink{0009-0004-4189-1433},
Ahmad~Mohammad~Saber\orcidlink{0000-0003-3115-2384},
Yiqun~Ma\orcidlink{0009-0002-0165-1393},
Ratinder~Kaur\orcidlink{0000-0003-1360-528X},
Mohammed~Al-Darwbi\orcidlink{0000-0002-5542-6231},
Daniela~Friedrich,
and
Deepa~Kundur\orcidlink{0000-0001-5999-1847}
\thanks{
Amr S. Mohamed is with the German University in Cairo, Cairo, Egypt.
}
\thanks{
Charlotte Fritz, Ahmad Mohammad Saber, and Deepa Kundur are with the University of Toronto, Toronto, ON, Canada.
}
\thanks{
Yiqun Ma is with the University of Illinois Urbana-Champaign, Champaign, IL, USA.
}
\thanks{
Ratinder Kaur, Mohammed Al-Darwbi, and Daniela Friedrich are with Siemens Canada, Canada.
}
\thanks{
Corresponding author: Deepa Kundur (e-mail: dkundur@ece.utoronto.ca)
}
}

\maketitle
\begin{abstract}
The increasing digitalization of industrial infrastructure and the convergence of information technology (IT) and operational technology (OT) have significantly expanded the cyberattack surface of modern industrial systems. To address the growing complexity of cyber threats, artificial intelligence (AI) and machine learning (ML) techniques are increasingly being deployed within industrial cybersecurity operations, particularly in Security Operations Centers (SOCs). While these approaches offer improved capabilities for anomaly detection, threat analysis, and automated response, their opaque decision-making processes present challenges for operational trust, regulatory compliance, and incident response.
EXplainable Artificial Intelligence (XAI) has emerged as a promising paradigm to improve the transparency and interpretability of AI-driven cybersecurity systems and decisions. This paper provides a comprehensive review of XAI techniques in the context of industrial cybersecurity. This survey focuses particularly on industrial SOC environments and operational industrial security workflows. The survey examines the role of AI in industrial SOC workflows, the types of operational data leveraged in industrial environments, and the benefits and limitations associated with AI-based threat detection. We then review major families of XAI approaches, including feature attribution methods, surrogate models, rule-based explanations, and visualization techniques, and analyze their applicability to industrial cybersecurity use cases.
In addition, the paper discusses the unique operational, regulatory, and safety requirements that distinguish industrial systems from traditional IT environments. Key challenges such as limited labeled datasets, model reliability, explainability-performance tradeoffs, and the integration of XAI tools into SOC workflows are examined. Finally, we identify open research directions and highlight opportunities for developing trustworthy, operationally viable, and domain-specific XAI-enabled cybersecurity solutions for industrial environments.
\end{abstract}


\maketitle

\section{Introduction}
\label{sec:introduction}

Industrial systems represent some of the most operationally critical environments in modern society. They sustain critical infrastructure and essential services across sectors such as energy, manufacturing, transportation, water treatment, and oil and gas. These industrial environments bring together physical processes, operational technologies, communication networks, computing platforms, and human operators to coordinate and sustain large-scale operations. As these crucial environments undergo rapid digital transformation, they are becoming increasingly interconnected through \gls{iiot} technologies, cloud services,  wireless communications, and the convergence of \gls{it} and \gls{ot}.

Within industrial environments, \gls{ot} broadly encompasses the technologies used to monitor and control physical processes, while \glspl{ics} refer more specifically to the computational and control systems responsible for real-time industrial monitoring and automation~\cite{canadian2022protect}; these systems include \glspl{plc}, \glspl{hmi}, \glspl{scada}, and \glspl{dcs}~\cite{bhamare2020cybersecurity}. Industrial systems are also increasingly characterized as \glspl{cps}, in which computational and communication systems interact closely with physical processes. In industrial contexts, these integrations are commonly referred to as \glspl{icps}, emphasizing the convergence of industrial operations, control systems, sensing infrastructure, and enterprise networks. Similarly, \gls{iiot} extends traditional industrial environments by enabling connected devices, sensors, and machinery to exchange operational data across distributed industrial infrastructures.

This complex industrial landscape is undergoing a rapid transformation driven by Industry~4.0 initiatives, increasing connectivity, and growing \gls{it}/\gls{ot} convergence~\cite{lasi2014industry, matt2023industrial, murray2017convergence, kayan2022cybersecurity}. Amongst these initiatives, industrial infrastructures are being increasingly interconnected through heterogeneous devices, distributed communication systems, cloud and edge computing platforms, and wireless industrial networks~\cite{singh_artificial_2020}. While these developments enable enhanced operational visibility and efficiency, and data-driven automation and predictive maintenance, they also increase system complexity and expand the cyberattack surface of industrial operations,  creating new vulnerabilities and challenges for industrial cybersecurity.

Consequently, industrial security has emerged as one of the most operationally critical and data-intensive cybersecurity domains, requiring continuous monitoring, anomaly detection, and incident response across converged \gls{it} and \gls{ot} environments~\cite{singh_artificial_2020}.

Traditionally, industrial cybersecurity relied heavily on expert systems and rule-based approaches that continuously monitored network traffic, device behavior, and operational events for predefined patterns or signatures of known threats~\cite{vc3_ai_cybersecurity_2023}. While effective for detecting previously identified attacks, these approaches depend heavily on prior knowledge and manually crafted detection rules. As industrial environments become increasingly interconnected, heterogeneous, and operationally complex, rule-based systems struggle to adapt to evolving attack strategies, unknown threats, and large-scale volumes of operational data. Consequently, novel or sophisticated attacks may evade detection when their signatures or behavioral patterns are absent from existing security rulesets.

To address the growing complexity of industrial cybersecurity operations, \gls{ai} and \gls{ml} are increasingly being adopted for threat detection, anomaly identification, and security decision-making~\cite{vc3_ai_cybersecurity_2023, abdul2018trends}. Unlike traditional rule-based systems that rely on predefined signatures and manually crafted heuristics, modern \gls{ml} approaches learn patterns directly from operational and security-relevant data to identify abnormal or potentially malicious behavior. Consequently, \gls{ml} is increasingly being integrated into industrial cybersecurity operations to support large-scale monitoring, improve threat detection capabilities, reduce analyst workload, and automate routine security tasks~\cite{khayat_empowering_2025, sailpoint_ai_cybersecurity_2023}.

However, many modern \gls{ml} models, particularly deep learning-based approaches, are often characterized as opaque or \textit{black-box} systems whose internal decision-making processes are difficult to interpret or justify~\cite{angelov2021explainable}. In safety-critical and compliance-regulated industrial environments, it is insufficient for \gls{ai} systems to merely produce accurate outputs; their decisions must also be understandable, trustworthy, auditable, and operationally actionable for security analysts, operators, engineers, and regulators. This challenge has driven growing interest in \gls{xai}, which seeks to improve the transparency, interpretability and accountability of \gls{ai}-driven cybersecurity systems by providing human-understandable explanations for model behavior and security decisions.

Despite growing interest in \gls{xai} across the broader cybersecurity domain, existing surveys have focused primarily on traditional \gls{it} environments and generalized cybersecurity applications. In contrast, industrial environments present distinct operational constraints, safety requirements, heterogeneous data characteristics, and regulatory considerations that fundamentally influence how \gls{ai} and \gls{xai} systems are designed, deployed, and evaluated. Furthermore, industrial cybersecurity operations are increasingly centralized around \glspl{soc}, where analysts, engineers, and operators must interpret and act upon \gls{ai}-driven security decisions in real time.

To address this gap, this paper presents a comprehensive review of \gls{xai} in the context of industrial security, with a particular focus on industrial operational environments and \glspl{soc}. Our main contributions are summarized as follows:

\begin{itemize}
    \item We present a comprehensive review of recent literature applying \gls{xai} to industrial cybersecurity, analyzing the underlying datasets, \gls{ml} models, \gls{xai} methods, visualization strategies, and evaluation practices used across existing works. We further organize existing \gls{xai} approaches into methodological categories and assess their suitability for industrial security datasets and operational environments.
    
    \item We characterize the operational roles, workflows, constraints, and security challenges that distinguish industrial cybersecurity from traditional IT-focused security. Through the lens of industrial \glspl{soc}, we examine how \gls{xai} methods can support stakeholder decision-making and outline the distinct requirements of explainability in industrial security environments.
    
    \item We examine the regulatory, safety, auditability, and accountability considerations associated with deploying AI and \gls{xai} in industrial environments, and identify key research gaps and future directions toward trustworthy, operationally viable, and domain-specific industrial \gls{xai} systems.
\end{itemize}

To the best of our knowledge, this review is the first to focus specifically on \gls{xai} for security in industrial contexts, as can be concluded from Table~\ref{tab:comparison}. We
observe that the relevant literature is primarily divided between two technical streams: \glspl{ids} applied to network traffic
data, and anomaly detection applied to industrial process and \gls{ot} data. This bifurcation reflects a lingering IT/OT divide in industrial
security research.

Given the increasing convergence of IT and OT systems, our review complements existing surveys that focus on adjacent cybersecurity domains. These domains remain highly relevant to industrial security, as vulnerabilities in enterprise systems, such as authentication or access control failures, can serve as entry points for attacks on industrial infrastructure.

\renewcommand{\arraystretch}{1.2}
\begin{table*}[!t]
\caption{Summary of related works}
\label{tab:comparison}
\centering
\begin{tabularx}{\textwidth}{c p{7.5cm} p{3.5cm} X X}
\hline
\bfseries Reference & \bfseries Description of Work & 
\bfseries Application Domain(s) & 
\multicolumn{2}{c}{\bfseries XAI Perspective} \\ 
\cline{4-5}
& & & \bfseries Regulatory Compliance & 
\bfseries Cyber Physical Systems (\gls{cps}) \\ 
\hline
~\cite{Zhang2022explainable}   
& Surveys \gls{xai} applications across malware detection, spam filtering, \gls{ids}, botnet detection, phishing, fraud prevention, and DoS attacks. Discussion of general \gls{xai} challenges, cybersecurity-specific concerns, and threats targeting \gls{xai} systems.
& Healthcare, smart cities, agriculture, finance, and transportation.
& \ding{115} Limited  
& \ding{115} Limited 
\\
\hline
~\cite{Capuano2022explainable_______}   
& Categorizes the literature by use case: IDS, malware detection, phishing and spam and botnet detection.
& Cybersecurity in Information and Communication Technology (ICT) systems.
& \ding{55} None
& \ding{115} Limited 
\\ 
\hline
~\cite{Rjoub2023survey} 
& Organizes their review around six core
security functions: intrusion detection, intrusion prevention, access control, privacy, authentication, and trust/reputation. Emphasizes the mathematical underpinnings of \gls{xai} methods and maps different \gls{xai} taxonomies (e.g., local vs. global, post-hoc vs. intrinsic) to these security functions.  
& Cybersecurity in networks and digital systems. 
& \ding{115} Limited 
&  \ding{55} None \\
\hline
~\cite{srivastava2022xai}   
& Takes a sectoral approach; highlights practical deployments of \gls{xai} at the research–industry interface. 
& Industry 4.0, smart transportation, healthcare, agriculture, governance, 5G, and finance. 
& \ding{115} Limited 
& \ding{115} Limited 
\\ 
\hline
~\cite{charmet2022explainable}    
& Organizes their review around key security properties:
fairness, integrity, privacy, confidentiality, and robustness, covering both the application of \gls{xai} to cybersecurity (e.g., \gls{ids}, malware) and the security of \gls{xai} itself (e.g., adversarial attacks and countermeasures). 
& Presents case study of fraud detection in online gambling. 
& \ding{115} Limited
&   \ding{55} None\\
\hline
~\cite{Moustafa2023explainable}  
& Reviews \gls{xai} for \gls{ids} in IoT environments. Discussion of the strengths and limitations of various \gls{xai} methods is particularly
informative.  
& IoT networks; categorizing research by deployment context(mobile, cloud, network, host-based), automotive.
& \ding{115} Limited 
& \ding{55} None\\
\hline
~\cite{alketbi2025comprehensive}   
& Focuses on insider threat detection.
& Malicious insider threats as a persistent and formidable challenge for organizations and Smart Healthcare Systems (SHS). 
& \ding{115} Limited 
& \ding{115} Limited 
\\ 
\hline
~\cite{saqib2024comprehensive} 
& Centers their review on malware hunting and offer a valuable comparative analysis of the time complexity of different \gls{xai} algorithms.
& Malware detection for critical infrastructure. 
& \ding{55} None
& \ding{115} Limited
\\ \hline
~\cite{sharma_2025}   
& Addresses challenges with AI blackbox models. 
& Smart healthcare, smart transportation, smart cities, smart agriculture and smart energy. 
& \ding{117} Moderate 
& \ding{115} Limited 
\\ 
\hline
This  work 
& 
Deeply explores  the challenges and opportunities associated with applying \gls{xai} in industrial security. 
& Industrial SOCs
& \ding{51} Comprehensive 
& \ding{51} Comprehensive 
\\ 
\hline
\end{tabularx}
\end{table*}
\renewcommand{\arraystretch}{1}

\begin{figure*}[t!] 
    \centering
    \includegraphics[width=0.75\textwidth,keepaspectratio]{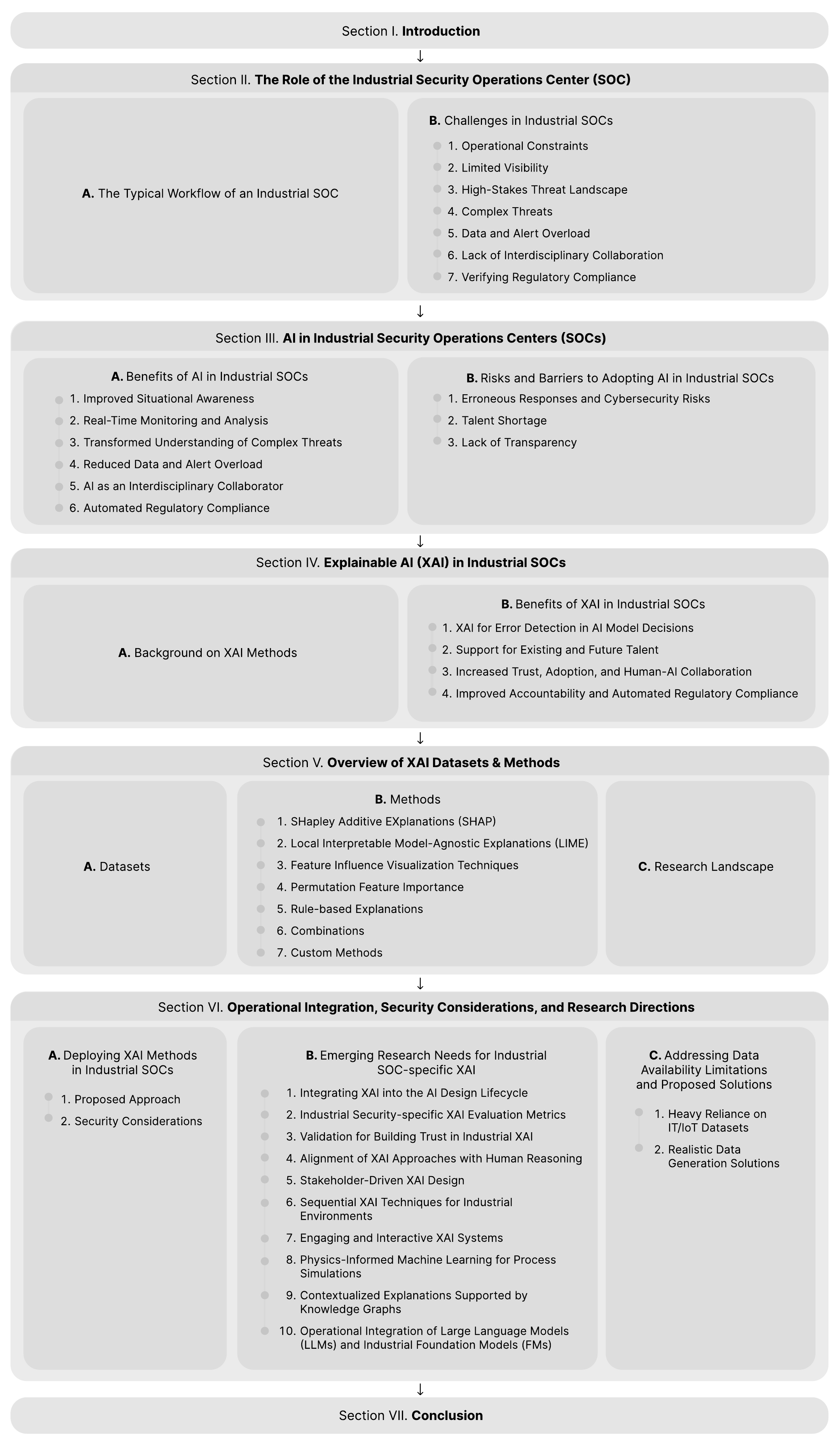}
    \caption{The outline of this survey.}
    \label{fig:fullpage}
\end{figure*}

As illustrated in Fig.~\ref{fig:fullpage}, the remainder of this paper is organized as follows. Section~\ref{sec:soc} examines the evolving role of industrial \glspl{soc}, highlighting how their workflows, operational priorities, and constraints differ from those of IT-focused \glspl{soc}. Section~\ref{sec:whyai} then discusses the benefits, risks, and operational challenges associated with adopting \gls{ai} in industrial cybersecurity environments. Section~\ref{sec:whyxai} introduces foundational concepts in \gls{xai}, including explanation taxonomies, evaluation considerations, and the role of explainability in industrial security operations. Section~\ref{sec:review} presents a detailed review of existing \gls{xai} datasets, methods, and applications in industrial cybersecurity. Section~\ref{sec:discussion} synthesizes insights from the literature, discusses operational integration challenges, security and regulatory considerations, and outlines emerging research directions for industrial \gls{xai}. Finally, Section~\ref{sec:conclusion} concludes the paper.

\section{The Role of the Industrial Security Operations Center}
\label{sec:soc}

While this survey is not limited to organizations with a dedicated \gls{soc}, we begin by discussing the roles, tasks, and responsibilities typically associated with an industrial \gls{soc}. This perspective is important because industrial security functions are often distributed across multiple teams, even in organizations without a formally established industrial \gls{soc}. In such cases, responsibilities may be shared among engineering, \gls{it}, and/or operations teams, with each team monitoring and responding to threats within their respective domains. 
Thus, the activities performed by individuals across these teams collectively reflect the functional roles of a \gls{soc}. Whether formally established or functionally distributed, the industrial \gls{soc} concept provides a useful lens for examining industrial security operations. Therefore, in this section, we examine the typical workflows, responsibilities, and challenges encountered within industrial security from the perspective of a \gls{soc}, providing a foundation for understanding how and where \gls{ai} and \gls{xai} can meaningfully augment security operations.

A \gls{soc} is a centralized unit that incorporates people, processes, technologies, and governance to identify, detect, and mitigate cyber threats~\cite{vielberth2020security}. \glspl{soc} have been a staple of enterprise \gls{it} environments, where their primary focus is on securing digital assets such as data, applications, and networks. Conventional \gls{it} \glspl{soc} are built around the principles of the C-I-A triad: confidentiality, integrity, and availability.

In contrast, \gls{ot} environments prioritize availability and safety above all. At the same time, the landscape of industrial cybersecurity is undergoing significant transformation. The historical isolation of \glspl{ics} through air-gapping is being eroded by increasing digitalization, remote accessibility, and \gls{it}/\gls{ot} convergence~\cite{singh_artificial_2020}, which introduces a broader and more complex threat landscape. Further, the consequences of a cybersecurity incident in \gls{ot} can be far more severe than in IT. Disruptions in industrial systems can lead to equipment failure, production downtime, environmental damage, or even threats to human life. As \gls{ot} networks become increasingly interconnected with \gls{it} systems and external networks, the need for specialized \gls{ot} \glspl{soc} capable of addressing the unique requirements of industrial security has become critical.

\subsection{The Typical Workflow of an Industrial SOC}

The workflow of an \gls{ot} \gls{soc} can be broadly categorized into three phases: {preparation}, {real-time operations}, and {post-incident activities}. 
The preparation phase includes proactive activities such as asset inventory management, behavioral baselining, and integration of OT-specific data sources, which lay the groundwork for effective monitoring and threat detection. The real-time phase involves continuous data collection, anomaly detection, alert triage, and incident response, typically under strict timing and safety constraints. The post-incident phase enables reflection, improvement, and compliance through root cause analysis, reporting and documentation, and iterative enhancements to detection capabilities and operational readiness. Insights gained during post-incident activities can further refine future preparation activities, reinforcing the continuous and adaptive nature of industrial security operations.

Table~\ref{tab:soc_workflow_phases} briefly outlines the core tasks within each phase.

\begin{table*}[!t]
\centering
\caption{Key tasks in each OT/ICS \gls{soc} workflow phase}
\begin{tabular}{>{\raggedright\arraybackslash}p{0.1\textwidth}  p{0.25\textwidth}  p{0.55\textwidth}}
\toprule
\textbf{Phase} & \textbf{Key Task} & \textbf{Description} \\
\midrule

\multirow{6}{*}{\textbf{Preparation}} 
& Asset and network mapping 
& Identify and document all assets (hardware and software), network topology, and communication protocols. \\
\cmidrule{2-3}
& Integration of \gls{ot} data sources and protocols 
& Ensure visibility into industrial protocols and legacy systems not natively supported by \gls{it} security tools. \\
\cmidrule{2-3}

& Vulnerability assessment and patch planning 
& Identify known vulnerabilities and schedule risk-informed patching prioritizing minimal operational disruption. \\
\cmidrule{2-3}

& Baseline behavior modeling 
& Establish normal operational and communication patterns to support anomaly detection. \\
\cmidrule{2-3}

& Configuration and policy management 
& Define secure configurations, access controls, and firewall rules for industrial environments. \\
\cmidrule{2-3}

& Playbook development and \gls{ir} plan testing 
& Create and simulate incident response workflows tailored for OT-specific security incidents. \\

\midrule

\multirow{7}{*}{\parbox{0.15\textwidth}{\raggedright\textbf{Real-Time\\Operations}}}

& Data collection and normalization 
& Gather and standardize logs, telemetry, and traffic from \gls{ot}/\gls{it} systems for central analysis. \\
\cmidrule{2-3}
& Threat detection 
& Identify suspicious behavior to detect potential security incidents. \\
\cmidrule{2-3}

& Alert triage, correlation, and prioritization 
& Filter noise, group related alerts, correlate alerts with threat intelligence, and triage incidents according to their urgency. \\
\cmidrule{2-3}

& Threat hunting 
& Conduct proactive searches for hidden threats based on hypotheses or weak signals in the data. \\
\cmidrule{2-3}

& Incident investigation and escalation 
& Analyze incidents to determine scope, origin, and potential impact; escalate to a higher \gls{soc} level if and when necessary. \\
\cmidrule{2-3}

& Response and mitigation 
& Implement containment and mitigation actions, often in coordination with plant engineers or safety operations teams. \\
\cmidrule{2-3}

& Collaboration with engineering teams 
& Ensure cross-functional communication between \gls{soc} analysts and other relevant teams, including engineers, to avoid unintended disruptions. \\

\midrule

\multirow{5}{*}{\textbf{Post-Incident}} 
& Post-incident review and root cause analysis 
& Document lessons learned and identify contributing factors for long-term improvement. \\
\cmidrule{2-3}

& Rule updates 
& Incorporate new \gls{ioc} and refine detection logic based on recent attacks. An \gls{ioc} is digital forensic evidence suggesting a system, network, or device may have been breached. \\
\cmidrule{2-3}

& Workflow refinement 
& Update playbooks, workflows, and team responsibilities based on incident outcomes. \\
\cmidrule{2-3}

& Training 
& Conduct regular tabletop or live-fire drills to enhance readiness and team coordination. \\
\cmidrule{2-3}

& Compliance reporting and audit documentation 
& Generate evidence and records required for regulatory compliance. \\

\bottomrule
\end{tabular}
\label{tab:soc_workflow_phases}
\end{table*}

\subsection{Challenges in Industrial SOCs}
\glspl{soc} in industrial environments face several domain-specific challenges:

\subsubsection{Operational Constraints} 
As previously stated, \gls{ot} environments prioritize system availability and operational safety. \gls{ot} environments function under strict operational constraints to ensure the uninterrupted operation of critical processes. Consequently, system shutdowns or reboots must be scheduled with significant lead time and coordinated planning~\cite{singh_artificial_2020}. This constrains the rapid deployment of software updates and security patches in contrast to \gls{it} environments where routine maintenance and updates can typically be performed with minimal impact on operations. As a result, \gls{ot} systems with unpatched software can remain exposed to known vulnerabilities for extended periods, increasing their susceptibility to cyber threats~\cite{idaho2016cyber}.

Additionally, threat response within \gls{ot} environments must be executed with caution to avoid interrupting essential services. 
Even during active cyber incidents, \gls{ot} systems might be kept operational, with plans to maintain the ability to operate \gls{ot} systems manually, while the cyber incident is handled in parallel~\cite{cisa2025}. 
Automated threat responses in \gls{ot} environments must be carefully scrutinized to avoid the risks of operation interruptions~\cite{stouffer_guide_2023}. Interruptions in industrial control systems may result in production loss, equipment damage, or regulatory violations. 

Furthermore, security monitoring in \gls{ot} must employ passive techniques to prevent interference with time-critical functions. This operational constraint limits the applicability of standard \gls{it} security tools, such as real-time or active antivirus or network scanners, which may degrade performance or introduce unpredictable behavior~\cite{falco2006using, stouffer_guide_2023}. Accordingly, the use of such tools in \gls{ot} environments is limited or absent. 

\subsubsection{Limited Visibility} 
A major challenge at the forefront of \gls{ot} security is the lack of comprehensive visibility within \gls{ot} environments, defined as the ability to identify, monitor, and collect data on all connected \gls{ot} assets, network traffic, and security events~\cite{dragos_why_ot_visibility}. Effective visibility is critical for managing asset inventories, identifying vulnerabilities, detecting intrusions, maintaining regulatory compliance, and supporting effective incident response and operational awareness.

This limitation is primarily due to industrial networks typically comprising a heterogeneous array of devices from various vendors, including legacy equipment that lacks modern security capabilities, vendor-specific operating systems, and proprietary communication protocols such as Siemens S7comm (S7 Communication)~\cite{s7comm}. \gls{ot} devices and their operating systems are often resource-constrained (such as PLCs and RTUs), lack built-in security features, and typically do not integrate with modern monitoring and security platforms. 
Many existing devices lack the basic logging functionality needed to enable security monitoring~\cite{skaronis2024cybersecurity}.

Furthermore, they lack the computational capacity to run complex attack detection models. In addition, the use of vendor-specific protocols makes it difficult to apply standard network monitoring tools, significantly impeding full visibility.

\subsubsection{High-Stakes Threat Landscape} 
Cyberattacks targeting industrial systems can result in environmental harm and economic loss, and endanger human safety. For operators of industrial systems, these incidents may also lead to equipment damage, regulatory non-compliance penalties, criminal liability, and significant reputational harm to the organization~\cite{singh_artificial_2020}.
Given the severity of these potential consequences, industrial environments face increasingly frequent and sophisticated cyber threats originating from a wide spectrum of well-resourced adversaries, including organized crime groups, terrorist organizations, hacktivists, anarchists, corporate competitors, and nation-states~\cite{idaho2016cyber, singh_artificial_2020}. 
The diverse nature of these threat actors has significantly heightened the risk landscape for industrial systems.

In response, regulatory requirements for industrial cybersecurity have become more stringent and continue to evolve to address emerging cyber threats. However, these regulations introduce substantial compliance costs to organizations, including workforce training, capital investments in equipment upgrades, strategic security planning, and regular auditing, as well as the risk of severe penalties for non-compliance. For instance, under the \gls{nerc} \gls{cip} framework for electric utilities in North America, organizations can face fines of up to one million U.S. dollars per day for each reliability standard violation~\cite{nerc2014sanctions}.

\subsubsection{Complex Threats} 
Threat actors targeting industrial environments are becoming increasingly sophisticated, employing techniques such as \glspl{apt}, zero-day exploits, and polymorphic malware. Insider threats also continue to pose significant risks~\cite{yuan2021deep}. These attack vectors are often difficult to detect using conventional expert systems or rule-based and signature-based detection methods.

Historically, security has relied on custom-written rules, such as signatures, or manually defined heuristics~\cite{saxe2018malware, sarker2020cybersecurity}. These approaches demand substantial manual effort to remain effective amid a rapidly evolving threat landscape. One of their most persistent limitations is the inability to detect novel or previously unseen attacks: although efficient at identifying known threats, such as documented \glspl{cve}, they depend heavily on the presence of easily identifiable\glsreset{ioc} \glspl{ioc}~\cite{nozomi_howAIisUsed}. As adversarial techniques grow more dynamic, there is an increasing need for defensive systems that can learn and adapt to emerging threat patterns.

\subsubsection{Data and Alert Overload}
A \gls{soc} processes an extensive volume of security alerts on a continuous basis. For each alert, a \gls{soc} analyst must determine whether it constitutes a true positive; and if so, the analyst must assess the alert's origin, scope, and urgency, group related events, and correlate them with relevant threat intelligence. This workflow is highly time- and resource-consuming~\cite{sarker2020cybersecurity} and repetitive, with a large proportion of alerts ultimately being false positives.

\subsubsection{Lack of Interdisciplinary Collaboration} 
Effective industrial security requires coordination and collaboration between \gls{ot} professionals, such as control engineers who understand the plant operation, and \gls{it} staff, such as cybersecurity analysts who manage threat detection and response~\cite{singh_artificial_2020}. Bridging this gap is critical for ensuring both security and operational continuity.

As a result of these factors, integrating \gls{it} and \gls{ot} security into unified \gls{it}-\gls{ot} \glspl{soc}, designed with a clear understanding of operational priorities and technological constraints, has become a pressing need. According to the SANS Institute 2024 State of \gls{ics}/\gls{ot} Cybersecurity survey, nearly 30\% of respondents reported having already merged their \gls{it} and \gls{ot} \glspl{soc}, signaling a strong trend toward convergence between these traditionally separate domains~\cite{sans2024state}. This shift underscores the demand for purpose-built tools and procedures that can secure critical infrastructure without disrupting the industrial processes they are intended to protect.

Additionally, the cybersecurity workforce shortage is particularly acute in the industrial sector~\cite{bcg2024cybersecurity}. In addition to this shortfall, bridging the gap between \gls{ot} and \gls{it} teams requires specialized knowledge of \gls{ot} systems alongside cybersecurity expertise~\cite{singh_artificial_2020}. Hiring and retaining qualified security analysts remains both challenging and costly.

\subsubsection{Verifying Regulatory Compliance} 
Industrial organizations operating critical infrastructure are increasingly subject to cybersecurity regulations and standards that require continuous monitoring, threat detection, incident reporting, and operational visibility. For instance, the \gls{nerc} \gls{cip} standards for electric utilities in North America include provisions to ``monitor connections, devices, and communications'' in order to ``improve the probability of detecting anomalous or unauthorized network activity ... and facilitate improved response and recovery from an attack''~\cite{nerc_cip015_1}. Complementary \gls{nerc} reliability standards mandate ``real-time monitoring and analysis capabilities to support reliable System operations''~\cite{nerc_top010_1}. 

Similarly, the \gls{eu} \gls{nis2} directive mandates the ``monitoring and analysing [of] cyber threats, vulnerabilities and incidents,'' including ``real-time or near real-time monitoring of network and information systems'' within critical infrastructure sectors~\cite{nis2_2022}. 

\vspace{0.2cm}
Collectively, this section underscores the growing difficulty of securing modern industrial environments using traditional manual and rule-based approaches alone. The increasing interconnectedness, operational complexity, and volume of security-relevant data in industrial systems motivate the adoption of \gls{ai} and \gls{xai} to support scalable monitoring, threat detection, and decision-making within industrial \glspl{soc}.

\section{AI in Industrial Security Operations Centers}
\label{sec:whyai}

\subsection{Benefits of AI in Industrial SOCs}
Generally, \gls{ai} offers the potential to enable real-time visibility, improve detection accuracy, reduce human workload, and support compliance in increasingly dynamic and heterogeneous operational environments.

\subsubsection{Improved Situational Awareness}
Industrial environments often operate with fragmented operational visibility due to incomplete telemetry, limited monitoring coverage, and the complexity of converged \gls{it}/\gls{ot} infrastructures. \gls{ai} systems can assist by analyzing large-scale industrial security data to identify anomalous behavior, correlate security-relevant events, and infer emerging threats from incomplete or noisy observations~\cite{sarker2020cybersecurity}. These capabilities can improve situational awareness and help analysts better understand the security state of complex industrial environments.

\subsubsection{Real-Time Monitoring and Analysis}

Modern industrial environments generate massive volumes of operational and security telemetry that must be continuously processed under strict timing, reliability, and safety constraints. \gls{ai} systems can support scalable high-stakes real-time monitoring and analysis by automating telemetry processing, event correlation, and continuous security monitoring across complex \gls{it}/\gls{ot} infrastructures.

\subsubsection{Transformed Understanding of Complex Threats}

Data-driven \gls{ai} introduces a transformative shift in security operations. Machine learning algorithms can be trained to extract insights from historical incident data, enabling the detection and prevention of evolving threats. For example, \gls{ai} can be used to identify malware, detect suspicious trends, or derive policy rules~\cite{sarker2020cybersecurity}. Moreover, \gls{ai} systems can learn the complex characteristics of normal industrial process behavior and detect deviations indicative of advanced threats, even in the absence of predefined signatures. These systems are also capable of identifying subtle patterns across heterogeneous data streams that may signal emerging security risks.

\subsubsection{Reduced Data and Alert Overload}

\gls{soc} teams experience notably high rates of burnout, driven by heavy workloads, limited staffing, inadequate automation, and pervasive alert fatigue~\cite{nozomi_howAIisUsed}. Overwhelming security alert generation can also lead to self-inflicted \gls{dos}, where the resources consumed by the security system impede effective security.

The imperative to rapidly analyze and correlate large volumes of data from diverse sources presents a compelling use case for the integration of data-driven \gls{ai}. In industrial networks, the sheer volume of communications and process variable data exceeds the capacity of human analysts to evaluate manually, especially within time frames necessary to mitigate serious threats~\cite{nozomi_howAIisUsed}. \gls{ai} technologies enable real-time processing and correlation of massive, heterogeneous data sets to detect anomalies, behavioral deviations, and emerging threats as they occur.

Additionally, \gls{ai} can automate the prioritization of alerts based on contextual relevance, historical trends, and dynamic risk scoring~\cite{oakley_cox_three_nodate}. This can contribute to significant improvements in key \gls{soc} performance metrics, including reduced false positive rates, shorter average analysis times, and lower mean times to detect security incidents~\cite{vielberth2020security}.

\subsubsection{AI as an  Interdisciplinary Collaborator}
\gls{ai} is helping to mitigate this limitation by automating routine tasks and enabling less experienced personnel to manage more complex security operations~\cite{angelo_ai_2024}. \gls{ai}-augmented decision-making can assist junior analysts, reduce the learning curve, and enhance operational resilience despite workforce constraints. 
Looking ahead, some believe \gls{ai} has the potential to reduce dependence on Tier 1 \gls{soc} analysts~\cite{nozomi_howAIisUsed}, while others contend that AI will fundamentally transform and elevate \gls{soc} analysts into strategic decision-makers while addressing workforce shortages. Tier 1 \gls{soc} analysts shift from initial alert triage to supervising AI-driven triage and handling edge cases, Tier 2 analysts move from reactive correlation work to proactive threat hunting with greater context, and Tier 3 analysts evolve toward guiding AI investigations, validating findings, and extracting high-level insights driving long-term SOC maturity~\cite{prophet_howAITransforms}.

\subsubsection{Automated Regulatory Compliance}
Industrial organizations are increasingly subject to stringent cybersecurity regulations and standards. Compliance with these frameworks requires continuous monitoring, logging, risk assessment, and incident documentation. \gls{ai} systems can automate log analysis and audit preparation, directly contributing to compliance workflows. 

Automated solutions for network baselining are essential to support effective anomaly detection. Given that normal behavior varies significantly across different \gls{ot} environments, manual monitoring is often impractical or insufficient. Scalable compliance therefore requires continuous internal monitoring powered by \gls{ai}.

The IEC/ISA 62443-aligned Cybersecurity Management System (CSMS) and the TSA Pipeline Security Guidelines also establish continuous monitoring requirements, which can be significantly enhanced through the application of \gls{ai}.

Notably, the \gls{eu} \gls{nis2} directive encourages member states to invest in ``the use [of] artificial intelligence [to] improve the detection and prevention of cyberattacks,'' including ``research and development [of] automated or semi-automated tools in cybersecurity''~\cite{nis2_2022}.

\gls{ai} can support security compliance through automation in the above security requirements. For example, the Dynatrace 2025 \textit{State of Observability} survey~\cite{dynatrace2025observability} reported that 98\% of surveyed security leaders are currently leveraging \gls{ai} to support compliance efforts.

\subsection{Risks and Barriers to Adopting \gls{ai} in Industrial SOCs}
\subsubsection{Erroneous Responses and Cybersecurity Risks}

While integrating \gls{ai} into \gls{ot} security offers significant benefits, it also introduces new risks. Poorly designed or inadequately implemented \gls{ai} systems can generate excessive false positives, overwhelming security teams and diverting attention from genuine threats, ultimately creating security vulnerabilities~\cite{angelo_ai_2024}. Erroneous responses, whether automated or mistakenly approved by human analysts, can disrupt operations. For example, an \gls{ai}-driven action might inadvertently block a legitimate engineering or safety procedure initiated by operators or programmable logic controllers.

During early stages of data-driven \gls{ai} adoption in \gls{ot}, before model accuracy, infrastructure maturity, and system alignment are fully established, \gls{ai} may inadvertently increase system risk. This can result in false positives, poor decision-making, operational delays, or even unnecessary safety hazards~\cite{dean_parsons_icsot_nodate}. Gartner has predicted that up to 85\% of AI projects may deliver erroneous outcomes due to bias in data, algorithms, or development processes~\cite{gartner2018ciosAI, Zolanvari2023trust}.

Additionally, leaked \gls{ai} model algorithms or training data could provide adversaries with new avenues to disrupt operations. It is therefore critical to ensure the confidentiality, integrity, and security of \gls{ai} models, algorithms, and data~\cite{dean_parsons_icsot_nodate}.

At the same time, \gls{ai}-enabled cyberattacks are becoming more prevalent. Threat actors are increasingly leveraging \gls{ai} to conduct sophisticated phishing campaigns, generate deepfakes, develop and modify malware, and launch \gls{ddos} attacks~\cite{shashkov2023adversarial, ibrar2025generative}. These capabilities underscore the dual-use nature of \gls{ai} in cybersecurity, both as a defensive tool and as an enabler of advanced threats.

\subsubsection{Talent Shortage}

The 2024 \textit{ISC2 Cybersecurity Workforce} Study identified \gls{ai} as one of the top five emerging security skills~\cite{isc2_2024report}. However, the global shortage of cybersecurity professionals~\cite{weforum2024cybersecurity}, coupled with a lack of specialized \gls{ai} expertise, poses a significant barrier to the widespread implementation of \gls{ai} in security operations~\cite{ibm2025isc2}.

In industrial contexts, the technical complexity of deploying and managing \gls{ai} within ICS/\gls{ot} environments further compounds this challenge. Successful integration requires domain-specific knowledge spanning control systems, operational constraints, and \gls{ai} model behavior. 
These demands can be difficult to meet in organizations facing underinvestment, limited staffing, or competing priorities within engineering and \gls{ot} security teams, where other initiatives may offer more immediate returns on investment than \gls{ai} adoption~\cite{dean_parsons_icsot_nodate}.

\subsubsection{Lack of Transparency}
While \gls{ai} holds significant promise for enhancing industrial cybersecurity, its effective adoption hinges on the ability of human operators, engineers, and decision-makers to understand, trust, and act upon \gls{ai}-driven decisions. 
In high-stakes industrial environments, where safety, reliability, and accountability are paramount, this lack of transparency can be a critical barrier to adoption. \gls{xai} seeks to address this challenge by providing human-interpretable justifications for \gls{ai} decisions, enabling responsible, trustworthy, and auditable use of \gls{ai} in security operations. 

\section{Explainable AI (XAI) in Industrial SOCs}
\label{sec:whyxai}

As succinctly outlined in~\cite{adadi2018peeking, spartalis2023balancing}, the objectives of \gls{xai} can be broadly categorized into four key functions: (1) providing justification for the decision-making processes of \gls{ai} systems, (2) extracting knowledge and insight by revealing data correlations and patterns embedded in the \gls{ai} system's learned strategies, (3) improving \gls{ai} systems by exposing biases and errors, and (4) supporting accountability and ongoing system maintenance. 
In the following section, we examine these objectives through the lens of industrial security, underscoring the imperative role of \gls{xai} in safeguarding and optimizing AI-driven operations within this domain.

\subsection{Background on XAI Methods}

\begin{figure*}[t!] 
    \centering
    \includegraphics[width=1\textwidth,keepaspectratio]{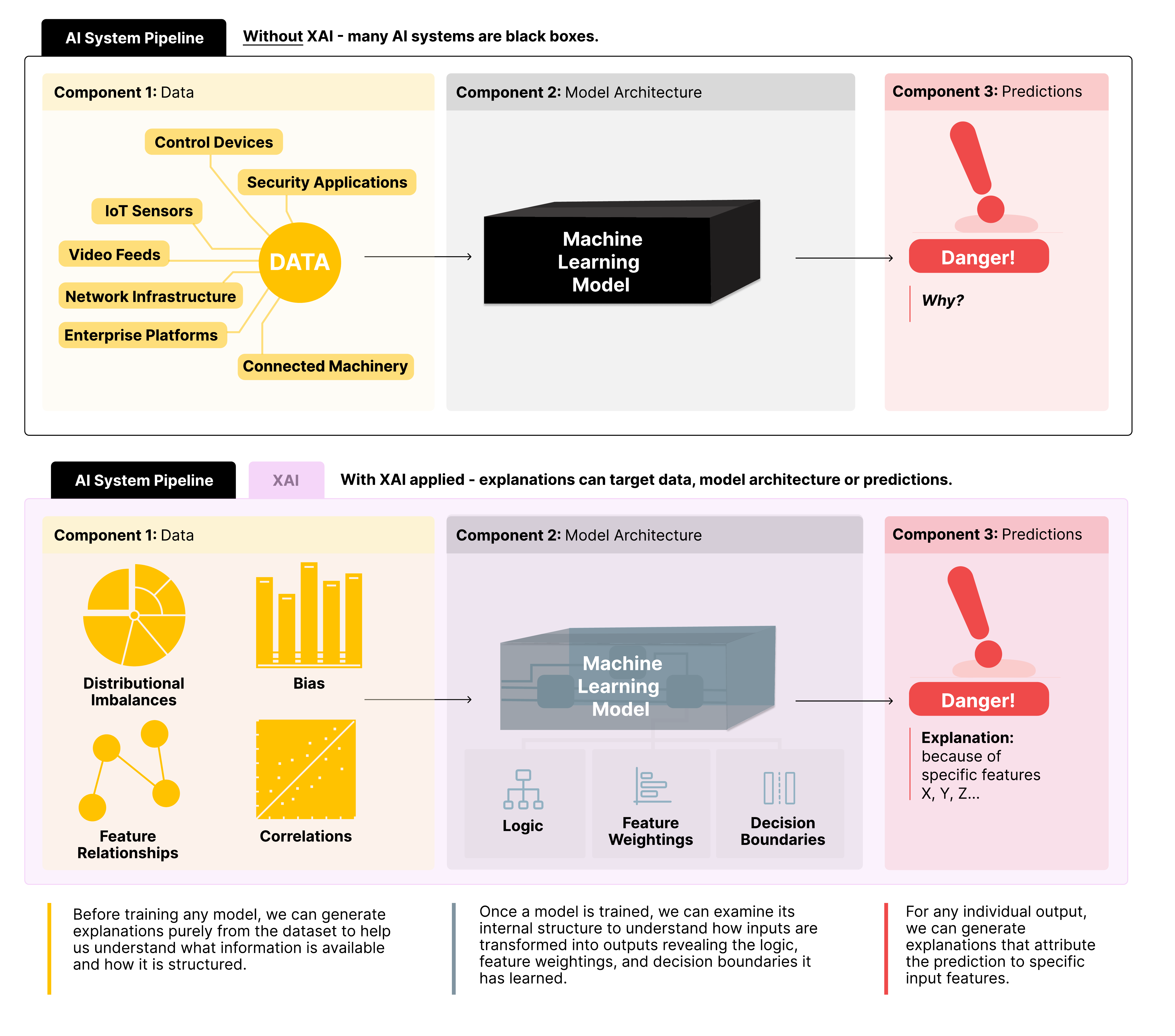}
    \caption{Procedural \gls{xai} tools aim to extract, approximate, or visualize model behavior in order to generate human-understandable insights into the \gls{ai} system's internal logic.}
    \label{fig:overview_xai}
\end{figure*}

\begin{figure*}[t!]
    \centering
    \includegraphics[width=1.0\linewidth]{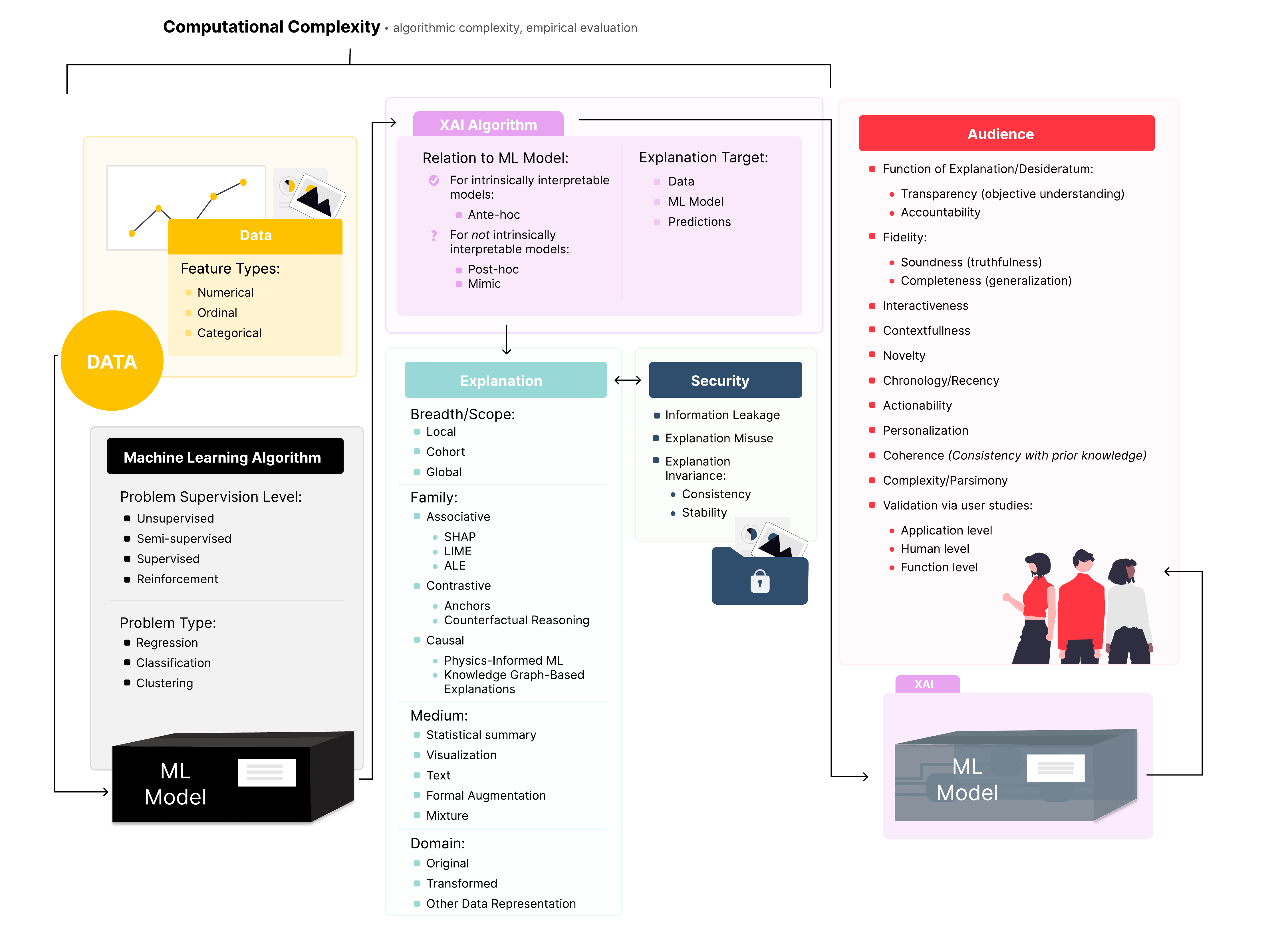}
    \caption{\gls{xai} Taxonomy}
    \label{fig:xai_taxonomy}
\end{figure*}

As noted by Adadi \textit{et al.}~\cite{adadi2018peeking}, \gls{xai} is not a formally defined concept, but rather a broader movement aimed at promoting transparency and trust in \gls{ai} systems. Consequently, there remains no unified definition of what constitutes an explanation under \gls{xai}, and the criteria for what makes an explanation effective are still active areas of research~\cite{confalonieri2021historical}. Gunning \textit{et al.}~\cite{gunning2019xai} emphasize that explanations are inherently contextual—shaped by the task at hand, the expectations and capabilities of the user, and the specific requirements of the application domain. Within efforts to formalize the notion of an explanation, Gilpin \textit{et al.}~\cite{gilpin2018explaining} broadly define an explanation as an interface between humans and AI systems.
Building on this perspective, and with a focus on industrial security, we define explainability as the process through which an \gls{ai} system's behavior can be clarified or revealed, either by leveraging the inherent transparency of the system itself or by applying external algorithms, methods, or techniques collectively referred to as \gls{xai} tools. For example, in the context of \gls{ai}-driven security alerts, an explanation should clarify the internal logic that led the system to flag a particular security event, thereby supporting human interpretation, validation, and response (see Fig.~\ref{fig:overview_xai}). We note that the field uses several terms interchangeably: transparency, interpretability, and explainability. 
\gls{xai} is therefore essential for fostering trust, accountability, and regulatory compliance.
We synthesize a taxonomy for XAI from prior analyses~\cite{langer2021we, sokol2020explainability, bellucci2021towards}, which we summarize in Fig.~\ref{fig:xai_taxonomy}. 

The \gls{xai} process typically begins with a dataset and a \gls{ml} model or algorithm trained to solve a specific problem type, such as regression, classification, or clustering. Explanations may target different aspects of this pipeline: the data itself, the model architecture, or the model's decisions or predictions. While \gls{xai} is most commonly associated with explaining predictions, understanding the data can reveal important characteristics such as feature relationships, correlations, and distributional imbalances. Similarly, explaining the model can provide insight into how inputs are processed into outputs, including the structure and logic of the model.

When the model is transparent, such as decision trees, linear models or rule-based systems the explanation is considered ante-hoc (i.e., intrinsically interpretable), meaning it is built into the model itself. In contrast, post-hoc \gls{xai} methods are used to explain the behavior of complex, opaque models, such as deep neural networks, after training. 

Explanations also vary in scope. Global explanations aim to describe the overall behavior of a \gls{ml} model across its entire input space, offering insights into how the model makes decisions in general. Local explanations, by contrast, focus on the rationale behind a specific prediction, typically in the immediate vicinity of a single data point. Cohort-level explanations occupy a middle ground, providing interpretability for a subset of the input space, such as a cluster of similar instances, without generalizing across the entire model.

Another important axis of categorization distinguishes between associative, contrastive, and causal explanations. Associative explanations highlight statistical relationships between input features and \gls{ml} model outputs. These form the basis of many widely used \gls{xai} techniques, including feature attribution methods such as \gls{shap}, \gls{lime}, \gls{ale}, \gls{pdp}, and saliency maps. Contrastive explanations, on the other hand, seek to clarify why one outcome occurred instead of another. Techniques such as anchors and counterfactual reasoning are well-established in this category. Both associative and contrastive methods are particularly useful when domain knowledge is limited or unavailable. 

In cases where domain knowledge is embedded into the \gls{ml} model, often through strong assumptions or structured priors (discussed further in Section~\ref{sec:discussion}), explanations can support causal interpretability. These approaches go beyond correlation to identify cause-and-effect relationships that drive model decisions. Examples include \gls{piml} and knowledge graph-based reasoning, which incorporate physical system dynamics, domain constraints, and structured relationships into the learning and explanation process.

Another common approach involves the use of surrogate or mimic models: simpler, transparent models trained to approximate the behavior of a more complex, opaque model.

Given the trade-off between interpretability and predictive performance, it is increasingly common to deploy surrogate
models in parallel with high-performing black-box models; as noted by~\cite{Zolanvari2023trust} this hybrid approach allows organizations
to retain the accuracy of complex models while providing interpretable explanations for oversight, validation, and compliance purposes.

Explanations can be communicated through various modalities, including visualizations, textual descriptions, and statistical summaries. In high-dimensional settings, dimensionality reduction techniques can be employed to project explanations into more interpretable 2D or 3D spaces, enhancing user comprehension and accessibility.

The chosen \gls{xai} method and its resulting explanations must satisfy a range of practical and contextual requirements. For example, in high-pressure, high-throughput, or time-sensitive environments, typical in industrial security operations, explanations must be generated rapidly, placing constraints on the computational complexity and latency of the \gls{xai} technique.

Because explanations are inherently user-centric, they must be tailored to the intended audience, taking into account their domain expertise, cognitive load, and decision-making context. As outlined in~\cite{sokol2020explainability}, several usability dimensions influence the acceptance and effectiveness of \gls{xai} outputs, including:
\begin{itemize}
    \item \textit{Coherence}: consistency with the explainee's prior knowledge;
    \item \textit{Novelty}: ability to highlight non-obvious or informative aspects;
    \item \textit{Actionability}: usefulness in guiding decisions, such as mitigation or root cause analysis;
    \item \textit{Complexity and Parsimony}: selectiveness and succinctness of the explanation;
    \item \textit{Interactivity and Personalization}: adaptability of the explanation and user interface to user needs and preferences.
\end{itemize}
\noindent Given the subjective nature of these criteria, researchers have proposed several objective metrics to evaluate and compare \gls{xai} methods. Table~\ref{tab:xai_metrics} outlines a selection of these metrics, including fidelity, stability, comprehensibility, and robustness, as they are relevant to the surveyed literature in Section~\ref{sec:review}.

In addition to usability and performance, explanations should be aligned with well-defined user design requirements. Some important properties include~\cite{langer2021we}:
\begin{itemize}
    \item \textit{Transparency}: enabling objective understanding of model behavior;
    \item \textit{Accountability}: providing mechanisms to determine responsibility for decisions;
    \item \textit{Debug-ability}: supporting the identification and correction of errors or misbehaviors.
\end{itemize}
These guiding principles are especially relevant in industrial and regulatory contexts, where multiple stakeholders with distinct desiderata, such as developers, engineers, analysts, auditors, and regulators, may rely on \gls{xai} outputs for oversight, validation, and operational decision-making.

\begin{table}
    \caption{XAI objective evaluation metrics~\cite{kadir2023evaluation, sokol2020explainability, coroama2022evaluation}}
    \centering
    \begin{tabularx}{\linewidth}{>{\raggedright\arraybackslash}p{0.2\linewidth} X}
        \toprule
        \textbf{Metric} & \textbf{Description}\\
        \midrule
        Soundness & Measures how truthful an explanation generated by a post-hoc or mimic explanation method is to the underlying predictive model. High soundness is desirable. \\
        \midrule
        Completeness & Measures how well an explanation generalizes and to what extent it covers the underlying predictive model. High completeness is desirable. \\
        \midrule
        Faithfulness & Tests whether features identified as important by the explanation actually influence the model's prediction. Typically computed by removing or masking the top-$k$ important features and observing the model's prediction change. A large deviation suggests high faithfulness; small deviation implies the explanation may not reflect actual model behavior. High faithfulness is desired. \\
        \midrule
        Complexity (or Simplicity) & Measures how easy the explanation is to understand, typically by quantifying its size or structural complexity. For example, by counting the number of nodes or tree depth in a decision tree, number of rules and average rule length in a rule-based method, or number of non-zero or top-ranked features in feature attribution. Lower complexity is desirable. \\
        \midrule
        Monotonicity & Ensures that increasing an input feature value should not decrease the model's decision output if the model is expected to behave monotonically with respect to that feature.\\
        \midrule
        Stability (Robustness to Perturbations) or Sensitivity & Measures whether small changes to input yield similar explanations. High stability and low sensitivity are desired. \\
        \bottomrule
    \end{tabularx}
    
    \label{tab:xai_metrics}
\end{table}

\subsection{Benefits of \gls{xai} in Industrial SOCs}
\subsubsection{XAI for Error Detection in AI Model Decisions}

Like all data-driven systems, \gls{ai} models are susceptible to misclassification errors arising from bias~\cite{roselli2019managing, schwartz2022towards, bove2022contextualization}, operational drift~\cite{sahiner2023data}, and adversarial manipulation~\cite{qiu2019review, bove2022contextualization}. In industrial environments, such errors may have operational or safety consequences, particularly when \gls{ai} systems are used to support real-time monitoring, anomaly detection, or automated response. Industrial data is also frequently noisy, incomplete, and subject to changing operational conditions, further increasing the risk of unreliable model behavior.

In the absence of transparency, it is difficult to detect when an \gls{ai} system is making systematically flawed decisions. \gls{xai} methods can surface the underlying logic and feature contributions behind model outputs, enabling stakeholders to identify and correct errors (for example, sources of bias~\cite{hofeditz2022applying}) to improve data quality and retrain models as needed.

Explainability also aids in managing the risks of false alerts: false negatives and false positives~\cite{da2023false}, both of which can be costly in industrial cybersecurity. False positives may lead to unnecessary operational disruption, while false negatives may allow critical threats to go undetected. \gls{xai} supports the refinement of detection strategies by revealing why such errors occur.

\subsubsection{Support for Existing and Future Talent}

Industrial \gls{soc} analysts often operate under significant pressure due to high alert volumes, limited staffing, and persistent alert fatigue. Raw security alerts frequently lack sufficient contextual information, forcing analysts to manually reconstruct attack activity and correlate events across multiple data sources, which can delay incident response.

Alert contextualization refers to the process of enriching raw security alerts with relevant, actionable information to help analysts better interpret and prioritize them. Rather than simply indicating that \textit{something suspicious happened}, contextualized alerts provide insight into why the alert matters, how severe the threat is, and what actions should be taken. This capability is especially critical for real-time \gls{soc} operations, supporting key tasks such as \textit{alert triage, correlation and prioritization}, \textit{incident investigation and escalation}, and \textit{post-incident review and root-cause analysis} (refer to Table~\ref{tab:soc_workflow_phases}). By offering explanatory depth, contextualization enables analysts to understand how an attack unfolded or why specific threats were detected or missed. Furthermore, the ability to generate structured explanations supports comprehensive incident documentation in alignment with security auditing standards such as \gls{nerc} \gls{cip}~\cite{nerc_cip015_1} and NIS2~\cite{nis2_2022}.

In parallel, the \gls{xai} research community has demonstrated the value of contextualizing explanations by incorporating domain knowledge, such as through Knowledge Graphs~\cite{lecue2020role, sarker2020wikipedia, bove2022contextualization} (more in Section~\ref{sec:discussion}) or by presenting relevant training examples~\cite{gomez2020vice, kim2016examples, bove2022contextualization}. For instance, Bove~\textit{et al.} conducted an experimental validation study showing that contextualized explanations significantly enhance users' objective understanding and satisfaction~\cite{bove2022contextualization}. From a social science perspective, contextualization is also recognized as a core component of effective explanation~\cite{miller2019explanation}.

Domain-grounded contextualization is paramount for security analysts. \gls{rnns} contextualize explanations of security-related events against the MITRE ATT\&CK Framework, a knowledge-based taxonomy of adversarial tactics and techniques. Further graph generation organizes security events into attack graphs that visualize attack sequences by attack phase. The benefit of applying a framework like MITRE ATT\&CK is that it provides security professionals a common taxonomy that they understand and use to discuss and share threat intelligence.

In the absence of contextual information, analysts are forced to manually investigate and reconstruct the relevant background, delaying response actions, increasing operational risk, and contributing to alert fatigue~\cite{Devry2025}. \gls{xai}, particularly when tailored to the needs of \glspl{soc}, offers a promising avenue for delivering the contextual insights required to support timely, informed, and confident decision-making in industrial cybersecurity.

\subsubsection{Increased Trust, Adoption, and Human-AI Collaboration}

Trust is a foundational prerequisite for the adoption of new technology; in this case, data-driven \gls{ai} in industrial environments~\cite{glikson2020human}. One of the most widely cited definitions of trust, proposed by Mayer \textit{et al.}~\cite{mayer1995integrative}, describes it as ``the willingness of a party to be vulnerable to the actions of another party based on the expectation that the other will perform a particular action important to the trustor, irrespective of the ability to monitor or control that other party.'' 
Within industrial \gls{soc} environments, analysts and operators are ultimately responsible for security outcomes, yet increasingly rely on \gls{ai}-driven recommendations to support threat detection, prioritization, and response decisions. These systems are typically integrated into human-in-the-loop or human-on-the-loop workflows, where personnel must interpret, validate, and act upon \gls{ai}-generated insights.

However, widespread adoption of machine learning models remains constrained by their limited ability to explain the rationale behind their predictions~\cite{bove2022contextualization}. 

Supporting this, KPMG's global 2025 study on \textit{Trust, attitudes and use of artificial intelligence} found that over half of AI users are wary about trusting AI~\cite{kpmg2025ai}.

This distrust reflects the pervasive ``fear of the unknown" across organizational levels: end users fear job displacement, while management worries about AI system's reliability and robustness~\cite{siemens_whitepaperXAI}.

Importantly, trust must be calibrated. Low trust in capable systems can lead to disuse and inefficiencies, while excessive trust in flawed systems may result in misuse and safety breaches~\cite{hoff2015trust, glikson2020human}. Explainability plays a critical role in fostering balanced trust and effective human-\gls{ai} collaboration. By enabling analysts to interrogate, validate, or challenge \gls{ai} outputs, explainability enhances situational awareness and decision quality while addressing concerns about system construction, limitations, and data transparency. In the context of industrial security incidents, transparent reasoning can accelerate response times and improve mitigation outcomes, ultimately boosting \gls{soc} performance. Sustaining this trust requires traceability and continuous feedback loops that enable iterative refinement of AI systems based on analyst insights.

Overall, research consistently shows that explainability is a key driver of trust. Users exhibit greater confidence in \gls{ai} systems when the underlying algorithms are transparent and their decisions are interpretable~\cite{shin2021effects, choung2023trust}.

\subsubsection{Improved Accountability and Automated Regulatory Compliance}

In recent years, governmental bodies (e.g.,~\cite{oecd_trustAI, eu_trustAI}), technology companies (e.g.,~\cite{microsoft_trustAI}), and professional associations (e.g.,~\cite{ieee_trustAI}) have introduced frameworks aimed at embedding trustworthiness into \gls{ai} systems~\cite{choung2023trust}.

For example, the \textit{\gls{oecd} AI Principles} include provisions on transparency and explainability, advocating for ``[transparent] and responsible disclosure around AI systems to ensure that people understand when they are engaging with them and can challenge outcomes,'' as well as accountability, which holds ``individuals developing, deploying or operating AI systems ... accountable for their proper functioning in line with the OECD's values-based principles for AI''~\cite{oecd_trustAI}. Voluntary since their release in 2019, the OECD principles have gained significant traction, with the OECD reporting over 1000 national policy initiatives aligned with these guidelines~\cite{oecd_trustAI}.

Similarly, the \textit{\gls{ai} \gls{rmf}}, published by the \gls{us} \gls{nist} on January 26, 2023, outlines voluntary characteristics to guide the integration of trustworthiness into the design, development, deployment, and evaluation of AI systems~\cite{rmf_what}. 
These considerations are particularly relevant for industrial \glspl{soc}, where \gls{ai} systems increasingly support operational monitoring, incident prioritization, and security decision-making under safety-critical conditions.
Among the key attributes of trustworthy AI identified in the framework are explainability, interpretability, accountability, and transparency.

The \textit{\gls{eu} \gls{ai} Act} will become fully enforceable by August 2026~\cite{eu_aiAct_article113}. Under the Act, risk management is fundamental to AI system development and testing, requiring organizations to identify, estimate, and evaluate risks, then either accept them or implement appropriate countermeasures. However, organizations struggle to accurately categorize their industrial AI systems because the risk classification demands extensive background knowledge, familiarity with regulatory definitions, and understanding of intersecting regulations. This classification is critical, as it determines which compliance obligations apply to AI developers or deployers. To address this challenge, automated classification tools aligned with EU requirements can streamline system categorization. Concurrently, the EU is developing harmonized technical standards that translate legal obligations into concrete, implementable controls and verifiable evidence requirements across the industry~\cite{eu_harmonizedstd}.

Depending on the assigned risk classification, the EU AI Act includes specific provisions for a class of high-risk AI systems, stating that ``[with] regards [to] the management and operation of critical infrastructure, it is appropriate to classify as high-risk the AI systems intended to be used as safety components in the management and operation of critical digital infrastructure''~\cite{eu_aiAct_recital55}. These provisions require that:
\begin{itemize}
    \item high-risk AI systems must be designed to be transparent, enabling users to understand and operate them correctly~\cite{eu_aiAct_article13};
    \item such systems must support effective human oversight~\cite{eu_aiAct_article14}; and
    \item accompanying instructions must explain how to interpret system outputs and maintain the system~\cite{eu_aiAct_article13}.
\end{itemize}

Although the act includes exclusions for components (e.g., \gls{ai} systems) used solely for cybersecurity purposes~\cite{eu_aiAct_recital55}, the act notes that these exclusions apply because such systems are generally intended to assist human operators rather than replace human decision-making~\cite{eu_aiAct_article6}. This distinction closely reflects current industrial \gls{soc} practice, where \gls{ai} systems are primarily deployed to augment analyst workflows, threat detection, and incident response activities. Supporting this, the Dynatrace 2025 \textit{State of Observability} survey reports that 99\% of respondents in \gls{ai} governance roles implement human-monitored validation measures, and that human users verify 69\% of \gls{ai}-powered decisions within their organizations~\cite{dynatrace2025observability}. As such, standalone \gls{ai} deployment in high-risk applications remains very low~\cite{Zolanvari2023trust}.

Nonetheless, the next wave of \gls{ai} adoption in security is expected to center on automation, where \gls{ai} may proactively eliminate threats before they escalate into breaches, particularly through modern \gls{soar} technologies~\cite{paloAlto_securityAutomation}. In such cases, where human oversight is reduced or absent, the \gls{eu} \gls{ai} Act's provisions for high-risk systems will apply, mandating explainability, transparency, and accountability in industrial security automation.

In China, the \textit{Action Plan for Global \gls{ai} Governance}, released on July 26, 2025, builds upon the earlier \textit{Global \gls{ai} Governance Initiative} introduced in October 2023~\cite{ansi2025chinaAIgovernance}. China's approach to AI governance is highly state-driven and regulation-intensive, with increasing emphasis on pre-deployment oversight, safety assessment, and standardization. This is evidenced by the reported removal of over 3,500 AI products from the market due to non-compliance with mandatory pre-deployment safety assessments for generative AI systems. Further highlighting the rapid expansion of this regulatory landscape, China issued more national \gls{ai} standards in the first half of 2025 alone than during the entire period from 2021 to 2024~\cite{tse_china_2025}. Among its governance proposals, a whitepaper on Trustworthy Artificial Intelligence outlines key principles of accountability, transparency, and explainability~\cite{caict2021_white, cset_chinaAI}.

Elsewhere, cybersecurity regulatory frameworks such as IEC 62443~\cite{iec_62443}, while not explicitly focused on \gls{ai}, impose expectations for accountability, traceability, auditability, and secure operational management. These requirements increasingly extend to \gls{ai}-enabled monitoring, detection, and response systems deployed within industrial \glspl{soc}.

Collectively, these frameworks position explainability as a foundational requirement for trustworthy industrial AI deployment. \gls{xai} offers the necessary mechanisms to demonstrate due diligence in \gls{ai}-assisted and \gls{ai}-augmented decision-making. By producing human-interpretable explanations, audit trails, and decision rationales, \gls{xai} can enable organizations to meet the demands of regulatory audits, internal oversight, and legal scrutiny. As per a 2023 industry survey that Siemens conducted with 36 key experts and decision-makers across various domains~\cite{siemens_whitepaperXAI}, \gls{xai} could emerge as a separate, dedicated component within formal certification processes, ensuring systematic evaluation of model transparency alongside performance metrics.
This is particularly critical in sectors involving critical infrastructure, where errors or misjudgments can have far-reaching operational and safety consequences.

\section{Overview of \gls{xai} Datasets and Methods}
\label{sec:review}

Data-driven \gls{ai} systems rely on high-quality, comprehensive datasets to function accurately. However, cybersecurity data remains scarce~\cite{cremer2022cyber}. Collecting such data poses significant challenges, particularly in industrial environments where legacy equipment and proprietary protocols limit visibility and interoperability~\cite{dean_parsons_icsot_nodate}. In addition, data from cyber incidents are inherently infrequent~\cite{okutan2018forecasting, zolanvari2018effect}, and concerns around privacy~\cite{khraisat2019survey} and data sensitivity often impede data sharing. For example, organizations that experience cyber incidents rarely disclose them publicly, due to concerns over reputation damage, loss of goodwill, or brand impact~\cite{eling2016we, cremer2022cyber, scala2019risk}.

\subsection{Datasets}

The majority of the research literature relies on \gls{ids} datasets to train and evaluate cybersecurity \gls{ml} models, and to develop \gls{xai} tools layered atop these models.

\gls{ids} datasets contain labeled instances of benign and malicious network traffic, with malicious traffic typically categorized into distinct attack types. For example, the NSL-KDD 2009 dataset~\cite{nslkdd}, widely used for benchmarking \gls{ids} models, organizes attacks into four primary categories: (1) \textit{\gls{dos}}: attacks that overwhelm system resources to render services unavailable to legitimate users; (2) \textit{U2R (User to Root)}: privilege escalation attacks where an attacker gains administrative access from a limited user account; (3) \textit{R2L (Remote to Local)}: attacks where an external actor attempts to gain unauthorized local access; and (4) \textit{Probe}: reconnaissance activities such as network scanning to gather information about hosts, ports, and services.
The dataset includes 23 specific attack labels across these categories.
Similarly, the UNSW-NB15 dataset~\cite{mostafa2015unswnb}, another popular benchmark, defines nine attack categories: Fuzzers, Analysis, Backdoors, Exploits, Generic, Reconnaissance, Shellcode, Worms, and \gls{dos}. \gls{ids} datasets typically include a rich set of features spanning packet-level (e.g., source/destination IP, port, protocol), flow-level (e.g., duration, byte count, packet count), and statistical metrics (e.g., mean packet size, inter-arrival time).

These datasets are generally synthetic, generated in controlled environments using tools such as TCPDump or Wireshark for traffic capture, and Snort or Zeek for traffic analysis. Synthetic generation enables partial automation of labeling, reducing the complexity, time, and error-prone nature of manual annotation. This labeling supports the use of supervised \gls{ml} techniques to train classifiers that learn attack patterns and categorize network traffic accordingly, forming the basis for experimentation with \gls{xai} methods applied to these trained models.

Many widely used \gls{ids} benchmark datasets are not fully OT/ICS-native: they may not cover industrial protocols (e.g., Modbus, DNP3, IEC 60870-5-104, OPC UA), may provide limited asset context (e.g., PLC/HMI/RTU roles), and often omit process-level variables such as setpoints, sensor readings, and actuator commands. In such cases, \gls{xai} methods can still provide faithful explanations with respect to the trained model and available features; however, the resulting insights might be less actionable for industrial defenders because they do not directly map to process behavior, safety implications, or engineering response decisions.

In contrast, a smaller subset of studies has explored anomaly detection using unsupervised \gls{ml} algorithms, particularly \gls{lstm}-based autoencoder architectures trained exclusively on normal data. Autoencoders learn to reconstruct input data by compressing it into a low-dimensional latent space and then decoding it. They become proficient at reconstructing normal operational patterns, but fail to do so for anomalous inputs. The resulting reconstruction error, i.e., the deviation between input signal and its reconstruction, serves as a signal for anomaly detection. Anomaly detection datasets typically comprise \gls{ot} timeseries data. These datasets remain scarce and are often generated within the same studies in which they are applied. Unlike \gls{ids} datasets, no anomaly detection dataset has yet attained widespread recognition or benchmarking status.

Table~\ref{tab:literature_by_dataset} summarizes the datasets employed across the surveyed literature. A comprehensive description and comparative analysis of most of these datasets is provided in the survey by~\cite{de2023survey}.

\renewcommand{\arraystretch}{1.2}
\begin{table}
    \caption{Literature survey: datasets}
    \centering
    \begin{tabularx}{\linewidth}{>{\raggedright\arraybackslash}p{0.1\linewidth}>{\raggedright\arraybackslash}p{0.1\linewidth}>{\raggedright\arraybackslash}p{0.3\linewidth} X}
        \toprule
        \textbf{Domain} & \textbf{Type} & \textbf{Dataset} & \textbf{References}\\
        \midrule
        \gls{ids} & Tabular & 
        NSL-KDD 2009~\cite{nslkdd} & \cite{Houda2022why, Zolanvari2023trust, Mahbooba_2021, Sharma_2024, Arreche2024xai1, arreche2024xai2, Sharma2023anomaly}\\
        \cline{3-4}
        & & UNSW-BW 2015~\cite{mostafa2015unswnb} & \cite{Houda2022why, Zolanvari2023trust, Kumar2023explainable, houda_2023, Sharma_2024}\\
        \cline{3-4}
        & & CIC-IDS 2017~\cite{sharafaldin2018toward} & \cite{Kumar2024blockchain, Patil2022explainable, Arreche2024xai1, arreche2024xai2}\\
        \cline{3-4}
        & & N-BaIoT 2018~\cite{meidan2018n} & \cite{Kalakoti2024improving, Kalakoti2023improving, Gummadi2024xai}\\
        \cline{3-4}
        & & ToN-IoT 2021~\cite{booij2021ton_iot} & \cite{Oseni2023explainable, Shtayat2023explainable}\\
        \cline{3-4}
        & & BOT-IoT 2019~\cite{koroniotis2019towards} & \cite{Kalakoti2024improving, Ben_2021}\\
        \cline{3-4}
        & & WUSTL-IIoT 2021~\cite{wustl2021} & \cite{Zolanvari2023trust}\\
        \cline{3-4}
        & & Survival 2018~\cite{han2018anomaly} & \cite{Lundberg2022experimental}\\
        \cline{3-4}
        & & MedBIoT 2022~\cite{guerra2020medbiot} & \cite{Kalakoti2024improving}\\
        \cline{3-4}
        & & NF-ToN-IoT-v2 2023~\cite{sarhan2023nf} & \cite{Gaitan-Cardenas2023explainable}\\
        \cline{3-4}
        & & WUSTL-EHMS-2020~\cite{hady2020intrusion} & \cite{Alani2023explainable}\\
        \cline{3-4}
        & & IoT-Flock~\cite{ghazanfar2020iot} & \cite{Gurbuz2023explainable}\\
        \cline{3-4}
        & & HVAC 2021 & \cite{Khan2022proactive}\\
        \cline{3-4}
        & & CIC-IoT 2023~\cite{neto2023ciciot2023} & \cite{Kaur_2024}\\
        \cline{3-4}
        & & RoEduNet-SIMARGL 2021~\cite{mihailescu2021proposition} & \cite{Arreche2024xai1, arreche2024xai2}\\
        \cline{2-4}
        & Sequential & 
        CIC-DDoS 2019~\cite{sharafaldin2019developing} & \cite{Javeed2024explainable}\\
        \cline{3-4}
        & & CIC-IDS 2017~\cite{sharafaldin2018toward} & \cite{Attique2024explainable}\\
        \cline{3-4}
        & & X-IIoTID 2021~\cite{al2021x} & \cite{Attique2024explainable}\\
        \cline{3-4}
        & & NSL-KDD 2009~\cite{nslkdd} & \cite{Keshk_2023, Sivamohan2023tea}\\
        \cline{3-4}
        & & UNSW-BW 2015~\cite{mostafa2015unswnb} & \cite{Keshk_2023}\\
        \cline{3-4}
        & & CSE-CIC-IDS 2018~\cite{sharafaldin2018toward} & \cite{Wali2025explainable, Sivamohan2023tea}\\
        \cline{3-4}
        & & CIC-IoT 2023~\cite{neto2023ciciot2023} & \cite{Wali2025explainable}\\
        \midrule
        Anomaly detection & Sequential & HAI 2020~\cite{haiDatasets} & \cite{Hwang2021e}\\
        \cline{3-4}
        & & Gas SCADA 2014 & \cite{Ha2022explainable}\\
        \bottomrule
    \end{tabularx}
    \label{tab:literature_by_dataset}
\end{table}
\renewcommand{\arraystretch}{1.0}

\subsection{Methods}
This section reviews the \gls{xai} methods most commonly applied in the surveyed industrial cybersecurity literature. While many studies evaluate these methods on \gls{ids} or IoT datasets, their relevance to industrial \glspl{soc} lies in how they can support alert triage, anomaly localization, model validation, analyst trust, and post-incident documentation. Accordingly, we discuss both the technical operation of each method and its potential operational value for industrial security workflows.

\subsubsection{SHapley Additive EXplanations (SHAP)}
Among surveyed literature, \gls{shap}~\cite{lundberg2017unified} emerged as the most widely applied \gls{xai} method, especially in intrusion detection. \gls{shap} is an associative, post-hoc, global, model-agnostic explanation technique grounded in cooperative game theory. It quantifies the contribution of each input feature to a model's prediction by assigning it a Shapley value, defined as the average marginal contribution of that feature across all possible feature subsets. Conceptually, \gls{shap} treats the model prediction as a collaborative game, where features act as players working together to produce the output. For each subset of features, the method evaluates how the prediction changes when a new feature (player) is added. The Shapley value is then computed as a weighted average of these changes across all subsets, weighted to ensure fairness and consistency across features. This approach provides a theoretically sound and equitable measure of feature importance, ensuring that features with equal influence receive equal attribution.

Although exact computation of Shapley values is computationally expensive, efficient approximations such as FastSHAP and TreeSHAP for tree-based models, can enable faster computations to support real-time explanations.

The \gls{shap} Python library offers a suite of visualization tools to support model interpretability. Research has commonly employed the following visualizations:
\begin{itemize}
    \item \textit{Force and Waterfall Plots}: These visualize the explanation of an individual prediction instance by showing how each feature pushes the model output away from a baseline value. In multi-class classification, the baseline represents the probability of predicting each class prior to observing any features. Each feature either increases or decreases the probability of a given class. Waterfall plots stretch features along the y-axis for improved readability, especially when many features are involved.

    \item \textit{Decision Plots}: These extend waterfall plots into line charts, allowing visualization of multiple prediction explanations simultaneously. The x-axis represents the evolving probability of a class prediction, starting from the baseline and incrementally adjusted by each feature's contribution.

    \item \textit{Bar (Summary) Plots}: These rank features by their mean absolute SHAP value across all instances, providing a global view of feature importance. While they do not indicate whether a feature increases or decreases the output, they are effective for identifying the most influential features overall.

    \item \textit{Beeswarm Plots}: These offer a richer visualization compared to bar plots by showing the distribution and directionality of SHAP values across all instances. Each point represents a SHAP value for a specific feature and instance, with color encoding the feature value. Beeswarm plots enable users to assess how feature values correlate with prediction probabilities and to identify impactful features with greater statistical nuance than bar plots.
\end{itemize}

In industrial \gls{soc} workflows, these visualizations can help analysts identify which network, device, or process variables contributed most strongly to an alert, supporting triage, escalation, and root-cause analysis.

Numerous studies have adopted \gls{shap} as the sole interpretability method for supervised learning models applied to tabular \gls{ids} datasets. These works span diverse model architectures and datasets, underscoring SHAP's versatility and accessibility in cybersecurity-focused \gls{xai} research.

Several studies utilized ensemble learning techniques such as \gls{rf}, XGBoost, and \gls{lgbm}, often paired with SHAP beeswarm, bar, or force plots. For instance, \cite{Siganos2023explainable} applied a \gls{rf} model for multi-class classification on the CIC-IoT 2022 and IEC 69870-5-104 datasets, visualizing global feature importance using SHAP bar plots and feature contributions locally using waterfall plots. Similarly, \cite{Ali2024explainable} trained XGBoost and \gls{rf} models on the CICIoT2023 dataset, employing beeswarm and force plots to highlight both global and local feature importance, respectively. \cite{Mallampati2024enhancing} used \gls{lgbm} for binary classification on UNSW-NB15 and CICIDS-2017, also leveraging beeswarm and force plots for interpretability.
A notable framework is presented by the authors in \cite{Wali2025explainable} who proposed an interpretable \gls{ids} integrating a lightweight \gls{rf} classifier with a SHAP-based Credibility Assessment Module (CAM). SHAP beeswarm plots provided global insights, while waterfall plots supported local decision analysis. To mitigate adversarial risks associated with increased interpretability, the authors introduced a secondary Transformer-based deep packet inspection module. The authors hypothesized that this dual-pipeline system, tested on CSE-CIC IDS 2018 and CIC-IoT 2023, operates on distinct feature spaces and architectures, reducing the likelihood of adversarial transferability.

Other studies explored diverse ensembles. \cite{Alani2023explainable} trained an ensemble of \gls{rf}, \gls{dt}, \gls{svm}, and XGBoost on the WUSTL-EHMS-2020 dataset, using SHAP beeswarm plots for global interpretability. Gaitán-Cárdenas et al. \cite{Gaitan-Cardenas2023explainable} applied \glspl{dt}, \gls{rf}, \gls{lr}, and \gls{dnn} to the NF-ToN-IoT-v2 dataset, explaining decisions with SHAP bar plots.
\cite{Hasan2023advanced} focused on detecting Advanced Persistent Threats (APTs) using boosting-based models including AdaBoost, Gradient Boosting, \gls{lgbm}, CatBoost, and XGBoost. SHAP force plots were used to interpret model decisions on the SCVIC-APT-2021 benchmark dataset, offering insights into subtle feature interactions characteristic of APT behavior.

SHAP has also been integrated into deep learning pipelines. \cite{Oseni2023explainable} trained 1D \glspl{cnn} on the ToN IoT dataset for both binary and multi-class classification. They used SHAP force plots for individual predictions, bar plots for feature importance, and beeswarm plots for global insights. \cite{Kumar2023explainable} trained a \gls{dnn} on UNSW-BW15, visualizing feature importance with SHAP bar plots.

Beyond tabular datasets, several studies have applied SHAP exclusively to interpret supervised learning models trained on sequential \gls{ids} data. These models typically leverage recurrent neural network architectures such as \gls{lstm} and \gls{gru} to capture temporal dependencies in network traffic or sensor data.

The authors of \cite{Attique2024explainable} developed a bi-\gls{lstm} model with a self-adaptive attention layer for multiclass classification on the CICIDS2017 and X-IIoTID datasets. SHAP waterfall and bar plots were employed to interpret the model's predictions.
\cite{Kumar2024blockchain} applied bi-directional \gls{gru} networks to the CIC-IDS2017 dataset, using SHAP decision and bar plots to explain classification outcomes. Similarly, \cite{Javeed2024explainable} trained a hybrid model combining bi-\gls{lstm} and \gls{gru} networks on the CICIDDoS2019 dataset, and used SHAP decision, waterfall, and bar plots to interpret model behavior.
\cite{Sivamohan2023tea} trained a bi-\gls{lstm} model on NSL-KDD and CICIDS 2018 datasets. Their interpretability approach included SHAP bar charts for feature importance and force plots for individual prediction analysis.

SHAP has also been applied to interpret unsupervised learning models for anomaly detection, which are particularly valuable in industrial settings where labeled attack data is scarce or costly to obtain. These models are typically trained on benign traffic and identify anomalies based on reconstruction error.

In \cite{Kalutharage2022explainable, Kalutharage2023explainable}, the authors trained fully connected autoencoders on the USB-IDS dataset to detect \gls{ddos} attacks. Treating the data as tabular, they used SHAP values to identify the most influential features contributing to anomalous predictions, helping users understand why certain traffic was flagged.

The work in \cite{Hwang2021e} extended SHAP interpretability to multivariate timeseries data in their bi-\gls{lstm}-based anomaly detection model trained on the H\gls{ai} Security Dataset. Each input sequence consisted of sensor readings over time, and the model predicted future values to detect deviations. To support rapid anomaly localization, the authors introduced custom SHAP-enhanced heatmaps that visualize both normalized prediction error and SHAP-based feature contributions across time steps. Warmer colors in the SHAP-enhanced heatmaps guided users toward inspecting specific sensor readings. Additionally, SHAP beeswarm plots were used to summarize feature importance globally.

Further, research has trained an \gls{lstm}-based autoencoder on a custom dataset capturing SCADA system traffic from a gas pipeline~\cite{Ha2022explainable, morris2014industrial}. The dataset included both network-level features (e.g., MODBUS packet lengths) and operational technology (\gls{ot}) signals (e.g., pump status, pipe pressure). SHAP beeswarm plots were used to visualize feature contributions to reconstruction error, aiding in anomaly localization.

These studies collectively illustrate SHAP's versatility in interpreting both supervised and unsupervised models operating on various datasets.
\gls{shap} was widely adopted as an explanation method across the majority of relevant studies. For works that employed \gls{shap} in conjunction with other \gls{xai} techniques, their detailed discussion is deferred to subsequent sections where combined methodologies are outlined.

\subsubsection{Local Interpretable Model-Agnostic Explanations (LIME)}

Local Interpretable Model-Agnostic Explanations (LIME)~\cite{ribeiro2016should} is an associative, local, model-agnostic explanation technique designed to explain individual predictions of complex models by approximating the model locally with an interpretable surrogate (or mimic) model. It works by generating perturbed samples around the instance to be explained, obtaining predictions from the black-box model, and fitting a simple model, typically linear or decision tree, to these samples. The surrogate model's feature weights then serve as a proxy for understanding the original model's decision in the local neighborhood of the instance. Visualizations from the LIME Python library often take the form of bar charts, highlighting the top contributing features for a single prediction and indicating whether each feature positively or negatively influenced the predicted class, similar to SHAP waterfall plots.
Thus, for industrial \glspl{soc}, \gls{lime} can be  relevant as a local explanation tool that provides rapid, instance-specific rationales for individual alerts, although its sampling-based nature may limit stability in high-stakes operational settings.

While \gls{lime} is primarily intended for local explanations, it can be extended to provide global insights through techniques such as Submodular Pick-LIME (SP-LIME). SP-LIME selects a diverse set of representative instances and generates local explanations for each, which are then aggregated to highlight globally important features or patterns. Although this approach does not yield a true global explanation in the strict sense (unlike \gls{shap} which averages feature contributions across all instances to produce more robust global insights) it offers a practical means of approximating global interpretability by summarizing multiple local views.

Several studies have applied LIME exclusively to interpret AI models in industrial security contexts. Sharma \textit{et al.}~\cite{Sharma2023anomaly} used LIME to interpret a \gls{dnn} trained for intrusion detection on the NSL-KDD dataset. Ben \textit{et al.}~\cite{Ben_2021} evaluated both black-box models (SVM, MLP, RF, Extra Trees) and white-box models (Decision Trees, kNN) on IoT traffic data infected with Mirai and Bashlite malware. They applied LIME post hoc to explain model decisions. The authors noted that while black-box models achieved higher accuracy, their explanations were less stable. In contrast, white-box models offered more consistent and transparent outputs, albeit with lower accuracy. 
In XAI, explanation stability refers to the consistency of the explanation across multiple queries and under similar conditions. An explanation is considered unstable if small changes in the instance to be explained or sampling lead to noticeably different feature importance rankings or decision rationales. This is particularly relevant for methods like LIME, which rely on randomly generated local perturbations, making them sensitive to sampling variance. As a result, repeated explanation queries may yield inconsistent explanations for the same instance, undermining trust and reproducibility. 

Al-Hawawreh \textit{et al.}~\cite{Al-Hawawreh2024explainable} proposed an explainable deep learning framework for attack intelligence in Industrial IoT, focusing on control loops in cyber-physical systems. Their architecture combined ML/DL models (LSTM, RF, XGBoost, DNN) with an intelligence profiling module that employed LIME and Submodular Pick-LIME (SP-LIME). Evaluated on a gas pipeline control-loop dataset with seven attack types, their results showed that no single AI model performed best across all scenarios; instead, combining AI models yielded robust detection, and LIME translated black-box predictions into actionable attack signatures. While LIME explained individual instances, SP-LIME aggregated explanations to derive generalized attack rules.

Many studies employed both LIME and SHAP to leverage their complementary strengths: LIME for local, instance-level explanations and SHAP for global feature importance. 
Khan \textit{et al.}~\cite{Khan2022proactive} trained decision trees, RF, AdaBoost, and XGBoost on an HVAC dataset~\cite{elnour2021application} featuring sensor readings, control signals, and power consumption. The attacks from the dataset are categorized into four types: \textit{(1) changing the control system's setpoints; (2) falsifying sensor readings; (3) falsifying control signals; and (4) modifying command signals}. The authors used SHAP beeswarm plots alongside LIME feature importance to interpret model behavior. 
Kaur \textit{et al.}~\cite{Kaur_2024} integrated XGBoost with SHAP and LIME to enhance IoT security in 6G environments. SHAP revealed global feature importance, while LIME provided local explanations. They used these insights to prune features via recursive elimination, improving model accuracy from 95.59\% to 97.02\%. This study demonstrated that \gls{xai} not only enhances interpretability but can also directly improve model performance. 
Kalakoti \textit{et al.}~\cite{Kalakoti2024improving} applied KG-Boost, LightGBM, Extra Trees, RF, and Gradient Boosting to botnet detection across N-BaIoT, MedBIoT, and BOT-IoT datasets. They used tree-rule-based LIME and SHAP force plots to explain decisions and compared the explainers across faithfulness, monotonicity, complexity, and sensitivity.
The authors found SHAP to outperform LIME across all four metrics.

Deep learning models have also been interpreted using both LIME and SHAP. 
Muna \textit{et al.}~\cite{Muna2023demystifying} proposed an XGBoost-based framework for IoT attack detection using the IoTID20 dataset, targeting binary, category, and sub-category classification levels. XGBoost was selected over deep learning models due to its relatively higher interpretability and faster training. To address high-dimensionality and class imbalance, \gls{pca}-based feature selection and SMOTE oversampling were applied. The authors used \gls{lime} and \gls{shap} force plots and bar plots to explain model decisions.
Abou El Houda \textit{et al.}~\cite{Houda2022why} trained a DNN-based \gls{ids} on NSL-KDD and UNSW-BW15 datasets, using SHAP bar plots, beeswarm plots, and force plots alongside LIME for feature attribution. 
Shtayat \textit{et al.}~\cite{Shtayat2023explainable} developed an ensemble of three CNNs for binary and multi-class classification on the ToN IoT dataset, visualizing feature importance using SHAP bar charts and LIME. 
Sharma \textit{et al.}~\cite{Sharma_2024} proposed a hybrid \gls{ids} combining DNN for anomaly detection and CNN for traffic classification, using SHAP beeswarm and force plots to visualize feature contributions and LIME bar plots for instance-level explanations on NSL-KDD and UNSW-NB15 datasets. 
Arreche \textit{et al.}~\cite{Arreche2024xai1, arreche2024xai2} trained RF, DNN, AdaBoost, kNN, SVM, and LightGBM models on CICIDS-2017, NSL-KDD, and RoEduNet-SIMARGL2021 datasets. They applied SHAP waterfall plots and LIME to generate local explanations. 
They also proposed textual augmentations to enhance visual explanations. 
Their evaluation spanned six \gls{xai} metrics: descriptive accuracy, sparsity, computational efficiency, stability, robustness, and completeness. 
Descriptive accuracy measures the drop in model performance when top influential features are removed. Sparsity assesses whether explanations concentrate influence on a few features. Stability evaluates consistency across repeated trials. Efficiency relates to the time required to generate explanations. Robustness examines resistance to adversarial manipulation, and completeness reflects the method's ability to explain diverse instances.
The authors' assessment was that while \gls{shap} performed well across all six evaluation metrics, \gls{lime} showed limitations in descriptive accuracy, sparsity, and completeness. They also found \gls{shap} to be more time-efficient for generating global explanations, whereas \gls{lime} was more efficient for local, instance-level interpretations.

\subsubsection{Feature Influence Visualization Techniques} \label{subsubsec: pdp-ale-ice}

\gls{pdp}~\cite{friedman2001greedy}, \gls{ice}~\cite{goldstein2015peeking}, and \gls{ale}~\cite{apley2020visualizing} are model-agnostic \gls{xai} techniques used to visualize the relationship between input features and model predictions. 
All three methods aim to interpret complex models by showing how changes in a feature influence the output, assuming other features are held constant or appropriately marginalized. 
PDP estimates the average effect of a feature by marginalizing over the joint distribution of other features, but this can lead to misleading interpretations when features are correlated. 
ICE addresses this limitation by plotting the effect of a feature for individual instances, thereby revealing heterogeneity and feature interactions that PDP may obscure. However, ICE plots can be noisy and difficult to summarize across populations. 
ALE improves upon both by estimating feature effects locally and accumulating them, making it more robust to feature correlations. 
The visualizations produced by all three methods are structurally similar in two dimensions: the x-axis represents the range of values for a specific feature, while the y-axis reflects the predicted likelihood for a specific class. In industrial security, these methods can help analysts understand how changes in operational variables, such as sensor readings, control signals, or traffic features, affect model predictions, thereby supporting anomaly interpretation and model validation.

Keshk \textit{et al.}~\cite{Keshk_2023} introduced an explainable \gls{ids} framework using an LSTM model for attack classification. The framework incorporated SHAP, ICE, and PDP to interpret model behavior. SHAP bar plots were used to visualize global feature importance across datasets, while force plots illustrated individual prediction breakdowns. The framework was evaluated on three benchmark datasets: NSL-KDD, UNSW-NB15, and ToN\_IoT.

Gummadi \textit{et al.}~\cite{Gummadi2024xai} trained a diverse set of AI models, including \gls{dt}, \gls{dnn}, AdaBoost, \gls{svm}, and ensemble methods such as \gls{rf}, bagging, blending, stacking, and voting—for anomaly detection on the N-BaIoT and MEMS datasets. The MEMS dataset, generated by the authors, captured machine condition data from a motor testbed using MEMS sensors under varying mechanical imbalance scenarios. Acceleration signals were recorded across the X, Y, and Z axes of the motor, and failure conditions were categorized as normal, near-failure, or failure. Among other interpretability techniques discussed in later sections, the authors employed \gls{ale} to analyze feature influence.

\subsubsection{Permutation Feature Importance}

\gls{pfi} is a model-agnostic global \gls{xai} technique used to estimate feature importance by measuring the impact of each feature on model performance. The method begins by evaluating the baseline performance of the trained model on the original dataset using metrics such as accuracy, F1-score, or precision. For each feature, its values are randomly shuffled across all samples, thereby disrupting its relationship with the target variable. The model is then re-evaluated on the modified dataset, and the change in performance is compared to the baseline. A significant drop in performance indicates that the model relied heavily on that feature, whereas minimal change suggests low importance. Repeating this process across multiple shuffling rounds allows for the computation of mean and standard deviation estimates for each feature's importance, providing a robust measure of sensitivity. For industrial deployments, \gls{pfi} can support model auditing by identifying which features the model depends on most, helping practitioners assess whether predictions are driven by meaningful industrial signals or by spurious dataset artifacts. Keshk \textit{et al.}~\cite{Keshk_2023} and Gummadi  applied \gls{pfi} to visualize global feature importance in their respective \gls{ids} frameworks.

\subsubsection{Rule-based Explanations}

Decision trees are inherently transparent \gls{ml} models whose interpretability stems directly from their structure. Each internal node represents a feature-based split, and each path from the root to a leaf forms a human-readable decision rule. This hierarchical logic allows users to trace predictions through a sequence of conditions, making decision trees particularly suitable for domains requiring clear, auditable reasoning. Feature importance plots can also be derived from decision trees by computing how much each feature contributes to reducing impurity (e.g., Gini index or entropy) across all splits in the tree. These scores are aggregated and normalized to reflect the global importance of each feature. Rule-based explanations are particularly relevant to \glspl{soc} because they can be translated into analyst-readable logic, detection rules, or audit documentation more readily than many feature-attribution outputs.

Despite their interpretability, single decision trees suffer from several limitations. They are prone to overfitting, especially when grown deep without pruning, and may struggle to capture complex, nonlinear relationships in high-dimensional data. Their expressiveness is limited, and they are highly sensitive to small changes in training data, which can lead to instability in both structure and decision logic. Ensemble methods such as random forests, bagging, and boosting are often preferred for improved accuracy, generalization, and robustness. However, these methods sacrifice transparency, as the logic is distributed across many trees, making it difficult to trace individual decisions.

RuleFit~\cite{friedman2008predictive} addresses this trade-off by combining the interpretability of decision tree rules with the predictive power of ensemble methods. It first trains an ensemble of decision trees and extracts decision rules from each tree. These rules are then encoded as binary features (indicating whether a rule is satisfied) and combined with the original features in a sparse linear model. Each rule corresponds to a condition derived from the tree ensemble, and the linear model assigns weights to these rules and features. L1 regularization is applied to select only the most informative rules, enhancing both sparsity and clarity.

RuleFit offers several interpretable visualizations. A common approach is to display rule importance using bar plots, similar to feature importance plots, where each bar reflects the magnitude of a rule's linear coefficient. These rules are presented in readable form, allowing users to understand the logic driving predictions. For individual predictions, RuleFit enables decomposition into rule contributions, offering a transparent breakdown similar in spirit to SHAP force plots but grounded in linear additive logic.

Mahbooba \textit{et al.}~\cite{Mahbooba_2021} designed a framework that uses a decision tree model as an \gls{xai} method to enhance trust management in \gls{ids}. Using the NSL-KDD dataset, they conducted experiments to rank feature importance, extract rule-based IF-THEN explanations from the decision tree, and compare its performance against \gls{svm} and logistic regression in binary classification tasks. Their results showed that the decision tree achieved comparable precision, recall, and F1-score while offering transparent rule sets that clearly illustrate how predictions are made.

Abou El Houda \textit{et al.}~\cite{Houda2022novel} trained a binary classification \gls{dnn} for \gls{iot}-based \gls{ids} on the UNSW-NB15 dataset and explained model decisions using RuleFit and \gls{shap} bar and force plots. They also used violin plots to visualize the distributions of top-ranked features.

Abou El Houda \textit{et al.}~\cite{houda_2023} proposed FedIoT, a hybrid framework that integrates \gls{xai} with Federated Learning and Blockchain to construct a secure, interpretable intrusion detection system for IoT environments. The blockchain component introduces a multi-layered reputation system to evaluate both local and global model contributions, helping to detect low-quality or malicious blockchain updates. FedIoT employs SHAP, LIME, and RuleFit to provide both local and global explanations. Feature importance plots generated using RuleFit and SHAP highlight critical features for detecting intrusions, SHAP force plots visually explain how individual features contribute to specific predictions, and LIME-based bar plots offer localized explanations for classified scenarios.

\subsubsection{Combinations}

Several studies have applied \gls{shap} alongside other \gls{xai} techniques. 
Gummadi \textit{et al.}~\cite{Gummadi2024xai} (refer to Section~\ref{subsubsec: pdp-ale-ice}) conducted a comprehensive feature importance analysis using a combination of global and local \gls{xai} methods. Their framework included global importance via \gls{shap}, \gls{loco}, \gls{pfi}, and \gls{ale}, and local importance using \gls{cem} and \gls{lime}.
\gls{loco} estimates global feature importance by measuring the drop in model performance when individual features are removed. The results are typically visualized using bar plots that rank features by their impact on predictive accuracy. \gls{cem} provides local contrastive explanations by identifying minimal perturbations needed to change a prediction (pertinent negatives) or preserve it (pertinent positives), visualized as highlighted feature subsets that delineate decision boundaries.

Kalakoti \textit{et al.}~\cite{kalakoti2025evaluating} explore explainability for network intrusion detection using IT network data from a university-deployed local SOC environment. The researchers trained an LSTM model that correctly classifies Suricata NIDS alerts as ``important" or ``irrelevant." To explain the model's alert prioritization decisions, they developed and evaluated four \gls{xai} methods: LIME, SHAP, Integrated Gradients, and DeepLIFT. DeepLIFT consistently outperformed the other \gls{xai} methods across a comprehensive evaluation framework assessing faithfulness, complexity, robustness, and reliability. The strong alignment between SOC analyst-identified features and those highlighted by the \gls{xai} methods validates practical applicability in real-world SOC operations, though the study is limited to a single organization's IT dataset with binary classification.

\subsubsection{Custom Methods}

A few studies proposed custom \gls{xai} frameworks tailored to specific interpretability goals.

Zolanvari \textit{et al.}~\cite{Zolanvari2023trust} introduced TRUST (Transparency Relying Upon Statistical Theory), a model-agnostic surrogate explanation method designed to provide statistical insights into model behavior. TRUST begins by applying factor analysis to reduce the original feature space into a smaller set of uncorrelated latent factors, each representing a linear combination of the original features. Mutual Information is then used to quantify the relevance of each latent factor to the model's class predictions. The most informative factors—termed representatives—are selected for explanation. For each class, TRUST fits a multimodal Gaussian distribution over these representatives. To explain a new sample, the framework computes the likelihood of the sample belonging to each class under the fitted distributions, and the class with the highest likelihood is selected as the predicted output. In practice, TRUST can generate statements such as: ``the sample's value of representative factor 1 increased its likelihood of belonging to the attack class,'' offering users a probabilistic rationale for classification.
The authors demonstrated TRUST on the WUSTL-IIoT, NSL-KDD, and UNSW-BW datasets and reported a 25-fold improvement in processing speed compared to LIME, highlighting its potential for real-time deployment. However, we anticipate that TRUST's reliance on latent factors introduces a key limitation: these factors are not directly human-interpretable, as they represent combinations of original features. Unlike SHAP and LIME, which explain predictions in terms of semantically meaningful input features, TRUST interprets model behavior through statistical abstractions, which may reduce transparency for non-technical users. Additionally, the framework assumes that data are sampled independently from Gaussian distributions, which may not hold in practice.

\subsection{Research Landscape}

The majority of the surveyed literature applied supervised tabular \gls{ml} models to \gls{ids} datasets. A smaller subset explored the use of sequential models for \gls{ids}, while only a few studies employed unsupervised \gls{ml} approaches for anomaly detection tasks. The prevalence of \gls{xai} applications in \gls{ids} research can be largely attributed to the availability of benchmark datasets, which facilitate the application of \gls{ml} techniques and enable comparative analysis. Among the most commonly used datasets in the reviewed papers are NSL-KDD (2009), UNSW-NB15 (2015), and CIC-IDS (2017).

Regarding \gls{xai} methods, \gls{shap} emerged as the most widely adopted technique. \gls{shap} bar plots were the predominant global explainability visualization, followed by \gls{shap} beeswarm plots. For local explanations, \gls{shap} force and waterfall plots, along with \gls{lime} bar plots, were frequently utilized. This trend highlights the dominant use of associative, feature-attribution-based explanation methods within the industrial security domain.

Table~\ref{tab:xai_method_explanations} summarizes the \gls{xai} methods employed in the surveyed literature, along with additional techniques referenced in the subsequent discussion. For each method, we outline the type of explanation it aims to provide, phrased as a question. These questions are further grouped into broader categories based on the taxonomy outlined in~\cite{liao2020questioning}:
\begin{itemize}
    \item \textit{How?:} Describes the process the model used to make its predictions.
    \item \textit{Why?:} Explains the rationale behind a specific prediction.
    \item \textit{Why not?:} Explains why an alternative prediction was not chosen.
    \item \textit{What if?:} Explores hypothetical changes to input features.
    \item \textit{How to be that?:} Suggests minimal changes needed to achieve a different prediction.
    \item \textit{How to still be this?:} Identifies conditions that preserve the current prediction.
    \item \textit{Performance:} Assesses model performance across some metric.
\end{itemize}

\begin{table*} [t]
    \caption{Uses of \gls{xai} methods}
    \centering
    \small
    \begin{tabularx}{\linewidth}{>{\raggedright\arraybackslash}p{0.15\linewidth} p{0.55\linewidth}X}
        \toprule
        \textbf{\gls{xai} Method} & \textbf{Explanation} & \textbf{References}\\
        \midrule
        GLOBAL\\
        \midrule
        \gls{shap} bar/beeswarm plots, SP-\gls{lime}, \gls{loco}, & \textit{Why?:} Which features are most influential in driving predictions across the dataset? &\cite{Oseni2023explainable, Houda2022why, Javeed2024explainable, Kumar2024blockchain, Kumar2023explainable, Shtayat2023explainable, Gaitan-Cardenas2023explainable, Gurbuz2023explainable, Attique2024explainable, Gummadi2024xai, Suhail2023enigma, houda_2023, Kaur_2024, Arreche2024xai1, Sivamohan2023tea}\\
        \midrule
        \gls{pfi} & \textit{Why?:} How strongly does each feature contribute to the model's predictions? & \cite{Gummadi2024xai, Ben_2021, Keshk_2023}\\
        \midrule
        \gls{shap} beeswarm plots & \textit{Why?:} How does each feature contribute to the model's predictions, and in what direction? & \cite{Oseni2023explainable, Houda2022why, Alani2023explainable, Khan2022proactive, Hwang2021e, Ha2022explainable, Keshk_2023, Sharma_2024, Wali2025explainable, Arreche2024xai1} \\
        \midrule
        Rulefit & \textit{How?:} Which combinations of feature conditions, expressed as decision rules, consistently influence model's predictions?\\ &
        \textit{Why?:} How strongly does each rule and original feature contributes to the model's predictions?& \cite{houda_2023}\\
        \midrule
        \gls{ale}, \gls{pdp} & \textit{What if?:} How does varying a specific feature influence the model's predictions on average across the dataset? & \cite{Suhail2023enigma, Keshk_2023} (PDP), \cite{Gummadi2024xai} (ALE) \\
        \toprule
        LOCAL\\
        \toprule
        \gls{shap} force/waterfall/ decision plots, \gls{lime} & \textit{Why?:} What specific feature contributions led to this prediction, and how do they interact to shape the outcome?\\
        & \textit{Why?:} Which features had the strongest influence on this prediction, and in what direction? & \cite{Oseni2023explainable, Houda2022why, Javeed2024explainable, Gurbuz2023explainable, Attique2024explainable, Kalakoti2024improving, houda_2023, Keshk_2023, Sharma_2024, Wali2025explainable, Arreche2024xai1, Sivamohan2023tea}, \cite{Javeed2024explainable, Kumar2024blockchain}\\
        \midrule
        \gls{lime} & \textit{Performance:} Is the model's behavior locally consistent and reliable around this instance? & \cite{Houda2022why, Shtayat2023explainable, Gurbuz2023explainable, Khan2022proactive, Gummadi2024xai, Kalakoti2024improving, houda_2023, Kaur_2024, Ben_2021, Sharma_2024, Arreche2024xai1, Sharma2023anomaly}\\
        \midrule
        \gls{ice} & \textit{What if?:} How does changing this specific feature affect the predictions for individual instances?\\
        & \textit{Performance:} Are there heterogeneous effects across samples, or does the feature behave consistently? & \cite{Keshk_2023}\\
        \midrule
        \gls{cem} & \textit{How to be that/Why not?:} What minimal changes would flip the model's prediction (pertinent negatives)? & \cite{Gummadi2024xai}\\
        & \textit{How to still be this/Why?:} What features must remain unchanged to preserve the prediction (pertinent positives)?\\
        \midrule
        Counterfactual &  \textit{How to be that/Why not?:} What is the closest alternative input that would change the prediction? & \cite{lewis2013counterfactuals}\\ 
        \midrule
         Anchors &  \textit{How to still be this/Why?:} What conditions (rules) guarantee the same prediction with high confidence? \\
        & \textit{Performance:}  How robust is the prediction to changes in other features? & \cite{ribeiro2018anchors} \\
        \toprule
    \end{tabularx}
    \label{tab:xai_method_explanations}
\end{table*}

\section{Operational Integration, Security Considerations, and Research Directions}\label{sec:discussion}

\subsection{Deploying \gls{xai} Methods in Industrial SOCs}\label{sec:discussion_xai}

While the surveyed literature demonstrates the broad applicability of existing \gls{xai} techniques to cybersecurity datasets and models, many works provide limited discussion of how these methods integrate into operational industrial security workflows. The following section therefore examines \gls{xai} from the perspective of industrial \glspl{soc}, focusing on deployment considerations, stakeholder requirements, and emerging research needs.

\subsubsection{Proposed Approach}
A recurring observation across the surveyed literature is that many studies provide limited discussion regarding which industrial security stakeholders or \gls{soc} tasks their proposed explanations are intended to support. This limits the operational applicability of \gls{xai} methods in industrial security environments, where workflows, responsibilities, and decision-making requirements are often highly specialized.

To help address this gap, we present a non-exhaustive mapping of \gls{xai} methods to representative \gls{soc} tasks, informed by our survey of the literature and our interpretation of how these techniques may integrate into the industrial \gls{soc} workflows summarized in Table~\ref{tab:soc_workflow_phases}.

The training of \gls{ai}/\gls{ml} models and the fitting of \gls{xai} explainers typically occur during the preparation phase. In real-time operations, data collection and normalization feed the \gls{ml} models, which perform threat detection. \gls{xai} methods are particularly valuable during alert triage, correlation, and prioritization, where they help analysts interpret model outputs, assess alert validity, and document findings for escalation or cross-team collaboration.

When the \gls{ml} model outputs alerts, it may include a confidence score (e.g., probability of prediction). Tier 1 \gls{soc} analysts can use this score to prioritize alert handling. 
\gls{xai} methods can further enhance this process by explaining why a prediction was made, which features contributed most, and whether the alert is likely a false positive.

For instance, a Tier 1 \gls{soc} analyst may use associative local \gls{xai} methods such as \gls{shap} waterfall plots or \gls{lime} to identify the specific data features that influenced the alert. Concurrently, contrastive methods like \gls{cem} or counterfactual explanations can highlight the minimal set of feature changes that would flip the prediction to benign, helping differentiate the alert from normal behavior. These insights help narrow down—or at least prioritize—the data artifacts and feature values that require validation, streamlining the analyst's investigation.

A {counterfactual example} identifies the minimal changes to an input instance that would alter the model's prediction. 
In contrast, {anchors} provide high-precision, rule-based conditions that, when satisfied, guarantee the model's prediction stays the same with high confidence. Each anchor is associated with two key metrics: {precision}, which quantifies the proportion of instances satisfying the anchor conditions that lead to the same prediction, and {coverage}, which measures how frequently such conditions occur across the dataset. Importantly, anchor precision can be {pre-specified} by the user (e.g., 95\%), and the algorithm will search for rules that meet or exceed this threshold. 
Together, counterfactual examples and anchors offer complementary insights: the former reveals sensitivity to change, while the latter highlights stable, interpretable regions of model behavior.

Based on the magnitude and direction of feature contributions, these \gls{xai} methods can also guide escalation decisions. For example, a benign counterfactual that is far from the current alert suggests the event is highly anomalous and may warrant escalation. Conversely, a close benign counterfactual may indicate borderline behavior, prompting further scrutiny into whether it is a false alarm before escalation. Similarly, high-impact features identified by \gls{shap} or \gls{lime} may signal that the alert is significant and should be escalated.

Anchor explanations can offer additional insight by producing rules with associated coverage values. A small coverage suggests the alert is rare or represents an edge case, meriting further scrutiny to detect false alarms. A large coverage, on the other hand, indicates that the rule generalizes well and could be incorporated into automated triage logic for future alerts.

For deeper investigations, sensitivity methods such as \gls{ale} and \gls{ice} can help analysts understand the anomalous value ranges of high-contributing features. These tools help not only identify which features matter but also clarify how their values deviate from expected norms, enhancing both reporting and root cause analysis.

Beyond triage, \gls{xai} insights can support rule updates, incident investigation, cross-functional communication, and compliance documentation. They can also reduce cognitive load and accelerate decision-making across \gls{soc} tiers.

\begin{table}[t!]
    \caption{Mapping of \gls{xai} methods to \gls{soc} tasks}
    \centering
    \begin{tabularx}{\linewidth}{>{\raggedright\arraybackslash}p{0.2\linewidth} X}
        \toprule
        \textbf{\gls{xai} Method} & \textbf{\gls{soc} Task}\\
        \midrule
        \multicolumn{2}{l}{\textbf{Alert Triage}}\\
        \midrule
        \gls{shap} force, waterfall, decision plot; \gls{lime} & Identify and rank feature contributions to an alert's prediction. Visualize how feature values cumulatively influence the model output, supporting root cause analysis and focused investigation.\\
        \midrule
        \gls{shap} decision plot & Trace the decision path across features and predictions. Useful for documenting why an alert was classified as malicious rather than benign, and for identifying false alarms.\\
        \midrule
        \gls{ice}, \gls{ale}, \gls{pdp} & Visualize how changes in high-impact features affect predictions, either on average or across individual samples, supporting deeper investigation.\\
        \midrule
        Anchors & Provide high-precision rules that explain predictions. Small coverage may indicate rare or edge-case alerts; large coverage suggests generalizable detection logic.\\
        \midrule
        \gls{cem}, Counterfactuals & Highlight what differentiates a malicious alert from benign behavior. Useful for root cause analysis and understanding decision boundaries to detect false alarms.\\
        \midrule
        \multicolumn{2}{l}{\textbf{Rule Updates}}\\
        \midrule
        RuleFit & Extract interpretable rules from \gls{ml} models to guide policy refinement, alert tuning, and SIEM configuration.\\
        \midrule
        Anchors & Anchors with broad coverage can inform robust detection rules. Useful for defining high-confidence alert boundaries.\\
        \midrule
        \gls{pfi}, \gls{pdp}, \gls{ale}, \gls{shap} bar/beeswarm & Reveal global feature importance and interactions, supporting feature engineering and model retraining.\\
        \midrule
        \multicolumn{2}{l}{\textbf{Training}}\\
        \midrule
        \gls{pfi}, \gls{shap} bar/beeswarm & Help train analysts by identifying which features consistently influence alert predictions. Useful for onboarding.\\
        \bottomrule
    \end{tabularx}
    \label{tab:xai_to_tasks}
\end{table}

\subsubsection{Security Considerations}

When \gls{xai} is applied to \gls{ai} systems operating exclusively within industrial \glspl{soc}, a critical security concern is the potential for explanation misinterpretation. This may arise from analysts misinterpreting explanations, overestimating their scope, or relying on explanations that are inconsistent, misleading, or ambiguous. Inaccurate or poorly scoped explanations can distract analysts from genuine threats, lead to wasted time in evaluating false rationales, justify incorrect predictions with seemingly plausible reasoning, or result in the erroneous implementation of security rules derived from flawed interpretations.

Each \gls{xai} method carries inherent limitations tied to its computational assumptions and design. For instance, some methods exhibit instability or inconsistency, producing different explanations for the same input and model across repeated runs. \gls{lime}, for example, relies on local sampling and assumes local linearity~\cite{ribeiro2016should}; a condition that may not hold in complex feature spaces. Consequently, \gls{lime} can yield divergent explanations for identical inputs. 
\gls{ml} model sensitivity can further compound explanation inconsistency. For example, \glspl{dt} are highly sensitive to training data variability; minor perturbations in input data can result in significantly different decision rules and explanations~\cite{hara2023average}. 

Next, while simple, low-complexity explanations are often preferred for user comprehension, they may obscure important model behaviors. 
Many widely used \gls{xai} methods---such as \gls{pdp}, \gls{pfi}, \gls{shap}, and \gls{loco}---rely on the assumption of feature independence, which can result in biased or misleading explanations when the input features are correlated.
As a result, simplified explanations may sacrifice fidelity for interpretability, potentially leading to incorrect conclusions.

Scalability also presents a security concern. Methods like \gls{shap}, while widely adopted, are computationally intensive. In time-sensitive environments, delayed explanations can hinder incident response. Moreover, resource-intensive \gls{xai} computations may compete with other essential monitoring and analytics functions, introducing performance bottlenecks or \gls{dos} risks. 

Lastly, interpreting local explanations as global can lead to flawed generalizations, which if coded as security rules can significantly hinder \gls{soc} workflow.
Given these security implications, it is essential that \gls{soc} personnel understand the intended scope, assumptions, and limitations of the \gls{xai} methods they employ. Misuse or overreliance on explanations without such understanding can compromise security.

\begin{figure*}[t!]
    \centering
    \includegraphics[width=0.9\linewidth]{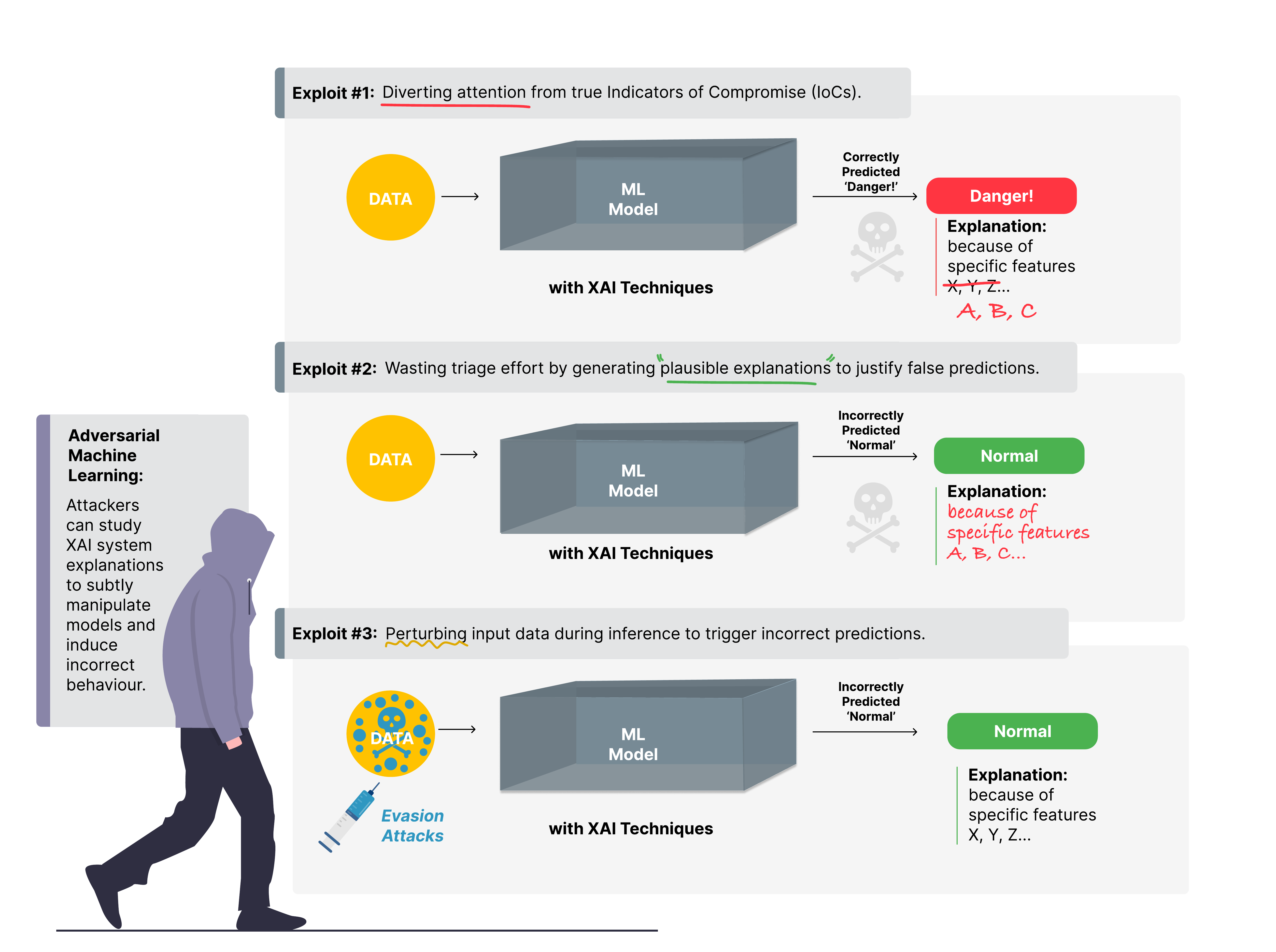}
    \caption{Illustration of three possible adversarial approaches to exploiting \gls{xai} systems. }
    \label{fig:adversarial_ml}
\end{figure*}

Additional security risks emerge if the \gls{xai} systems are exposed intentionally or inadvertently to adversaries, including malicious insiders. Explanations can reveal sensitive information about the underlying data or \gls{ml} model behavior. This vulnerability is central to the field of adversarial \gls{ml} (see Fig.~\ref{fig:adversarial_ml}), which explores how attackers can subtly manipulate models to induce incorrect behavior~\cite{zhang2020interpretable, pachl2025view}. Adversaries may exploit \gls{xai} in several ways~\cite{pachl2025view}:
\begin{enumerate}
    \item Manipulating explanations without altering predictions, thereby misleading analysts and diverting attention from true \glspl{ioc};
    \item Generating plausible explanations to justify false predictions, increasing the likelihood of false positives or wasted triage effort;
    \item Crafting adversarial inputs that produce distinct predictions with similar explanations.
\end{enumerate}

The last can facilitate evasion attacks, which involve perturbing input data during inference to trigger incorrect predictions~\cite{zhang2020interpretable}. By analyzing \gls{xai} outputs, adversaries can infer decision boundaries and craft inputs that exploit model weaknesses, either to overwhelm the \gls{soc} with false positives or to evade detection (false negatives). Contrastive explanation methods, such as counterfactuals and \gls{cem}, are particularly susceptible, as they reveal how minimal input changes affect model decisions~\cite{pawlicki2024explainability, spartalis2023balancing}. If these methods draw from real data, they may also leak sensitive operational information.

Thus, while \gls{xai} methods are designed to enhance analyst understanding, they can equally empower adversaries to reverse-engineer and exploit \gls{ml} models. To mitigate this risk, it is critical to treat \gls{ml} models, training data, and \gls{xai} systems as sensitive assets, protected by strong access controls~\cite{pachl2025view}. Organizations should also prepare contingency measures, including incident response and data breach plans, to address potential exposure.
Additional safeguards include hardening \gls{ai} systems by simulating attacks during development to improve robustness against such explanation-based exploits.

We observe a gap in the literature regarding the security implications of proposed \gls{xai} methods. Addressing this gap requires a systematic evaluation of each method’s scope, assumptions, and vulnerabilities, as well as the discussion of secure deployment practices tailored to industrial \gls{soc} environments.

\subsection{Emerging Research Needs for Industrial SOC-specific XAI}
\subsubsection{Integrating \gls{xai} into the AI Design Lifecycle}
When developing AI to enhance SOC capabilities, explainability must be specified from the start. Explainability requirements shape model selection, architecture, and evaluation metrics. If postponed, it can force costly redesigns or produce explanations that are misleading or irrelevant to stakeholders. Moreover, the utility of \gls{xai} methods depends on input data and a mismatch between data design and chosen explanation technique can yield technically correct model predictions but fail to guide operational decisions. For example, integrating \gls{xai} methods early in the design lifecycle establishes an iterative feedback mechanism between model development and data curation. \gls{xai} techniques systematically identify influential data points which can be checked for data quality, data bias, and labeling errors\cite{siemens_whitepaperXAI}. This iterative refinement cycle yields concurrent enhancements to data integrity, predictive performance and actionable explanations.

\subsubsection{Industrial Security-specific \gls{xai} Evaluation Metrics} 
Despite the growing interest in \gls{xai} for industrial security, the literature remains limited in its evaluation of both the methods and their presentation. Among the surveyed works, only a few, namely Kalakoti \textit{et al.}~\cite{Kalakoti2023improving, Kalakoti2024improving} and Arreche \textit{et al.}~\cite{Arreche2024xai1, arreche2024xai2}, quantitatively assessed their \gls{xai} methods. 

A key challenge in \gls{xai} evaluation is the absence of universally accepted evaluation criteria of explainability~\cite{spartalis2023balancing} and the absence of standardized absolute thresholds and consistent terminology for existing quantitative metrics. This ambiguity forces researchers to rely on relative comparisons (e.g., comparing SHAP and LIME under identical settings) or human-centered validation methods such as user studies~\cite{kadir2023evaluation, seth2025bridging}. While these approaches offer valuable insights, they introduce subjectivity and hinder efforts to automate \gls{xai} evaluation.

In alignment with Pietilä \textit{et al.}~\cite{pietila2023explanation}, we do not advocate for universal metrics. However, it is both feasible and necessary to define domain-specific metrics for industrial security, specify their definitions, and establish absolute thresholds that determine whether an explanation or explainer is operationally fit for use.

\subsubsection{Validation for Building Trust in Industrial XAI}
Only Lundberg \textit{et al.}~\cite{Lundberg2022experimental} conducted a user study to evaluate presentation quality, and even then, just 7 out of 30 participants were cybersecurity professionals.
Regarding user validation, although Lundberg \textit{et al.}~\cite{Lundberg2022experimental} reported increased trust in their visual explanations, the results were not statistically significant. This underscores the need for larger, more rigorous user studies to assess the impact of visualization-based \gls{xai} on trust, especially in high-stakes domains like industrial security~\cite{wang2021explanations}.

User validation itself presents several challenges. First, evaluations must involve the actual stakeholders who will use the explanations, as their goals, expertise, and operational contexts vary significantly. Substituting stakeholders or conducting studies outside their working environments risks generating misleading conclusions. The effectiveness of an explanation is not solely a function of its quality, but also of the user's cognitive capacity, domain knowledge, alertness, and available time~\cite{Lundberg2022experimental, langer2021we}. Experts often rely on tacit knowledge gained through experience, which allows them to process complex information more efficiently~\cite{wang2021explanations}. This expertise also shapes their preferences, experts may favor local explanations for individual predictions, while non-experts may find such explanations confusing or misleading~\cite{bove2022contextualization}.

Therefore, explanation design must be tailored to the specific stakeholders who will use them, with attention to their interpretability needs and operational constraints. Moreover, the relationship between users and explanations is dynamic. As users gain experience or as models evolve, trust can fluctuate. Even a single erroneous AI decision can significantly erode trust, and rebuilding it may require sustained effort~\cite{glikson2020human}. This implies that \gls{xai} evaluation should be treated as a continuous process, with emphasis on how explanations help maintain or restore trust over time.

Another challenge lies in designing and administering stakeholder interviews and questionnaires that yield statistically significant and reliable insights. This process demands dedicated resources and careful attention to phrasing, sampling, and response integrity.

Despite the high cost and limited availability of industrial security stakeholders, their participation in application-level validation, where explanations are tested on real-world tasks, is critical to ensuring the practical usability of \gls{xai} methods~\cite{sokol2020explainability}. These stakeholders include the core \gls{soc} team who will interact with AI systems or their outputs in their daily workflows: \gls{soc} analysts across tier 1 (alert triage), tier 2 (incident response), and tier 3 (threat hunting), \gls{soc} managers, as well as auditors~\cite{vielberth2020security}. External stakeholders also play a vital role, such as \gls{it} teams with expertise in network infrastructure and \gls{it} systems, process engineers with deep knowledge of industrial processes and \gls{ot} systems, and \gls{ml} operations (MLOps) engineers responsible for deploying and maintaining the \gls{ml} models used within the SOC. 

Fostering open, cross-disciplinary dialogue among these groups is essential for advancing human-AI collaboration and shaping the development of \gls{ai} and \gls{xai} tools in industrial security environments.

\subsubsection{Alignment of \gls{xai} Approaches with Human Reasoning} We note that most surveyed papers employed associative \gls{xai} methods, with only a few applying a contrastive approach: \gls{cem}~\cite{Gummadi2024xai}, counterfactual explanations~\cite{lewis2013counterfactuals}, and anchors~\cite{ribeiro2018anchors}. Yet, social science literature suggests that humans naturally use contrastive reasoning when explaining decisions~\cite{byrne2019counterfactuals, miller2019explanation, wang2021explanations}. Contrastive explanations might align more closely with human explanatory preferences, indicating a research gap in the application of contrastive methods to industrial security.

\subsubsection{Stakeholder-Driven \gls{xai} Design} 
The large majority of surveyed papers primarily applied well-established, general-purpose \gls{xai} methods and visualizations without tailoring them to the specific needs of industrial security stakeholders. This highlights a critical gap in current research: the lack of stakeholder-centered design of \gls{xai} tools.

Close engagement with stakeholders in industrial cybersecurity is essential to drive the development of specialized \gls{xai} methods and visualizations that align with their unique usability desiderata. The overarching goal is to foster effective human-AI collaboration grounded in trust. 
Empirical research demonstrates that personalization and intuitive interaction with \gls{ai} systems significantly improve user trust, satisfaction, and adoption~\cite{glikson2020human, choung2023trust}. These qualities are especially critical in light of findings from the \textit{SANS Institute 2024 \gls{soc} Survey}, which reported that \gls{ai} and \gls{ml} technologies ranked lowest in user satisfaction among a wide range of evaluated security tools~\cite{crowley2024sansSOC}.

To achieve this, it is crucial to understand the specific knowledge gaps that stakeholders aim to fill using \gls{ai} within their workflows. \gls{xai} should be designed not merely to explain model behavior in abstract terms, but to directly and precisely address these gaps~\cite{lombrozo2007simplicity, sokol2020explainability}. As emphasized by~\cite{sokol2020explainability}, explanation usability depends on multiple criteria, including the coherence (alignment with prior knowledge), novelty, recency, and fidelity (faithfulness to the model's reasoning) of the explanations. 

Redirecting research toward stakeholder-driven \gls{xai} design in industrial security will likely yield new, specialized methods and visualizations that are tightly coupled to the types of data that these stakeholders observe, the workflows they follow, the documentation and communication they maintain, and the regulatory compliance they must uphold. 

\subsubsection{Sequential \gls{xai} Techniques for Industrial Environments
} 
While several of the surveyed papers employed sequential \gls{ml} models to capture time-dependent patterns in the data, there remains a significant gap in the application of \gls{xai} techniques specifically tailored for sequential data. \gls{xai} methods in these papers were applied at the level of static feature importance, akin to how they are applied to tabular data, without accounting for the temporal dynamics in the time-series data.

In industrial environments, data is inherently sequential, and anomalies often evolve over time. Certain incidents may only become apparent when comparing feature values across multiple time steps, rather than inspecting a single snapshot in isolation. 

Although sequential \gls{ml} models are capable of detecting such temporal anomalies, it is equally important for \gls{xai} methods to provide explanations that reflect this temporal structure. This includes identifying which time steps and transitions contributed most to the model's prediction, and how the evolution of feature values over time influenced the outcome. Without such temporal interpretability, analysts are left with static explanations that may obscure the true cause of time-dependent incidents.

Advancing \gls{xai} for sequential data in industrial security will require new methods that can attribute importance not only to features, but also to their temporal context---enabling explanations that align with how anomalies manifest and propagate in real-world operational systems.

\subsubsection{Engaging and Interactive \gls{xai} Systems} 

Most existing \gls{xai} methods provide static explanations, including all those reviewed in this paper. Static \gls{xai} refers to techniques that generate a fixed, one-time explanation without supporting further user interaction or iterative exploration~\cite{zhang2025may}. However, a well-established body of research highlights the cognitive benefits of allowing explainees to ask questions. Question-asking enables users to articulate and deepen their understanding, connect explanations to prior knowledge, and actively engage with the reasoning process~\cite{taboada2006contributions, yu2009scaffolding, chin2002student, rqi2025research}.

Empirical studies have shown that user comprehension, trust, and collaboration improve significantly when \gls{xai} systems support free-form dialogue or allow flexible exploration of explanations~\cite{zhang2025may, bove2022contextualization, cheng2019explaining}. Rohlfing \textit{et al.}~\cite{rohlfing2020explanation} argue that explanations should be co-constructed through interaction between explainer and explainee, facilitating the establishment of common ground, addressing knowledge gaps, and tailoring explanations to the user's background and needs.
In a user study by Zhang \textit{et al.}~\cite{zhang2025may}, participants frequently asked questions about the functioning of the AI and \gls{xai} methods, sought clarification on visualizations, and generated alternative explanations through exploratory dialogue.

Recent advances in conversational agents, particularly in question answering over \glspl{kg} and \glspl{llm}~\cite{ni2023recent, ouyang2022training, chakraborty2021introduction}, offer promising avenues for developing interactive XAI. In such systems, the agent can incorporate contextual information about the \gls{soc} task, the deployed \gls{ml} models and \gls{xai} techniques, the visualization techniques, and the underlying data. Further, the agent can be programmed to generate explanations based on user questions. This enables dynamic, user-driven conversations where stakeholders can pose specific questions and receive tailored, context-aware responses.

\subsubsection{Physics-Informed Machine Learning for Process Simulations}

In industrial security contexts, where \gls{ot} and process data are governed by well-defined physical dynamics, there is a compelling opportunity to apply \gls{piml}~\cite{sophiya2025comprehensive, karniadakis2021physics}. \gls{piml} integrates domain-specific physical laws into the training of \gls{ml} models. This hybrid approach enables models to learn both data-driven patterns and physically consistent behaviors.
By embedding prior physical dynamics knowledge and constraints, \gls{piml} can provide consistent predictions and explanations, and enhance interpretability by ensuring that predictions align with domain expertise~\cite{karniadakis2021physics}. 

In industrial security, \gls{piml} can be used to model and simulate expected process behavior based on governing physics. 
Compared to traditional \gls{ml}, \gls{piml} can offer several advantages: anomaly detection is based on violations of physical laws rather than statistical outliers; attack classification leverages physics-based propagation modeling rather than pattern recognition; and explanations are grounded in physical mechanisms rather than abstract feature importance. As such, \gls{piml} decisions can be traced to specific physical violations, enabling model-based reasoning and intuitive, mechanistic explanations.

While the application of \gls{piml} in industrial security remains nascent, early work such as~\cite{zideh2023physics} provides a preliminary survey of its potential, particularly in power systems.

\subsubsection{Contextualized Explanations Supported by Knowledge Graphs}

\begin{figure}[t!]
    \centering
    \includegraphics[width=1\linewidth]{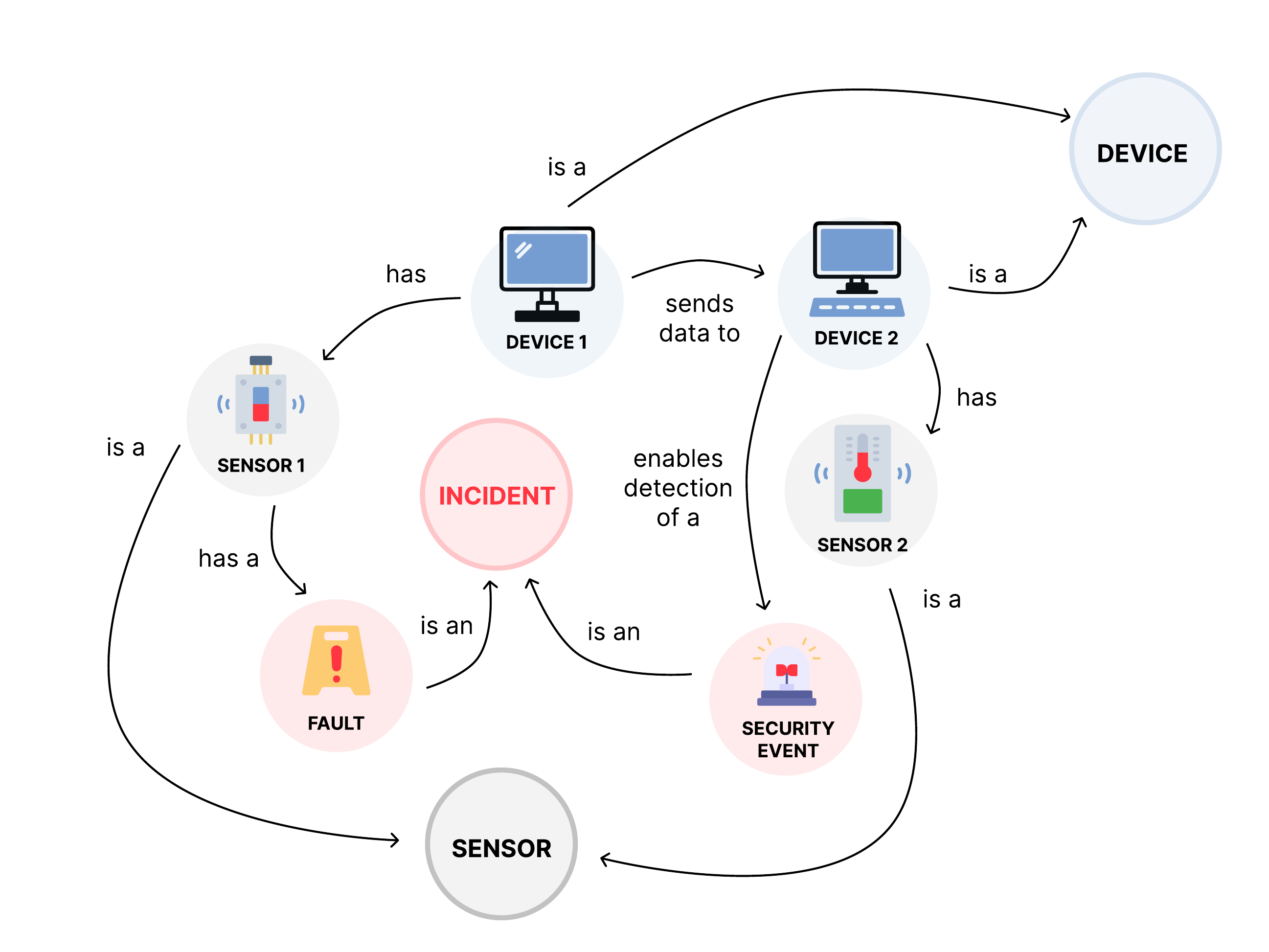}
    \caption{A high-level sample knowledge graph for an industrial security environment.}
    \label{fig:knowledge_graph}
\end{figure}

\gls{kg} combined with \gls{gnn} can offer significant contextual advantages over traditional feature-based machine learning approaches in industrial security.
A \gls{kg} is a structured representation of entities (nodes) and their relationships (edges), where nodes may represent devices, sensors, faults, or security events, and edges encode relationships such as causality, connectivity, or spatial containment. These graphs, as illustrated conceptually in Fig.~\ref{fig:knowledge_graph}, can be constructed from expert knowledge, structured databases, or extracted from unstructured text. In \gls{kg}-\gls{gnn}, the graph structure serves as input for learning node and edge embeddings that capture system roles, dependencies, and vulnerabilities~\cite{ye2022comprehensive}.

This graph-based approach enables anomaly detection as deviations from expected structural patterns, rather than purely statistical outliers. Explanations are grounded in graph semantics, allowing stakeholders to trace causal chains and link alerts to root causes. Such explanations are inherently more interpretable, as they reflect system topology and operational logic, and can be expressed as human-readable narratives derived from graph relationships.

\glspl{kg} can be applied to \gls{ot} data to detect component failures and to trace anomalies back to their originating attack vectors. In \gls{it} contexts, they can support more contextualized explanations of attack classifications based on network and device activity.

Furthermore, \glspl{kg} can be integrated with \gls{llm}-based chat agents within interactive \gls{xai} interfaces. Through \gls{rag}, \glspl{llm} can query \glspl{kg} to retrieve structured facts and relationships that enrich generated responses~\cite{shu2024knowledge, abu2024knowledge}. This fusion enhances the factual grounding and contextual relevance of conversational explanations (as discussed in Section \ref{sec:discussion}).

Recent research has also explored the integration of \glspl{kg} with \gls{piml}~\cite{liu2023physics, ashraf2024physics, thangamuthu2022unravelling}. In such hybrid systems, the knowledge graph provides semantic context---capturing system roles, configurations, and dependencies---while \gls{piml} enforces physical consistency through embedded governing equations. This combination can enable localized, context-sensitive predictions and support multi-modal reasoning: symbolic inference from the \gls{kg} and numerical simulation from the \gls{piml}. 

\subsubsection{Operational Integration of Large Language Models (LLMs) and Industrial Foundation Models (FMs)}
Industry is increasingly integrating LLMs into cybersecurity products and SOC pipelines to provide enrichment (summaries and metadata), recommendations, and decision support. These systems are commonly combined with retrieval (RAG), fine-tuning, or human feedback to align outputs with business needs.

For comprehensive Industrial AI, a complementary approach combining traditional ML/DL and LLMs is essential. Traditional ML/DL models excel at processing numerical data (such as sensor readings, visual defects, numerical time-series), delivering the precision and determinism required for forecasting, anomaly detection, and physics-grounded prediction. LLMs, conversely, manage text-based data (or document-retreived or document-oriented or otherwise) such as manuals, documentation, logs, and code. LLMs provide contextual enrichment by explaining trends and identifying patterns in sequential data, and while they offer valuable decision support and interpretability, they cannot replace purpose-built models for the time-critical, numerically-intensive tasks that demand strict latency guarantees and domain constraints.

However, emerging research on industrial \gls{fms} extends LLM strengths to sensor-driven, time-series, and multimodal tasks. Rather than training from scratch, manufacturers can adapt pre-trained FMs with parameter-efficient fine-tuning, freezing most weights and updating small adapter modules or a few layers---greatly reducing data and compute while preserving performance. It has been demonstrated by researchers at IBM that by training a transformer-based FM on industrial time-series data, the deployment time of models can be cut in half, while also resulting in improved model forecasting accuracy\cite{ayyat2025opportunities}.

Explanations for LLMs/FMs outputs take multiple forms, each with distinct trade-offs between fidelity and usability. The first form, chain-of-thought (CoT) prompting and rationale generation, exposes reasoning traces by decomposing model decisions into intermediate steps. Supervised fine-tuning and reinforcement learning from human feedback (RLHF) can be utilized to shape LLM behavior toward producing a desired output. Additionally, retrieval and attribution methods ground outputs in source documents, improving traceability and trustworthiness~\cite{cambria2024xai}. However, while these techniques enhance interpretability, they introduce computational overhead and may not fully capture the underlying mechanisms driving predictions.

We do not provide in-depth treatment of LLMs/FMs explainability in this survey; however, we acknowledge it as an active research area within the broader and rapidly-evolving landscape of AI transparency and trustworthiness in industrial systems.

\subsection{Addressing Data Availability Limitations and Proposed Solutions}
\subsubsection{Heavy Reliance on IT/IoT Datasets}
\begin{figure}[t!]
    \centering
    \includegraphics[width=1\linewidth]{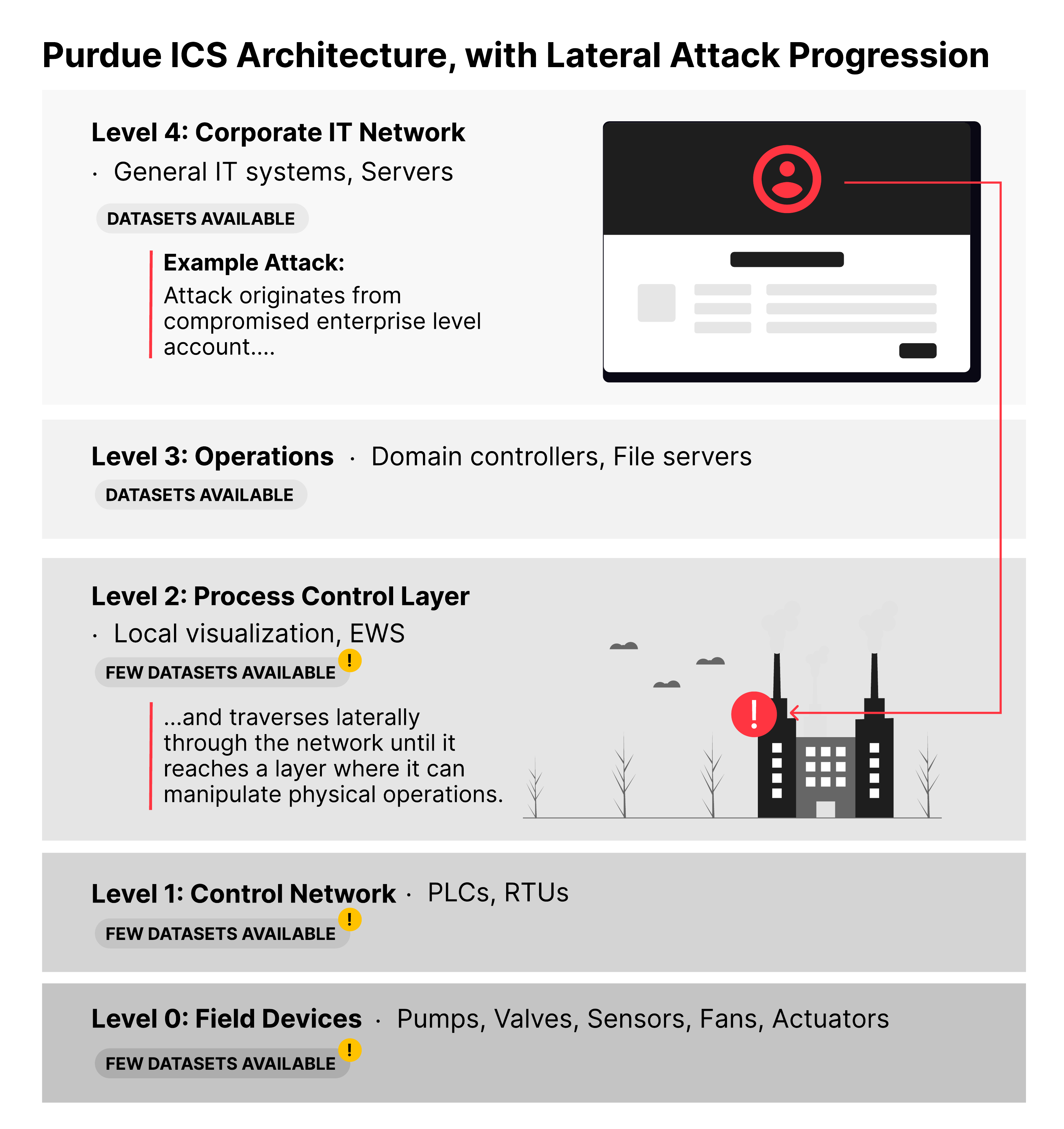}
    \caption{Purdue ICS Architecture with Lateral Attack Progression}
    \label{fig:purdue_attack}
\end{figure}

A critical observation from this survey is that the majority of the surveyed research relied on publicly available benchmark IDS datasets to train and evaluate ML models, and demonstrate XAI methods. 
These datasets offer several advantages: they enable rapid experimentation,
facilitate reproducibility, and allow direct ML performance
comparisons across different research efforts. However, with the focus of these datasets being on IT/IoT environments, much of the research in XAI for industrial security has focused on network traffic analysis.

In contrast, industrial systems' networks are typically divided into hierarchical, segmented levels, such as those defined by the Purdue ICS Architecture. Fig. 6 illustrates the Purdue ICS Architecture, whose primary goal is to isolate physical OT (levels 3 and below) from enterprise IT networks (level 4). A demilitarized zone (level 3.5) acts as a secure, isolated buffer to safely pass data between the OT and IT networks. Within this architecture, successful attacks against industrial systems may need to laterally move across these layers. For example, the BlackEnergy 3 trojan compromised enterprise-level user computers to pivot into the OT network, where it subsequently manipulated physical control equipment to cut power to roughly 230,000 people during the 2015 Ukraine Distribution Grid Attack~\cite{ukrainegrid2024}.

The aim of addressing this scarcity gap in data is to enable the development of XAI for industrial security that identifies and traces cross-level industrial attacks, and contextualizes and correlates IoCs across IT and OT data to better support operational decision-making in
industrial security.

\subsubsection{Realistic Data Generation Solutions}

Many benchmark datasets used in industrial cybersecurity research are synthetic and limited in scope. Typically, researchers employ network simulators to generate traffic and collect packets; a method that is both cost-effective and easier to set up as it avoids the complexities of integrating physical equipment~\cite{de2023survey}. However, synthetic data often struggles to capture the full spectrum of realistic network behavior during both benign and malicious operational conditions.

These datasets also fail to represent the scale and heterogeneity of operational environments. Most \gls{iot} datasets, for instance, involve a relatively small number of devices and lack the diversity of communication protocols and configurations found in real industrial systems~\cite{de2023survey}. As a result, models trained on such datasets may exhibit poor generalization and limited applicability when deployed in live environments. Moreover, synthetic attack vectors may not reflect the complexity, diversity, or tactics of real-world adversaries. 

Another limitation is that \gls{ml} and \gls{xai} models are rarely evaluated in large-scale, real-world settings. Their ability to detect and explain attacks across diverse system components, equipment types, and operational contexts remains largely untested. This gap undermines the reliability and interpretability of \gls{xai} in practical security operations.

Beyond advances in network simulation, several emerging technologies offer promising avenues for improving data realism. First, digital twins, which are dynamic, real-time virtual replicas of physical systems~\cite{jiang2021industrialv2}, can significantly enhance data quality. Digital twins generate time-synchronized data across \gls{ics} layers, including network traffic, sensor readings, control commands, and physical responses. This supports multi-modal, context-aware detection models that learn from both cyber and physical indicators. Unlike high-fidelity simulation testbeds, digital twins support bidirectional communication and are continuously updated with live operational data. Researchers can inject attacks into the twin without endangering real infrastructure~\cite{angelo_ai_2024, eckhart2019digital}, enabling controlled generation of labeled datasets for training and evaluating \gls{ml} security models. Additionally, digital twins support human-in-the-loop experimentation~\cite{cacm2023digitaltwins}, facilitating interactive \gls{xai} and decision support.

Second, \gls{ics} honeypots~\cite{lupia2023ics, franco2021survey} can be used to collect realistic attack data. Honeypots act as digital decoys, luring attackers into interaction while capturing detailed system data during the attackers' engagements. This enables the generation of authentic attack traces from real adversaries. 
The authors in~\cite{krishnaveni2022network, Sivamohan2023optimized} deployed honeypots on AWS cloud infrastructure to log up-to-date cyber incident data.
Advances in honeypot design, including algorithms that increase deception and prolong attacker engagement~\cite{mohamed2025resilient}, can further improve the quality of collected data.  The combination of honeypots and digital twins~\cite{nintsiou2023threat, eckhart2019digital} offers a powerful framework for enhancing both data fidelity and attack realism.

Finally, in the absence of real attacker engagement, \gls{rl} can be used to simulate sophisticated attack behaviors within simulation environments and digital twin testbeds~\cite{nguyen2021deep, mohamed2023reinforcement, mohamed2023use}. RL agents can iteratively learn to generate increasingly complex and adaptive attack strategies, providing a rich source of synthetic data for model training and evaluation.

\section{Conclusion}
\label{sec:conclusion}

This paper presented a comprehensive review of \gls{xai} in the context of industrial cybersecurity. As industrial systems continue to undergo rapid digital transformation along with a deeper \gls{it}/\gls{ot} integration, \gls{ai} and \gls{ml} are increasingly being adopted to support cybersecurity monitoring, threat detection, and incident response. However, the opaque nature of many \gls{ai} models introduces significant challenges for operational trust, safety, and regulatory compliance, particularly in industrial settings where incorrect decisions may have serious consequences.
To address these challenges, this review examined the role of \gls{xai} as an enabling framework for improving the transparency and interpretability of AI-driven cybersecurity systems. The paper first discussed the characteristics of industrial cybersecurity environments and the operational role of \glspl{soc} in monitoring and protecting industrial infrastructure. We then analyzed how AI techniques are currently used to process industrial data sources, detect anomalies, and assist cybersecurity analysts in identifying potential threats.
Next, the survey examined the strengths, limitations, and suitability of these XAI methods and visualizations for industrial cybersecurity applications, with particular attention to operational deployment requirements.
In addition, the paper highlighted the importance of explainability in supporting analyst decision-making, facilitating incident investigation, and meeting emerging regulatory and governance requirements for trustworthy AI systems.
Despite the progress in both AI-based cybersecurity and explainable machine learning, several challenges remain. These include the scarcity of labeled industrial cybersecurity datasets, the difficulty of balancing model accuracy with interpretability, the integration of explanation tools into existing SOC workflows, and the need for standardized evaluation metrics for explainability in safety-critical environments. Addressing these challenges will require interdisciplinary collaboration between cybersecurity researchers, industrial engineers, AI specialists, and regulatory stakeholders.
Overall, \gls{xai} represents a critical step toward enabling trustworthy and operationally viable AI systems for industrial cybersecurity. Continued research in this area will be essential to ensure that AI-driven security solutions not only provide strong detection capabilities but also deliver explanations that are meaningful, reliable, and actionable for human operators responsible for protecting industrial infrastructure.

\bibliographystyle{IEEEtran}
\bibliography{_references}

@article{Zhang2022explainable,
  title = {Explainable Artificial Intelligence Applications in Cyber Security: State-of-the-Art in Research},
  year = {2022},
  journal = {IEEE Access},
  issn = {2169-3536     VO  - 10},
  volume = {10},
  pages = {93104-93139},
  author = {Z. Zhang and H. A. Hamadi and E. Damiani and C. Y. Yeun and F. Taher},
  doi = {10.1109/ACCESS.2022.3204051}
}

@article{Capuano2022explainable_______,
  title = {Explainable Artificial Intelligence in CyberSecurity: A Survey},
  year = {2022},
  journal = {IEEE Access},
  issn = {2169-3536     VO  - 10},
  volume = {10},
  pages = {93575-93600},
  author = {N. Capuano and G. Fenza and V. Loia and C. Stanzione},
  doi = {10.1109/ACCESS.2022.3204171}
}

@article{Houda2022why,
  title = {“Why Should I Trust Your IDS?”: An Explainable Deep Learning Framework for Intrusion Detection Systems in Internet of Things Networks},
  year = {2022},
  journal = {IEEE Open Journal of the Communications Society},
  issn = {2644-125X     VO  - 3},
  volume = {3},
  pages = {1164-1176},
  author = {El Houda, Zakaria Abou and B. Brik and L. Khoukhi},
  doi = {10.1109/OJCOMS.2022.3188750}
}

@article{Moustafa2023explainable,
  title = {Explainable Intrusion Detection for Cyber Defences in the Internet of Things: Opportunities and Solutions},
  year = {2023},
  journal = {IEEE Communications Surveys \& Tutorials},
  issn = {1553-877X     VO  - 25},
  volume = {25},
  number = {3},
  pages = {1775-1807},
  author = {N. Moustafa and N. Koroniotis and M. Keshk and A. Y. Zomaya and Z. Tari},
  doi = {10.1109/COMST.2023.3280465}
}

@article{Oseni2023explainable,
  title = {An Explainable Deep Learning Framework for Resilient Intrusion Detection in IoT-Enabled Transportation Networks},
  year = {2023},
  journal = {IEEE Transactions on Intelligent Transportation Systems},
  issn = {1558-0016     VO  - 24},
  volume = {24},
  number = {1},
  pages = {1000-1014},
  author = {A. Oseni and N. Moustafa and G. Creech and N. Sohrabi and A. Strelzoff and Z. Tari and I. Linkov},
  doi = {10.1109/TITS.2022.3188671}
}

@article{Zolanvari2023trust,
  title = {TRUST XAI: Model-Agnostic Explanations for {AI} With a Case Study on IIoT Security},
  year = {2023},
  journal = {IEEE Internet of Things Journal},
  issn = {2327-4662     VO  - 10},
  volume = {10},
  number = {4},
  pages = {2967-2978},
  author = {M. Zolanvari and Z. Yang and K. Khan and R. Jain and N. Meskin},
  doi = {10.1109/JIOT.2021.3122019}
}

@article{Javeed2024explainable,
  title = {An Explainable and Resilient Intrusion Detection System for Industry 5.0},
  year = {2024},
  journal = {IEEE Transactions on Consumer Electronics},
  issn = {1558-4127     VO  - 70},
  volume = {70},
  number = {1},
  pages = {1342-1350},
  author = {D. Javeed and T. Gao and P. Kumar and A. Jolfaei},
  doi = {10.1109/TCE.2023.3283704}
}

@article{Hwang2021e,
  title = {E-SFD: Explainable Sensor Fault Detection in the ICS Anomaly Detection System},
  year = {2021},
  journal = {IEEE Access},
  issn = {2169-3536     VO  - 9},
  volume = {9},
  pages = {140470-140486},
  author = {C. Hwang and T. Lee},
  doi = {10.1109/ACCESS.2021.3119573}
}

@article{Rjoub2023survey,
  title = {A Survey on Explainable Artificial Intelligence for Cybersecurity},
  year = {2023},
  journal = {IEEE Transactions on Network and Service Management},
  issn = {1932-4537     VO  - 20},
  volume = {20},
  number = {4},
  pages = {5115-5140},
  author = {G. Rjoub and J. Bentahar and O. Abdel Wahab and R. Mizouni and A. Song and R. Cohen and H. Otrok and A. Mourad},
  doi = {10.1109/TNSM.2023.3282740}
}

@article{Shtayat2023explainable,
  title = {An Explainable Ensemble Deep Learning Approach for Intrusion Detection in Industrial Internet of Things},
  year = {2023},
  journal = {IEEE Access},
  issn = {2169-3536     VO  - 11},
  volume = {11},
  pages = {115047-115061},
  author = {M. M. Shtayat and M. K. Hasan and R. Sulaiman and S. Islam and A. U. R. Khan},
  doi = {10.1109/ACCESS.2023.3323573}
}

@article{Lundberg2022experimental,
  title = {Experimental Analysis of Trustworthy In-Vehicle Intrusion Detection System Using eXplainable Artificial Intelligence (XAI)},
  year = {2022},
  journal = {IEEE Access},
  issn = {2169-3536     VO  - 10},
  volume = {10},
  pages = {102831-102841},
  author = {H. Lundberg and N. I. Mowla and S. F. Abedin and K. Thar and A. Mahmood and M. Gidlund and S. Raza},
  doi = {10.1109/ACCESS.2022.3208573}
}

@article{Kalakoti2024improving,
  title = {Improving IoT Security With Explainable AI: Quantitative Evaluation of Explainability for IoT Botnet Detection},
  year = {2024},
  journal = {IEEE Internet of Things Journal},
  issn = {2327-4662     VO  - 11},
  volume = {11},
  number = {10},
  pages = {18237-18254},
  author = {R. Kalakoti and H. Bahsi and S. Nõmm},
  doi = {10.1109/JIOT.2024.3360626}
}

@article{Kumar2024blockchain,
  title = {Blockchain-Based Authentication and Explainable {AI} for Securing Consumer IoT Applications},
  year = {2024},
  journal = {IEEE Transactions on Consumer Electronics},
  issn = {1558-4127     VO  - 70},
  volume = {70},
  number = {1},
  pages = {1145-1154},
  author = {R. Kumar and D. Javeed and A. Aljuhani and A. Jolfaei and P. Kumar and A. K. M. N. Islam},
  doi = {10.1109/TCE.2023.3320157}
}

@article{Gummadi2024xai,
  title = {XAI-IoT: An Explainable {AI} Framework for Enhancing Anomaly Detection in IoT Systems},
  year = {2024},
  journal = {IEEE Access},
  issn = {2169-3536     VO  - 12},
  volume = {12},
  pages = {71024-71054},
  author = {A. Namrita Gummadi and J. C. Napier and M. Abdallah},
  doi = {10.1109/ACCESS.2024.3402446}
}

@article{Attique2024explainable,
  title = {Explainable and Data-Efficient Deep Learning for Enhanced Attack Detection in IIoT Ecosystem},
  year = {2024},
  journal = {IEEE Internet of Things Journal},
  issn = {2327-4662     VO  - 11},
  volume = {11},
  number = {24},
  pages = {38976-38986},
  author = {D. Attique and W. Hao and W. Ping and D. Javeed and P. Kumar},
  doi = {10.1109/JIOT.2024.3384374}
}

@article{Gurbuz2023explainable,
  title = {Explainable AI-Based Malicious Traffic Detection and Monitoring System in Next-Gen IoT Healthcare},
  year = {2023},
  journal = {2023 International Conference on Smart Applications, Communications and Networking (SmartNets)},
  issn = {     VO  - },
  pages = {1-6},
  author = {E. G\"urb\"uz and \"O. Turgut and I. K\"ok},
  doi = {10.1109/SmartNets58706.2023.10215896}
}

@article{Gaitan-Cardenas2023explainable,
  title = {Explainable AI-Based Intrusion Detection Systems for Cloud and IoT},
  year = {2023},
  journal = {2023 32nd International Conference on Computer Communications and Networks (ICCCN)},
  issn = {2637-9430     VO  - },
  pages = {1-7},
  author = {M. C. Gaitan-Cardenas and M. Abdelsalam and K. Roy},
  doi = {10.1109/ICCCN58024.2023.10230177}
}

@article{Alani2023explainable,
  title = {Explainable Ensemble-Based Detection of Cyber Attacks on Internet of Medical Things},
  year = {2023},
  journal = {2023 IEEE Intl Conf on Dependable, Autonomic and Secure Computing, Intl Conf on Pervasive Intelligence and Computing, Intl Conf on Cloud and Big Data Computing, Intl Conf on Cyber Science and Technology Congress (DASC/PiCom/CBDCom/CyberSciTech)},
  issn = {2837-0740     VO  - },
  pages = {0609-0614},
  author = {M. M. Alani and A. Mashatan and A. Miri},
  doi = {10.1109/DASC/PiCom/CBDCom/Cy59711.2023.10361448}
}

@article{Kalakoti2023improving,
  title = {Improving Transparency and Explainability of Deep Learning Based IoT Botnet Detection Using Explainable Artificial Intelligence (XAI)},
  year = {2023},
  journal = {2023 International Conference on Machine Learning and Applications (ICMLA)},
  issn = {1946-0759     VO  - },
  pages = {595-601},
  author = {R. Kalakoti and S. Nõmm and H. Bahsi},
  doi = {10.1109/ICMLA58977.2023.00088}
}

@article{Patil2022explainable,
  title = {Explainable artificial intelligence for intrusion detection system},
  year = {2022},
  journal = {Electronics},
  author = {Patil, Shruti and Vijayakumar Varadarajan and Siddiqui Mohd Mazhar and Abdulwodood Sahibzada and Nihal Ahmed and Onkar Sinha and Satish Kumar and Kailash Shaw and Ketan Kotecha},
  publisher = {mdpi.com}
}

@article{Suhail2023enigma,
  title = {ENIGMA: An explainable digital twin security solution for cyber–physical systems},
  year = {2023},
  journal = {Computers in Industry},
  author = {Suhail, S and Iqbal, M and Hussain, R and Jurdak, R},
  publisher = {Elsevier}
}

@article{Wali2025explainable,
  title = {Explainable AI and Random Forest Based Reliable Intrusion Detection system},
  journal = {Computers \& Security},
  volume = {157},
  year = {2025},
  doi = {10.1016/j.cose.2025.104542},
  author = {Syed Wali and Yasir Ali Farrukh and Irfan Khan},
}

@article{Arreche2024xai1,
  title = {Xai-ids: Toward proposing an explainable artificial intelligence framework for enhancing network intrusion detection systems},
  year = {2024},
  journal = {Applied Sciences},
  author = {Arreche, O and Guntur, T and Abdallah, M},
  publisher = {mdpi.com}
}

@article{Sivamohan2023tea,
  title = {TEA-EKHO-IDS: An intrusion detection system for industrial CPS with trustworthy explainable {AI} and enhanced krill herd optimization},
  year = {2023},
  journal = {Peer-to-Peer Networking and Applications},
  author = {Sivamohan, S and Sridhar, SS and Krishnaveni, S},
  publisher = {Springer},
  doi = {10.1007/s12083-023-01507-8}
}

@article{Hasan2023advanced,
  title = {Advanced persistent threat identification with boosting and explainable AI},
  year = {2023},
  journal = {SN Computer Science},
  author = {Hasan, MM and Islam, MU and Uddin, J},
  publisher = {Springer},
  doi = {10.1007/s42979-023-01744-x}
}

@article{Al-Hawawreh2024explainable,
  title = {Explainable deep learning for attack intelligence and combating cyber–physical attacks},
  year = {2024},
  journal = {Ad Hoc Networks},
  author = {Al-Hawawreh, M and Moustafa, N},
  publisher = {Elsevier}
}

@article{Kumar2023explainable,
  title = {Explainable artificial intelligence envisioned security mechanism for cyber threat hunting},
  year = {2023},
  journal = {Security and Privacy},
  author = {Kumar, Pankaj and Wazid, Mohammad and Singh, D. P. and Singh, Jaskaran and Das, Ashok Kumar and Park, Youngho and Rodrigues, Joel J. P. C.},
  publisher = {Wiley Online Library},
  doi = {10.1002/spy2.312}
}

@article{Muna2023demystifying,
  title = {Demystifying machine learning models of massive IoT attack detection with Explainable {AI} for sustainable and secure future smart cities},
  year = {2023},
  journal = {Internet of Things},
  author = {Rabeya Khatun Muna and Muhammad Iqbal Hossain and Md. Golam Rabiul Alam and Mohammad Mehedi Hassan and Michele Ianni and Giancarlo Fortino},
  publisher = {Elsevier}
}

@article{Kalutharage2022explainable,
  title = {Explainable {AI} and deep autoencoders based security framework for IoT network attack certainty},
  year = {2022},
  journal = {Attacks and Defenses for the Internet-of-Things},
  author = {Kalutharage, CS and Liu, X and Chrysoulas, C},
  publisher = {Springer},
  doi = {10.1007/978-3-031-21311-3_8}
}

@article{Sharma2023anomaly,
  title = {Anomaly-based DNN model for intrusion detection in IoT and model explanation: Explainable artificial intelligence},
  year = {2023},
  journal = {Proceedings of Second International Conference on Computational Electronics for Wireless Communications},
  author = {Sharma, B and Sharma, L and Lal, C},
  publisher = {Springer},
  doi = {10.1007/978-981-19-6661-3_28}
}

@article{Mallampati2024enhancing,
  title = {Enhancing intrusion detection with explainable ai: A transparent approach to network security},
  year = {2024},
  journal = {Cybernetics and Information Technologies},
  author = {Mallampati, SB and Seetha, H},
  publisher = {sciendo.com},
  doi = {10.2478/cait-2024-0006}
}

@article{Ali2024explainable,
  title = {Explainable artificial intelligence enabled intrusion detection in the internet of things},
  year = {2024},
  journal = {International Symposium on Intelligent Computing and Networking},
  author = {Ali, M and Zhang, J},
  publisher = {Springer},
  doi = {10.1007/978-3-031-67447-1_30}
}

@article{Kalutharage2023explainable,
  title = {Explainable AI-Based DDOS Attack Identification Method for IoT Networks},
  year = {2023},
  journal = {Computers (Basel)},
  issn = {1886-1881},
  volume = {12},
  pages = {32},
  author = {Kalutharage, Chathuranga Sampath and Liu, Xiaodong and Chrysoulas, C. and Pitropakis, Nikolaos and Papadopoulos, Pavlos},
  doi = {10.3390/computers12020032}
}

@article{Sivamohan2023optimized,
  title = {An optimized model for network intrusion detection systems in industry 4.0 using XAI based Bi-LSTM framework},
  year = {2023},
  journal = {Neural Computing \&  Applications},
  volume = {35},
  pages = {11459-11475},
  author = {Sivamohan, S. and Sridhar, S.},
  doi = {10.1007/s00521-023-08319-0},
}

@article{Houda2022novel,
  title = {A Novel IoT-Based Explainable Deep Learning Framework for Intrusion Detection Systems},
  year = {2022},
  journal = {IEEE Internet of Things Magazine},
  issn = {2576-3180},
  volume = {5},
  pages = {20-23},
  author = {Houda, Zakaria Abou El and Brik, Bouziane and Senouci, Sidi-Mohammed},
  doi = {10.1109/IOTM.005.2200028}
}

@article{Ha2022explainable,
  title = {Explainable Anomaly Detection for Industrial Control System Cybersecurity},
  year = {2022},
  journal = {ArXiv},
  issn = {2405-8963},
  author={Do Thu Ha and Nguyen Xuan Hoang and Nguyen Viet Hoang and Nguyen Huu Du and Truong Thu Huong and Kim Phuc Tran},
  doi = {10.48550/arXiv.2205.01930}
}

@article{Khan2022proactive,
  title = {A Proactive Attack Detection for Heating, Ventilation, and Air Conditioning (HVAC) System Using Explainable Extreme Gradient Boosting Model (XGBoost)},
  year = {2022},
  journal = {Sensors (Basel, Switzerland)},
  issn = {1424-8220},
  volume = {22},
  author = {Khan, Irfan Ullah and Aslam, Nida and AlShedayed, Rana and AlFrayan, Dina and AlEssa, Rand and AlShuail, Noura A. and Safwan, Alhawra Al},
  doi = {10.3390/s22239235}
}

@article{Siganos2023explainable,
  title = {Explainable AI-based Intrusion Detection in the Internet of Things},
  year = {2023},
  journal = {Proceedings of the 18th International Conference on Availability, Reliability and Security},
  issn = {2554-2656},
  author = {Siganos, M. and Radoglou-Grammatikis, Panagiotis I. and Kotsiuba, I. and Markakis, E. and Moscholios, I. and Goudos, Sotirios K and Sarigiannidis, Panos},
  doi = {10.1145/3600160.3605162}
}

@misc{nslkdd,
  author       = {{Canadian Institue of Cybersecurity}},
  title        = {NSL-KDD Dataset},
  year         = {2009},
  howpublished = {\url{https://www.unb.ca/cic/datasets/nsl.html}},
  note         = {Accessed: 2025-06-19}
}

@INPROCEEDINGS{mostafa2015unswnb,
  author={Moustafa, Nour and Slay, Jill},
  booktitle={2015 Military Communications and Information Systems Conference (MilCIS)}, 
  title={UNSW-NB15: a comprehensive data set for network intrusion detection systems (UNSW-NB15 network data set)}, 
  year={2015},
  volume={},
  number={},
  pages={1-6},
  doi={10.1109/MilCIS.2015.7348942}
}

@article{elnour2021application,
  title={Application of data-driven attack detection framework for secure operation in smart buildings},
  author={Elnour, Mariam and Meskin, Nader and Khan, Khaled and Jain, Raj},
  journal={Sustainable Cities and Society},
  volume={69},
  pages={102816},
  year={2021},
  publisher={Elsevier}
}

@inproceedings{morris2014industrial,
  title={Industrial control system traffic data sets for intrusion detection research},
  author={Morris, Thomas and Gao, Wei},
  booktitle={Critical Infrastructure Protection VIII: 8th IFIP WG 11.10 International Conference, ICCIP 2014, Arlington, VA, USA, March 17-19, 2014, Revised Selected Papers 8},
  pages={65--78},
  year={2014},
  organization={Springer}
}

@inproceedings{murray2017convergence,
  title={The convergence of {IT} and {OT} in critical infrastructure},
  author={Murray, Glenn and Johnstone, Michael N and Valli, Craig},
  booktitle = {Proceedings of the 15th Australian Information Security Management Conference},
  year={2017},
  pages = {149--155},
  publisher = {Edith Cowan University}
}

@article{matt2023industrial,
  title={Industrial digitalization. A systematic literature review and research agenda},
  author={Matt, Dominik T and Pedrini, Giulio and Bonfanti, Angelo and Orzes, Guido},
  journal={European Management Journal},
  volume={41},
  number={1},
  pages={47--78},
  year={2023},
  publisher={Elsevier}
}

@article{lasi2014industry,
  title={Industry 4.0},
  author={Lasi, Heiner and Fettke, Peter and Kemper, Hans-Georg and Feld, Thomas and Hoffmann, Michael},
  journal={Business \& information systems engineering},
  volume={6},
  pages={239--242},
  year={2014},
  publisher={Springer}
}

@incollection{singh_artificial_2020,
	address = {Cham},
	title = {Artificial {Intelligence} and {Security} of {Industrial} {Control} {Systems}},
	isbn = {978-3-030-38557-6},
	booktitle = {Handbook of {Big} {Data} {Privacy}},
	publisher = {Springer International Publishing},
	author = {Singh, Suby and Karimipour, Hadis and HaddadPajouh, Hamed and Dehghantanha, Ali},
	editor = {Choo, Kim-Kwang Raymond and Dehghantanha, Ali},
	year = {2020},
	doi = {10.1007/978-3-030-38557-6_7},
	pages = {121--164},
}

@article{khayat_empowering_2025,
	title = {Empowering {Security} {Operation} {Center} {With} {Artificial} {Intelligence} and {Machine} {Learning}—{A} {Systematic} {Literature} {Review}},
	volume = {13},
	issn = {2169-3536},
	doi = {10.1109/ACCESS.2025.3532951},
	urldate = {2025-03-25},
	journal = {IEEE Access},
	author = {Khayat, Mohamad and Barka, Ezedin and Adel Serhani, Mohamed and Sallabi, Farag and Shuaib, Khaled and Khater, Heba M.},
	year = {2025},
	pages = {19162--19197},
}

@article{vielberth2020security,
  title={Security operations center: A systematic study and open challenges},
  author={Vielberth, Manfred and B{\"o}hm, Fabian and Fichtinger, Ines and Pernul, G{\"u}nther},
  journal={Ieee Access},
  volume={8},
  pages={227756--227779},
  year={2020},
  publisher={IEEE}
}

@article{bhamare2020cybersecurity,
  title={Cybersecurity for industrial control systems: A survey},
  author={Bhamare, Deval and Zolanvari, Maede and Erbad, Aiman and Jain, Raj and Khan, Khaled and Meskin, Nader},
  journal={Computers \& Security},
  volume={89},
  pages={101677},
  year={2020},
  publisher={Elsevier}
}

@article{kayan2022cybersecurity,
  title={Cybersecurity of industrial cyber-physical systems: A review},
  author={Kayan, Hakan and Nunes, Matthew and Rana, Omer and Burnap, Pete and Perera, Charith},
  journal={ACM Computing Surveys (CSUR)},
  volume={54},
  number={11s},
  pages={1--35},
  year={2022},
  publisher={ACM New York, NY}
}

@misc{idaho2016cyber,
	title        = {Cyber Threat and Vulnerability Analysis of the US Electric Sector},
	author       = {{Idaho National Laboratory}},
	year         = 2016,
	publisher    = {Idaho National Laboratory Idaho Falls}
}

@techreport{falco2006using,
  title={Using host-based anti-virus software on industrial control systems: Integration guidance and a test methodology for assessing performance impacts},
  author={Falco, Joseph A and Hurd, Steve and Teumim, Dave},
  institution = {National Institute of Standards and Technology (NIST)},
  type = {NIST Special Publication 1058},
  year={2006}
}

@techreport{stouffer_guide_2023,
	address = {Gaithersburg, MD},
	title = {Guide to {Operational} {Technology} ({OT}) security},
	url = {https://nvlpubs.nist.gov/nistpubs/SpecialPublications/NIST.SP.800-82r3.pdf},
	language = {en},
	number = {NIST SP 800-82r3},
	urldate = {2025-06-26},
	institution = {National Institute of Standards and Technology (U.S.)},
	author = {Stouffer, Keith and Pease, Michael and Tang, CheeYee and Zimmerman, Timothy and Pillitteri, Victoria and Lightman, Suzanne and Hahn, Adam and Saravia, Stephanie and Sherule, Aslam and Thompson, Michael},
	month = sep,
	year = {2023},
	doi = {10.6028/NIST.SP.800-82r3},
}

@misc{cisa2025,
  author       = {{Cybersecurity and Infrastructure Security Agency (CISA)}},
  title        = {Primary Mitigations to Reduce Cyber Threats to Operational Technology},
  year         = {2025},
  url          = {https://www.cisa.gov/resources-tools/resources/primary-mitigations-reduce-cyber-threats-operational-technology},
  urldate = {2025-07-01}
}

@misc{dragos_why_ot_visibility,
	title = {Why {OT} Visibility is Crucial for Industrial Cybersecurity},
	url = {https://www.dragos.com/resources/guide/https-hub-dragos-com-guide-why-ot-visibility-is-crucial-for-industrial-cyber-security/},
	urldate = {2025-07-01},
    year={2025},
    author = {Dragos}
}

@misc{skaronis2024cybersecurity,
	author = {Skaronis, Peter},
    title = {Cybersecurity in the {Manufacturing} {Industry}},
	url = {https://insecm.ca/en/newsletter/cybersecurity-in-the-manufacturing-industry/},
	urldate = {2025-07-01},
	journal = {The Canadian Cybersecurity Cluster (In-Sec-M)},
    year = 2024,
    month = jul,
}

@misc{nerc2014sanctions,
    title = {Sanction Guidelines of the North American Electric Reliability Corporation},
	url = {https://www.nerc.com/pa/Stand/Resources/Documents/Appendix\_4B\_of \_the\_Rules\_of\_Procedure\_Sanction\_Guidelines.pdf},
	urldate = {2025-07-01},
	author = {{North American Electric Reliability Corporation (NERC)}},
    year = 2014,
    month = jul,
}

@misc{dean_parsons_icsot_nodate,
	title = {{ICS}/{OT} {Cybersecurity} \& {AI}: {Considerations} for {Now} and the {Future} ({Part} {I})},
	url = {https://www.sans.org/blog/ics-ot-cybersecurity-ai-considerations-for-now-the-future-part-i/},
	urldate = {2025-07-02},
	journal = {SANS Institute},
	author = {{Dean Parsons}},
    year={2024},
    month={5},
}

@misc{nozomi_howAIisUsed,
	title = {How {Is} {AI} {Used} in {OT}/{ICS} {Cybersecurity}?},
	url = {https://www.nozominetworks.com/cybersecurity-faqs/how-is-ai-used-in-ot-ics-cybersecurity},
	language = {en},
	year = {2025},
	journal = {Nozomi Networks},
	author ={{Nozomi Networks}},
}

@misc{angelo_ai_2024,
	title = {{AI} in {OT} {Security} — {Balancing} {Industrial} {Innovation} and {Cyber} {Risk}},
	url = {https://www.paloaltonetworks.com/blog/2024/08/ai-in-ot-security/},
	language = {en-US},
	urldate = {2025-07-02},
	journal = {Palo Alto Networks Blog},
	author = {Angelo, Dena De},
	month = aug,
	year = {2024},
}

@misc{oakley_cox_three_nodate,
	title = {Three {Ways} {AI} {Secures} {OT} \& {ICS} from {Cyber} {Attacks}},
	url = {https://www.darktrace.com/blog/three-ways-ai-secures-operational-technology-ot-industrial-control-systems-ics-from-cyber-attacks},
	language = {en-US},
	urldate = {2025-07-02},
	author = {{Oakley Cox}},
}

@article{yuan2021deep,
  title={Deep learning for insider threat detection: Review, challenges and opportunities},
  author={Yuan, Shuhan and Wu, Xintao},
  journal={Computers \& Security},
  volume={104},
  pages={102221},
  year={2021},
  publisher={Elsevier}
}

@misc{bcg2024cybersecurity,
	title = {2024 CYBERSECURITY WORKFORCE REPORT: Bridging the Workforce  Shortage and Skills Gap},
	url = {https://web-assets.bcg.com/61/d3/705fbd684d70b0e5f98cdcf7cf47/2024-cybersecurity-workforce-report.pdf},
	urldate = {2025-07-02},
	author = {{Boston Consulting Group (BCG)}},
}

@misc{nerc_top010_1,
  author       = {{North American Electric Reliability Corporation (NERC)}},
  title        = {Reliability Standard TOP-010-1(i): Real-time Reliability Monitoring and Analysis Capabilities},
  year         = {2016},
  url = {https://www.nerc.com/pa/Stand/Reliability\%20Standards/TOP-010-1(i).pdf},
  note         = {Accessed: 2025-07-02},
}

@misc{nerc_cip015_1,
  author       = {{North American Electric Reliability Corporation (NERC)}},
  title        = {Reliability Standard CIP-015-1: Cyber Security — Communications Between Control Centers},
  year         = {2024},
  url = {https://www.nerc.com/pa/Stand/Reliability\%20Standards/CIP-015-1.pdf},
  note         = {Accessed: 2025-07-02},
}

@misc{nis2_2022,
  author       = {{European Parliament and Council of the European Union}},
  title        = {Directive (EU) 2022/2555 of the European Parliament and of the Council of 14 December 2022 on measures for a high common level of cybersecurity across the Union (NIS2 Directive)},
  year         = {2022},
  url = {https://eur-lex.europa.eu/legal-content/EN/TXT/PDF/?uri=CELEX:32022L2555},
  note         = {Accessed: 2025-07-02},
}

@misc{iec_62443,
  author       = {{International Electrotechnical Commission (IEC)}},
  title        = {IEC 62443 Series: Industrial communication networks – Network and system security for industrial automation and control systems},
  year         = {2018},
  url = {{https://www.isa.org/standards-and-publications/isa-standards/isa-iec-62443-series-of-standards}},
  note         = {Accessed: 2025-07-02}
}

@misc{houda_2023,
  author       = {El Houda, Zakaria Abou and Moudoud, Hajar and Brik, Bouziane and Khoukhi, Lyes},
  title        = {Securing Federated Learning through Blockchain and Explainable {AI} for Robust Intrusion Detection in IoT Networks},
  year         = {2023},
  howpublished = {IEEE INFOCOM WKSHPS: ICCN 2023},
  url          = {https://ieeexplore.ieee.org/document/10225769}
}

@misc{Kaur_2024,
  author={Kaur, Navneet and Gupta, Lav},
  title={Enhancing IoT Security in 6G Environment With Transparent AI: Leveraging XGBoost, SHAP and LIME}, 
  year={2024},
  howpublished={2024 IEEE 10th International Conference on Network Softwarization (NetSoft)}, 
  url={https://ieeexplore.ieee.org/document/10588922} 
}

@misc{Ben_2021,
  author={Ben Rabah, Nourhène and Le Grand, Bénédicte and Pinheiro, Manuele Kirsch},
  title={IoT Botnet Detection using Black-box Machine Learning Models: the Trade-off between Performance and Interpretability}, 
  year={2021},
  howpublished={2021 IEEE 30th International Conference on Enabling Technologies: Infrastructure for Collaborative Enterprises (WETICE)}, 
  url={https://ieeexplore.ieee.org/document/9680493}
}

@misc{Mahbooba_2021,
  author={Basim Mahbooba and Mohan Timilsina and Radhya Sahal and Martin Serrano},
  title={Explainable Artificial Intelligence (XAI) to Enhance Trust Management in Intrusion Detection Systems Using Decision Tree Model},
  year={2021},
  booktitle={Complexity},
  url={https://onlinelibrary.wiley.com/doi/10.1155/2021/6634811}
}

@misc{Keshk_2023,
  author={Keshk, Marwa and Koroniotis, Nickolaos and Pham, Nam and Moustafa, Nour and Turnbull, Benjamin and Zomaya, Albert Y.},
  title={An explainable deep learning-enabled intrusion detection framework in IoT networks},
  year={2023},
  journal={Information Sciences},
  url={https://www.sciencedirect.com/science/article/pii/S0020025523005856}
}

@misc{Sharma_2024,
  author={Bhawana Sharma and Lokesh Sharma and Chhagan Lal and Satyabrata Roy},
  title={Explainable artificial intelligence for intrusion detection in IoT networks: A deep learning based approach},
  year={2024},
  journal={Expert Systems with Applications},
  publisher={Elsevier},
  url={https://www.sciencedirect.com/science/article/pii/S0957417423022534}
}

@article{glikson2020human,
  title={Human trust in artificial intelligence: Review of empirical research},
  author={Glikson, Ella and Woolley, Anita Williams},
  journal={Academy of management annals},
  volume={14},
  number={2},
  pages={627--660},
  year={2020},
  publisher={Briarcliff Manor, NY}
}

@article{choung2023trust,
  title={Trust in {AI} and its role in the acceptance of {AI} technologies},
  author={Choung, Hyesun and David, Prabu and Ross, Arun},
  journal={International Journal of Human--Computer Interaction},
  volume={39},
  number={9},
  pages={1727--1739},
  year={2023},
  publisher={Taylor \& Francis}
}

@article{langer2021we,
  title={What do we want from Explainable Artificial Intelligence (XAI)?--A stakeholder perspective on XAI and a conceptual model guiding interdisciplinary XAI research},
  author={Langer, Markus and Oster, Daniel and Speith, Timo and Hermanns, Holger and K{\"a}stner, Lena and Schmidt, Eva and Sesing, Andreas and Baum, Kevin},
  journal={Artificial intelligence},
  volume={296},
  pages={103473},
  year={2021},
  publisher={Elsevier}
}

@article{mayer1995integrative,
  title={An integrative model of organizational trust},
  author={Mayer, Roger C and Davis, James H and Schoorman, F David},
  journal={Academy of management review},
  volume={20},
  number={3},
  pages={709--734},
  year={1995},
  publisher={Academy of Management Briarcliff Manor, NY 10510}
}

@article{hoff2015trust,
  title={Trust in automation: Integrating empirical evidence on factors that influence trust},
  author={Hoff, Kevin Anthony and Bashir, Masooda},
  journal={Human factors},
  volume={57},
  number={3},
  pages={407--434},
  year={2015},
  publisher={Sage Publications Sage CA: Los Angeles, CA}
}

@article{shin2021effects,
  title={The effects of explainability and causability on perception, trust, and acceptance: Implications for explainable AI},
  author={Shin, Donghee},
  journal={International journal of human-computer studies},
  volume={146},
  pages={102551},
  year={2021},
  publisher={Elsevier}
}

@misc{oecd_trustAI,
	title = {{OECD} {AI} {Principles} {Overview}},
	url = {https://oecd.ai/en/principles},
	language = {en},
	urldate = {2025-08-14},
	author = {{Organisation for Economic Co-operation and Development (OECD)}},
	month = may,
	year = {2024},
}

@misc{microsoft_trustAI,
	title = {Microsoft {Trustworthy} {AI}: {Unlocking} human potential starts with trust},
	shorttitle = {Microsoft {Trustworthy} {AI}},
	url = {https://blogs.microsoft.com/blog/2024/09/24/microsoft-trustworthy-ai-unlocking-human-potential-starts-with-trust/},
	language = {en-US},
	urldate = {2025-08-14},
	journal = {The Official Microsoft Blog},
	author = {Numoto, Takeshi},
	month = sep,
	year = {2024},
}

@misc{ieee_trustAI,
	title = {{IEEE} {Standards} {Association} {Announces} {Joint} {Specification} {V1}.0 for the {Assessment} of the {Trustworthiness} of {AI} {Systems}},
	url = {https://standards.ieee.org/news/joint-specification-trustworthy-ai-systems/},
	language = {en},
    author={{{IEEE} Standards Association}},
	urldate = {2025-08-14},
	journal = {IEEE Standards Association},
	month = nov,
	year = {2024},
}

@misc{eu_trustAI,
	title = {Ethics guidelines for trustworthy {AI} {\textbar} {Shaping} {Europe}’s digital future},
	url = {https://digital-strategy.ec.europa.eu/en/library/ethics-guidelines-trustworthy-ai},
	language = {en},
	urldate = {2025-08-14},
	author = {{High-Level Expert Group on Artificial Intelligence, European Commission}},
}

@misc{eu_aiAct_article113,
	title = {The {EU} Artificial Intelligence Act. Article 113},
	url = {https://artificialintelligenceact.eu/article/113/},
	language = {en},
	year = {2025},
	author = {{Future of Life Institute}},
}

@misc{eu_aiAct_recital55,
	title = {The {EU} Artificial Intelligence Act. Recital 55.},
	url = {https://artificialintelligenceact.eu/recital/55/},
	language = {en},
	year = {2025},
	author = {{Future of Life Institute}},
}

@misc{eu_aiAct_article6,
	title = {The {EU} Artificial Intelligence Act. Article 6.},
	url = {https://artificialintelligenceact.eu/article/6/},
	language = {en},
	year = {2025},
	author = {{Future of Life Institute}},
}

@misc{eu_aiAct_article13,
	title = {The {EU} Artificial Intelligence Act. Article 13.},
	url = {https://artificialintelligenceact.eu/article/13/},
	language = {en},
	year = {2025},
	author = {{Future of Life Institute}},
}

@misc{eu_aiAct_article14,
	title = {The {EU} Artificial Intelligence Act. Article 14.},
	url = {https://artificialintelligenceact.eu/article/14/},
	language = {en},
	year = {2025},
	author = {{Future of Life Institute}},
}

@misc{paloAlto_securityAutomation,
	title = {What is Security Automation?},
	url = {https://www.paloaltonetworks.com/cyberpedia/what-is-security-automation},
	language = {en},
	urldate = {2025-08-14},
    month={2},
    year={2023},
	author = {{Palo Alto Networks}},
}

@misc{rmf_what,
	title = {AI Risk Management Framework},
	url = {https://airc.nist.gov/airmf-resources/airmf/},
	year = {2025},
	author = {{Trustworthy \& Responsible {AI} Resource Center, NIST}},
}

@misc{cset_chinaAI,
	title = {An Analysis of China’s {AI} Governance Proposals},
	url = {https://cset.georgetown.edu/article/an-analysis-of-chinas-ai-governance-proposals/},
	year={2025},
    month={9},
	author = {{Center for Security and Emerging Technology (CSET)}},
}

@article{caict2021_white,
    title={White Paper on Trustworthy Artificial Intelligence},
    author={{China Academy of Information and Communications Technology (CAICT)}, {JD Explore Academy}},
    journal={China's {AI} Governance Proposals},
    pages={25},
    month={7},
    year={2021},
}

@misc{tse_china_2025,
	title = {China {Is} {Taking} {AI} {Safety} {Seriously}. {So} {Must} the {U}.{S}.},
	url = {https://www.msn.com/en-us/technology/artificial-intelligence/china-is-taking-ai-safety-seriously-so-must-the-u-s/ar-AA1Krxbc},
	language = {en-US},
	urldate = {2025-08-19},
	author = {Tse, Brian},
	month = {8},
	year = {2025},
}

@inproceedings{roselli2019managing,
  title={Managing bias in AI},
  author={Roselli, Drew and Matthews, Jeanna and Talagala, Nisha},
  booktitle={Companion proceedings of the 2019 world wide web conference},
  pages={539--544},
  year={2019}
}

@phdthesis{schwartz2022towards,
  title={Towards a standard for identifying and managing bias in artificial intelligence},
  author={Schwartz, Reva and Vassilev, Apostol and Greene, Kristen and Perine, Lori and Burt, Andrew and Hall, Patrick},
  year={2022},
  school={National Institute of Standards and Technology}
}

@article{sahiner2023data,
  title={Data drift in medical machine learning: implications and potential remedies},
  author={Sahiner, Berkman and Chen, Weijie and Samala, Ravi K and Petrick, Nicholas},
  journal={The British Journal of Radiology},
  volume={96},
  number={1150},
  pages={20220878},
  year={2023},
  publisher={Oxford University Press}
}

@article{qiu2019review,
  title={Review of artificial intelligence adversarial attack and defense technologies},
  author={Qiu, Shilin and Liu, Qihe and Zhou, Shijie and Wu, Chunjiang},
  journal={Applied Sciences},
  volume={9},
  number={5},
  pages={909},
  year={2019},
  publisher={MDPI}
}

@article{hofeditz2022applying,
  title={Applying XAI to an AI-based system for candidate management to mitigate bias and discrimination in hiring},
  author={Hofeditz, Lennart and Clausen, S{\"u}nje and Rie{\ss}, Alexander and Mirbabaie, Milad and Stieglitz, Stefan},
  journal={Electronic Markets},
  volume={32},
  number={4},
  pages={2207--2233},
  year={2022},
  publisher={Springer}
}

@article{da2023false,
  title={False positive identification in intrusion detection using XAI},
  author={da Silveira Lopes, Ricardo and Duarte, Julio Cesar and Goldschmidt, Ronaldo Ribeiro},
  journal={IEEE Latin America Transactions},
  volume={21},
  number={6},
  pages={745--751},
  year={2023}
}

@article{bellucci2021towards,
  title={Towards a terminology for a fully contextualized XAI},
  author={Bellucci, Matthieu and Delestre, Nicolas and Malandain, Nicolas and Zanni-Merk, Cecilia},
  journal={Procedia Computer Science},
  volume={192},
  pages={241--250},
  year={2021},
  publisher={Elsevier}
}

@inproceedings{bove2022contextualization,
  title={Contextualization and exploration of local feature importance explanations to improve understanding and satisfaction of non-expert users},
  author={Bove, Clara and Aigrain, Jonathan and Lesot, Marie-Jeanne and Tijus, Charles and Detyniecki, Marcin},
  booktitle={Proceedings of the 27th international conference on intelligent user interfaces},
  pages={807--819},
  year={2022}
}

@inproceedings{wang2021explanations,
  title={Are explanations helpful? a comparative study of the effects of explanations in ai-assisted decision-making},
  author={Wang, Xinru and Yin, Ming},
  booktitle={Proceedings of the 26th International Conference on Intelligent User Interfaces},
  pages={318--328},
  year={2021}
}

@article{lecue2020role,
  title={On the role of knowledge graphs in explainable AI},
  author={Lecue, Freddy},
  journal={Semantic Web},
  volume={11},
  number={1},
  pages={41--51},
  year={2020},
  publisher={SAGE Publications Sage UK: London, England}
}

@inproceedings{sarker2020wikipedia,
  title={Wikipedia knowledge graph for explainable AI},
  author={Sarker, Md Kamruzzaman and Schwartz, Joshua and Hitzler, Pascal and Zhou, Lu and Nadella, Srikanth and Minnery, Brandon and Juvina, Ion and Raymer, Michael L and Aue, William R},
  booktitle={Iberoamerican Knowledge Graphs and Semantic Web Conference},
  pages={72--87},
  year={2020},
  organization={Springer}
}

@inproceedings{gomez2020vice,
  title={Vice: Visual counterfactual explanations for machine learning models},
  author={Gomez, Oscar and Holter, Steffen and Yuan, Jun and Bertini, Enrico},
  booktitle={Proceedings of the 25th international conference on intelligent user interfaces},
  pages={531--535},
  year={2020}
}

@article{kim2016examples,
  title={Examples are not enough, learn to criticize! criticism for interpretability},
  author={Kim, Been and Khanna, Rajiv and Koyejo, Oluwasanmi O},
  journal={Advances in Neural Information Processing Systems},
  volume={29},
  year={2016}
}

@article{miller2019explanation,
  title={Explanation in artificial intelligence: Insights from the social sciences},
  author={Miller, Tim},
  journal={Artificial intelligence},
  volume={267},
  pages={1--38},
  year={2019},
  publisher={Elsevier}
}

@online{Devry2025,
  author       = {Jane Devry},
  title        = {How to Add Context to Threat Alerts: Quick Guide for SOCs},
  year         = {2025},
  url          = {https://www.cybersecurity-insiders.com/how-to-add-context-to-threat-alerts-quick-guide-for-socs/},
  note         = {Accessed: August 21, 2025},
  organization = {Cybersecurity Insiders}
}

@inproceedings{sokol2020explainability,
  title={Explainability fact sheets: A framework for systematic assessment of explainable approaches},
  author={Sokol, Kacper and Flach, Peter},
  booktitle={Proceedings of the 2020 conference on fairness, accountability, and transparency},
  pages={56--67},
  year={2020}
}

@article{shashkov2023adversarial,
  title={Adversarial agent-learning for cybersecurity: a comparison of algorithms},
  author={Shashkov, Alexander and Hemberg, Erik and Tulla, Miguel and O’Reilly, Una-May},
  journal={The Knowledge Engineering Review},
  volume={38},
  pages={e3},
  year={2023},
  publisher={Cambridge University Press}
}

@article{ibrar2025generative,
  title={Generative AI: a double-edged sword in the cyber threat landscape},
  author={Ibrar, Werisha and Mahmood, Danish and Al-Shamayleh, Ahmad Sami and Ahmed, Ghufran and Alharthi, Salman Z and Akhunzada, Adnan},
  journal={Artificial Intelligence Review},
  volume={58},
  number={9},
  pages={285},
  year={2025},
  publisher={Springer}
}

@misc{canadian2022protect,
	title = {Protect your operational technology (ITSAP.00.051)},
	url = {https://www.cyber.gc.ca/en/guidance/protect-your-operational-technology-itsap00051},
	author = {{Canadian Center for Cyber Security}},
	month = {7},
	year = {2022},
}

@article{khraisat2019survey,
  title={Survey of intrusion detection systems: techniques, datasets and challenges},
  author={Khraisat, Ansam and Gondal, Iqbal and Vamplew, Peter and Kamruzzaman, Joarder},
  journal={Cybersecurity},
  volume={2},
  number={1},
  pages={1--22},
  year={2019},
  publisher={Springer}
}

@article{cremer2022cyber,
  title={Cyber risk and cybersecurity: a systematic review of data availability},
  author={Cremer, Frank and Sheehan, Barry and Fortmann, Michael and Kia, Arash N and Mullins, Martin and Murphy, Finbarr and Materne, Stefan},
  journal={The Geneva papers on risk and insurance. Issues and practice},
  volume={47},
  number={3},
  pages={698},
  year={2022}
}

@article{eling2016we,
  title={What do we know about cyber risk and cyber risk insurance?},
  author={Eling, Martin and Schnell, Werner},
  journal={The Journal of Risk Finance},
  volume={17},
  number={5},
  pages={474--491},
  year={2016},
  publisher={Emerald Group Publishing Limited}
}

@article{okutan2018forecasting,
  title={Forecasting cyberattacks with incomplete, imbalanced, and insignificant data},
  author={Okutan, Ahmet and Werner, Gordon and Yang, Shanchieh Jay and McConky, Katie},
  journal={Cybersecurity},
  volume={1},
  number={1},
  pages={15},
  year={2018},
  publisher={Springer}
}

@article{scala2019risk,
  title={Risk and the five hard problems of cybersecurity},
  author={Scala, Natalie M and Reilly, Allison C and Goethals, Paul L and Cukier, Michel},
  journal={Risk Analysis},
  volume={39},
  number={10},
  pages={2119--2126},
  year={2019},
  publisher={Wiley Online Library}
}

@article{sarker2020cybersecurity,
  title={Cybersecurity data science: an overview from machine learning perspective},
  author={Sarker, Iqbal H and Kayes, ASM and Badsha, Shahriar and Alqahtani, Hamed and Watters, Paul and Ng, Alex},
  journal={Journal of Big data},
  volume={7},
  number={1},
  pages={41},
  year={2020},
  publisher={Springer}
}

@techreport{isc2_2024report,
    title={2024 ISC2 Cybersecurity Workforce Study: Global Cybersecurity Workforce Prepares for an AI-Driven World},
    author={{International Information System Security Certification Consortium (ISC2)}},
    year={2024},
    institution = {International Information System Security Certification Consortium (ISC2)},
    url={https://www.isc2.org/Insights/2024/10/ISC2-2024-Cybersecurity-Workforce-Study},
}

@misc{weforum2024cybersecurity,
	title = {The cybersecurity industry has an urgent talent shortage. Here’s how to plug the gap},
	url = {https://www.weforum.org/stories/2024/04/cybersecurity-industry-talent-shortage-new-report/},
	author = {Meineke, Michelle},
	month = {4},
	year = {2024},
    publisher={World Economic Forum: Center for Cybersecurity},
}

@misc{ibm2025isc2,
	title = {ISC2 Cybersecurity Workforce Study: Shortage of {AI} skilled workers},
	url = {https://www.ibm.com/think/insights/isc2-cybersecurity-workforce-study-shortage-ai-skilled-workers},
	author = {Poremba, Sue},
	month = {1},
	year = {2025},
    publisher={IBM},
}

@inproceedings{kadir2023evaluation,
  title={Evaluation metrics for xai: A review, taxonomy, and practical applications},
  author={Kadir, Md Abdul and Mosavi, Amir and Sonntag, Daniel},
  booktitle={2023 IEEE 27th International Conference on Intelligent Engineering Systems (INES)},
  pages={000111--000124},
  year={2023},
  organization={IEEE}
}

@inproceedings{coroama2022evaluation,
  title={Evaluation metrics in explainable artificial intelligence (XAI)},
  author={Coroama, Loredana and Groza, Adrian},
  booktitle={International conference on advanced research in technologies, information, innovation and sustainability},
  pages={401--413},
  year={2022},
  organization={Springer}
}

@inproceedings{zolanvari2018effect,
  title={Effect of imbalanced datasets on security of industrial IoT using machine learning},
  author={Zolanvari, Maede and Teixeira, Marcio A and Jain, Raj},
  booktitle={2018 IEEE international conference on intelligence and security informatics (ISI)},
  pages={112--117},
  year={2018},
  organization={IEEE}
}

@misc{sans2024state,
	title = {SANS 2024 State of ICS/OT Cybersecurity},
	url = {https://www.sans.org/white-papers/sans-2024-state-ics-ot-cybersecurity},
	author = {Christopher, Jason D.},
	month = {10},
	year = {2024},
    publisher={SANS Institute},
}

@techreport{crowley2024sansSOC,
  author       = {Christopher Crowley},
  title        = {SANS 2024 SOC Survey: Facing Top Challenges in Security Operations},
  institution  = {SANS Institute},
  year         = {2024},
  url          = {https://www.sans.org/white-papers/sans-2024-soc-survey-facing-top-challenges-security-operations},
  note         = {Accessed October 30, 2025}
}

@article{arreche2024xai2,
  title={E-xai: Evaluating black-box explainable {AI} frameworks for network intrusion detection},
  author={Arreche, Osvaldo and Guntur, Tanish R and Roberts, Jack W and Abdallah, Mustafa},
  journal={IEEE Access},
  volume={12},
  pages={23954--23988},
  year={2024},
  publisher={IEEE}
}

@article{booij2021ton_iot,
  title={ToN\_IoT: The role of heterogeneity and the need for standardization of features and attack types in IoT network intrusion data sets},
  author={Booij, Tim M and Chiscop, Irina and Meeuwissen, Erik and Moustafa, Nour and Den Hartog, Frank TH},
  journal={IEEE Internet of Things Journal},
  volume={9},
  number={1},
  pages={485--496},
  year={2021},
  publisher={IEEE}
}

@misc{wustl2021,
	title = {WUSTL-IIOT-2021 Dataset for IIoT Cybersecurity Research},
	url = {http://www.cse.wustl.edu/~jain/iiot2/index.html},
	author = {Zolanvari, M and Teixeira, M. A. and Gupta, L. and Khan, K. M. and Jain, R.},
	month = {10},
	year = {2021},
    publisher={Washington University in St. Louis, USA},
}

@article{han2018anomaly,
  title={Anomaly intrusion detection method for vehicular networks based on survival analysis},
  author={Han, Mee Lan and Kwak, Byung Il and Kim, Huy Kang},
  journal={Vehicular communications},
  volume={14},
  pages={52--63},
  year={2018},
  publisher={Elsevier}
}

@article{meidan2018n,
  title={N-baiot—network-based detection of iot botnet attacks using deep autoencoders},
  author={Meidan, Yair and Bohadana, Michael and Mathov, Yael and Mirsky, Yisroel and Shabtai, Asaf and Breitenbacher, Dominik and Elovici, Yuval},
  journal={IEEE Pervasive Computing},
  volume={17},
  number={3},
  pages={12--22},
  year={2018},
  publisher={IEEE}
}

@inproceedings{guerra2020medbiot,
  title={MedBIoT: Generation of an IoT botnet dataset in a medium-sized IoT network.},
  author={Guerra-Manzanares, Alejandro and Medina-Galindo, Jorge and Bahsi, Hayretdin and N{\~o}mm, Sven},
  booktitle={ICISSP},
  pages={207--218},
  year={2020}
}

@article{koroniotis2019towards,
  title={Towards the development of realistic botnet dataset in the internet of things for network forensic analytics: Bot-iot dataset},
  author={Koroniotis, Nickolaos and Moustafa, Nour and Sitnikova, Elena and Turnbull, Benjamin},
  journal={Future Generation Computer Systems},
  volume={100},
  pages={779--796},
  year={2019},
  publisher={Elsevier}
}

@article{sharafaldin2018toward,
  title={Toward generating a new intrusion detection dataset and intrusion traffic characterization.},
  author={Sharafaldin, Iman and Lashkari, Arash Habibi and Ghorbani, Ali A and others},
  journal={ICISSp},
  volume={1},
  number={2018},
  pages={108--116},
  year={2018}
}

@misc{sarhan2023nf,
    title={NF-ToN-IoT-v2},
    author={Sarhan, Mohanad and Layeghy, Siamak and Portmann, Marius},
    year={2023},
    howpublished = {Dataset hosted at the University of Queensland Research Data Repository},
    institution={The University of Queensland},
    url={https://doi.org/10.48610/38a2d07},
}

@article{hady2020intrusion,
  title={Intrusion detection system for healthcare systems using medical and network data: A comparison study},
  author={Hady, Anar A and Ghubaish, Ali and Salman, Tara and Unal, Devrim and Jain, Raj},
  journal={IEEE Access},
  volume={8},
  pages={106576--106584},
  year={2020},
  publisher={IEEE}
}

@inproceedings{ghazanfar2020iot,
  title={Iot-flock: An open-source framework for iot traffic generation},
  author={Ghazanfar, Syed and Hussain, Faisal and Rehman, Atiq Ur and Fayyaz, Ubaid U and Shahzad, Farrukh and Shah, Ghalib A},
  booktitle={2020 International Conference on Emerging Trends in Smart Technologies (ICETST)},
  pages={1--6},
  year={2020},
  organization={IEEE}
}

@article{neto2023ciciot2023,
  title={CICIoT2023: A real-time dataset and benchmark for large-scale attacks in IoT environment},
  author={Neto, Euclides Carlos Pinto and Dadkhah, Sajjad and Ferreira, Raphael and Zohourian, Alireza and Lu, Rongxing and Ghorbani, Ali A},
  journal={Sensors},
  volume={23},
  number={13},
  pages={5941},
  year={2023},
  publisher={MDPI}
}

@article{mihailescu2021proposition,
  title={The proposition and evaluation of the roedunet-simargl2021 network intrusion detection dataset},
  author={Mihailescu, Maria-Elena and Mihai, Darius and Carabas, Mihai and Komisarek, Miko{\l}aj and Pawlicki, Marek and Ho{\l}ubowicz, Witold and Kozik, Rafa{\l}},
  journal={Sensors},
  volume={21},
  number={13},
  pages={4319},
  year={2021},
  publisher={MDPI}
}

@inproceedings{sharafaldin2019developing,
  title={Developing realistic distributed denial of service (DDoS) attack dataset and taxonomy},
  author={Sharafaldin, Iman and Lashkari, Arash Habibi and Hakak, Saqib and Ghorbani, Ali A},
  booktitle={2019 international carnahan conference on security technology (ICCST)},
  pages={1--8},
  year={2019},
  organization={IEEE}
}

@article{al2021x,
  title={X-IIoTID: A connectivity-agnostic and device-agnostic intrusion data set for industrial Internet of Things},
  author={Al-Hawawreh, Muna and Sitnikova, Elena and Aboutorab, Neda},
  journal={IEEE Internet of Things Journal},
  volume={9},
  number={5},
  pages={3962--3977},
  year={2021},
  publisher={IEEE}
}

@misc{haiDatasets,
    author={Shin, Hyeok-Ki and Lee, Woomyo and Choi, Seungoh and Yun, Jeong-Han and Min, Byung-Gi},
    title={HAI security datasets},
    year={2023},
    url={https://github.com/icsdataset/hai},
 }

@article{lundberg2017unified,
  title={A unified approach to interpreting model predictions},
  author={Lundberg, Scott M and Lee, Su-In},
  journal={Advances in neural information processing systems},
  volume={30},
  year={2017}
}

@inproceedings{ribeiro2016should,
  title={" Why should i trust you?" Explaining the predictions of any classifier},
  author={Ribeiro, Marco Tulio and Singh, Sameer and Guestrin, Carlos},
  booktitle={Proceedings of the 22nd ACM SIGKDD international conference on knowledge discovery and data mining},
  pages={1135--1144},
  year={2016}
}

@article{krishnaveni2022network,
  title={Network intrusion detection based on ensemble classification and feature selection method for cloud computing},
  author={Krishnaveni, Sivamohan and Sivamohan, Sivanandam and Sridhar, Subramanian and Prabhakaran, Subramani},
  journal={Concurrency and Computation: Practice and Experience},
  volume={34},
  number={11},
  pages={e6838},
  year={2022},
  publisher={Wiley Online Library}
}

@article{friedman2001greedy,
  title={Greedy function approximation: a gradient boosting machine},
  author={Friedman, Jerome H},
  journal={Annals of statistics},
  volume={29},
  number={5},
  pages={1189--1232},
  year={2001},
  publisher={JSTOR}
}

@article{goldstein2015peeking,
  title={Peeking inside the black box: Visualizing statistical learning with plots of individual conditional expectation},
  author={Goldstein, Alex and Kapelner, Adam and Bleich, Justin and Pitkin, Emil},
  journal={Journal of Computational and Graphical Statistics},
  volume={24},
  number={1},
  pages={44--65},
  year={2015},
  publisher={Taylor \& Francis}
}

@article{apley2020visualizing,
  title={Visualizing the effects of predictor variables in black box supervised learning models},
  author={Apley, Daniel W and Zhu, Jingyu},
  journal={Journal of the Royal Statistical Society: Series B (Statistical Methodology)},
  volume={82},
  number={4},
  pages={1059--1086},
  year={2020},
  publisher={Wiley Online Library}
}

@article{friedman2008predictive,
  title={Predictive learning via rule ensembles},
  author={Friedman, Jerome H and Popescu, Bogdan E},
  journal={The Annals of Applied Statistics},
  volume={2},
  number={3},
  pages={916--954},
  year={2008},
  publisher={Institute of Mathematical Statistics}
}

@article{seth2025bridging,
  title={Bridging the Gap in XAI-Why Reliable Metrics Matter for Explainability and Compliance},
  author={Seth, Pratinav and Sankarapu, Vinay Kumar},
  journal={arXiv preprint arXiv:2502.04695},
  year={2025}
}

@inproceedings{pietila2023explanation,
  title={When an explanation is not enough: an overview of evaluation metrics of explainable {AI} systems in the healthcare domain},
  author={Pietil{\"a}, Essi and Moreno-S{\'a}nchez, Pedro A},
  booktitle={Mediterranean Conference on Medical and Biological Engineering and Computing},
  pages={573--584},
  year={2023},
  organization={Springer}
}

@inproceedings{byrne2019counterfactuals,
  title={Counterfactuals in explainable artificial intelligence (XAI): Evidence from human reasoning.},
  author={Byrne, Ruth MJ},
  booktitle={IJCAI},
  pages={6276--6282},
  year={2019},
  organization={California, CA}
}

@misc{liao2020questioning,
    author       = {Liao, Vera},
    title        = {Questioning the AI: Towards Human Centered Explainable {AI} (XAI)},
    year         = {2020},
    howpublished = {\url{https://www.zurich.ibm.com/africa/whats-next-ai/pdf/18-3-liao.pdf}},
    organization = {IBM Research},
    note         = {Accessed: 2025-10-15}
}

@article{lombrozo2007simplicity,
  title={Simplicity and probability in causal explanation},
  author={Lombrozo, Tania},
  journal={Cognitive psychology},
  volume={55},
  number={3},
  pages={232--257},
  year={2007},
  publisher={Elsevier}
}

@article{zhang2025may,
  title={May I Ask a Follow-up Question? Understanding the Benefits of Conversations in Neural Network Explainability},
  author={Zhang, Tong and Yang, X Jessie and Li, Boyang},
  journal={International Journal of Human--Computer Interaction},
  volume={41},
  number={9},
  pages={5623--5647},
  year={2025},
  publisher={Taylor \& Francis}
}

@article{taboada2006contributions,
  title={Contributions of student questioning and prior knowledge to construction of knowledge from reading information text},
  author={Taboada, Ana and Guthrie, John T},
  journal={Journal of literacy research},
  volume={38},
  number={1},
  pages={1--35},
  year={2006},
  publisher={SAGE Publications Sage CA: Los Angeles, CA}
}

@article{yu2009scaffolding,
  title={Scaffolding student-generated questions: Design and development of a customizable online learning system},
  author={Yu, Fu-Yun},
  journal={Computers in Human Behavior},
  volume={25},
  number={5},
  pages={1129--1138},
  year={2009},
  publisher={Elsevier}
}

@article{chin2002student,
  title={Student-generated questions: A meaningful aspect of learning in science},
  author={Chin, Christine and Brown, David E},
  journal={International Journal of Science Education},
  volume={24},
  number={5},
  pages={521--549},
  year={2002},
  publisher={Taylor \& Francis}
}

@misc{rqi2025research,
    author       = {{Right Question Institute}},
    title        = {Research on the Impact of Student Questions on Learning},
    year         = {2025},
    howpublished = {\url{https://rightquestion.org/resources/research-on-the-impact-of-student-questions-on-learning/}},
    organization = {Right Question Institute},
    note         = {Accessed: 2025-10-21}
}

@inproceedings{cheng2019explaining,
  title={Explaining decision-making algorithms through UI: Strategies to help non-expert stakeholders},
  author={Cheng, Hao and Wang, Ruotong and Zhang, Zheng and O'connell, Fiona and Gray, Terrance and Harper, F Maxwell and Zhu, Haiyi},
  booktitle={Proceedings of the 2019 chi conference on human factors in computing systems},
  pages={1--12},
  year={2019}
}

@article{rohlfing2020explanation,
  title={Explanation as a social practice: Toward a conceptual framework for the social design of {AI} systems},
  author={Rohlfing, Katharina J and Cimiano, Philipp and Scharlau, Ingrid and Matzner, Tobias and Buhl, Heike M and Buschmeier, Hendrik and Esposito, Elena and Grimminger, Angela and Hammer, Barbara and H{\"a}b-Umbach, Reinhold and others},
  journal={IEEE Transactions on Cognitive and Developmental Systems},
  volume={13},
  number={3},
  pages={717--728},
  year={2020},
  publisher={IEEE}
}

@article{ni2023recent,
  title={Recent advances in deep learning based dialogue systems: A systematic survey},
  author={Ni, Jinjie and Young, Tom and Pandelea, Vlad and Xue, Fuzhao and Cambria, Erik},
  journal={Artificial intelligence review},
  volume={56},
  number={4},
  pages={3055--3155},
  year={2023},
  publisher={Springer}
}

@article{ouyang2022training,
  title={Training language models to follow instructions with human feedback},
  author={Ouyang, Long and Wu, Jeffrey and Jiang, Xu and Almeida, Diogo and Wainwright, Carroll and Mishkin, Pamela and Zhang, Chong and Agarwal, Sandhini and Slama, Katarina and Ray, Alex and others},
  journal={Advances in neural information processing systems}, 
  pages = {27730--27744},
  volume = {35},
  year={2022},
}

@article{chakraborty2021introduction,
  title={Introduction to neural network-based question answering over knowledge graphs},
  author={Chakraborty, Nilesh and Lukovnikov, Denis and Maheshwari, Gaurav and Trivedi, Priyansh and Lehmann, Jens and Fischer, Asja},
  journal={Wiley Interdisciplinary Reviews: Data Mining and Knowledge Discovery},
  volume={11},
  number={3},
  pages={e1389},
  year={2021},
  publisher={Wiley Online Library}
}

@article{karniadakis2021physics,
  title={Physics-informed machine learning},
  author={Karniadakis, George Em and Kevrekidis, Ioannis G and Lu, Lu and Perdikaris, Paris and Wang, Sifan and Yang, Liu},
  journal={Nature Reviews Physics},
  volume={3},
  number={6},
  pages={422--440},
  year={2021},
  publisher={Nature Publishing Group UK London}
}

@article{sophiya2025comprehensive,
  title={A comprehensive analysis of PINNs: Variants, Applications, and Challenges},
  author={Sophiya, Afila Ajithkumar and Nair, Akarsh K and Maleki, Sepehr and Krishnababu, Senthil K},
  journal={arXiv preprint arXiv:2505.22761},
  year={2025}
}

@article{zideh2023physics,
  title={Physics-informed machine learning for data anomaly detection, classification, localization, and mitigation: A review, challenges, and path forward},
  author={Zideh, Mehdi Jabbari and Chatterjee, Paroma and Srivastava, Anurag K},
  journal={IEEE Access},
  volume={12},
  pages={4597--4617},
  year={2023},
  publisher={IEEE}
}

@article{thangamuthu2022unravelling,
  title={Unravelling the performance of physics-informed graph neural networks for dynamical systems},
  author={Thangamuthu, Abishek and Kumar, Gunjan and Bishnoi, Suresh and Bhattoo, Ravinder and Krishnan, NM and Ranu, Sayan},
  journal={Advances in Neural Information Processing Systems},
  volume={35},
  pages={3691--3702},
  year={2022}
}

@article{liu2023physics,
  title={Physics-informed graph neural network for spatial-temporal production forecasting},
  author={Liu, Wendi and Pyrcz, Michael J},
  journal={Geoenergy Science and Engineering},
  volume={223},
  pages={211486},
  year={2023},
  publisher={Elsevier}
}

@inproceedings{ashraf2024physics,
  title={Physics-informed graph neural networks for water distribution systems},
  author={Ashraf, Inaam and Strotherm, Janine and Hermes, Luca and Hammer, Barbara},
  booktitle={Proceedings of the AAAI Conference on Artificial Intelligence},
  volume={38},
  number={20},
  pages={21905--21913},
  year={2024}
}

@article{shu2024knowledge,
  title={Knowledge graph large language model (KG-LLM) for link prediction},
  author={Shu, Dong and Chen, Tianle and Jin, Mingyu and Zhang, Chong and Du, Mengnan and Zhang, Yongfeng},
  journal={arXiv preprint arXiv:2403.07311},
  year={2024}
}

@inproceedings{abu2024knowledge,
  title={Knowledge graphs as context sources for llm-based explanations of learning recommendations},
  author={Abu-Rasheed, Hasan and Weber, Christian and Fathi, Madjid},
  booktitle={2024 IEEE Global Engineering Education Conference (EDUCON)},
  pages={1--5},
  year={2024},
  organization={IEEE}
}

@article{ye2022comprehensive,
  title={A comprehensive survey of graph neural networks for knowledge graphs},
  author={Ye, Zi and Kumar, Yogan Jaya and Sing, Goh Ong and Song, Fengyan and Wang, Junsong},
  journal={IEEE Access},
  volume={10},
  pages={75729--75741},
  year={2022},
  publisher={IEEE}
}

@article{de2023survey,
  title={A survey of public IoT datasets for network security research},
  author={De Keersmaeker, Fran{\c{c}}ois and Cao, Yinan and Ndonda, Gorby Kabasele and Sadre, Ramin},
  journal={IEEE Communications Surveys \& Tutorials},
  volume={25},
  number={3},
  pages={1808--1840},
  year={2023},
  publisher={IEEE}
}

@article{eckhart2019digital,
  title={Digital twins for cyber-physical systems security: State of the art and outlook},
  author={Eckhart, Matthias and Ekelhart, Andreas},
  journal={Security and Quality in Cyber-Physical Systems Engineering: With Forewords by Robert M. Lee and Tom Gilb},
  pages={383--412},
  year={2019},
  publisher={Springer}
}

@misc{cacm2023digitaltwins,
    author       = {Williams, Alex},
    title        = {The Power of Digital Twins in Cybersecurity},
    year         = {2025},
    month={8},
    howpublished = {\url{https://cacm.acm.org/blogcacm/the-power-of-digital-twins-in-cybersecurity/}},
    note         = {Accessed: 2025-10-22}
}

@article{mohamed2025resilient,
  title={Resilient Cyber-Physical System Honeypots for Cyberattacker Engagement},
  author={Mohamed, Amr S and Kundur, Deepa},
  journal={IEEE Transactions on Industrial Informatics},
  year={2025},
  publisher={IEEE}
}

@article{mohamed2023use,
  title={On the use of reinforcement learning for attacking and defending load frequency control},
  author={Mohamed, Amr S and Kundur, Deepa},
  journal={IEEE Transactions on Smart Grid},
  volume={15},
  number={3},
  pages={3262--3277},
  year={2023},
  publisher={IEEE}
}

@inproceedings{mohamed2023reinforcement,
  title={Reinforcement Learning for Supply Chain Attacks Against Frequency and Voltage Control},
  author={Mohamed, Amr S and Lee, Sumin and Kundur, Deepa},
  booktitle={2023 International Conference on Machine Learning and Applications (ICMLA)},
  pages={369--375},
  year={2023},
  organization={IEEE}
}

@article{nguyen2021deep,
  title={Deep reinforcement learning for cyber security},
  author={Nguyen, Thanh Thi and Reddi, Vijay Janapa},
  journal={IEEE Transactions on Neural Networks and Learning Systems},
  volume={34},
  number={8},
  pages={3779--3795},
  year={2021},
  publisher={IEEE}
}

@inproceedings{lupia2023ics,
  title={Ics honeypot interactions: A latitudinal study},
  author={Lupia, Francesco and Lucchese, Marco and Merro, Massimo and Zannone, Nicola},
  booktitle={2023 IEEE International Conference on Big Data (BigData)},
  pages={3025--3034},
  year={2023},
  organization={IEEE}
}

@article{franco2021survey,
  title={A survey of honeypots and honeynets for internet of things, industrial internet of things, and cyber-physical systems},
  author={Franco, Javier and Aris, Ahmet and Canberk, Berk and Uluagac, A Selcuk},
  journal={IEEE Communications Surveys \& Tutorials},
  volume={23},
  number={4},
  pages={2351--2383},
  year={2021},
  publisher={IEEE}
}

@inproceedings{nintsiou2023threat,
  title={Threat intelligence using Digital Twin honeypots in Cybersecurity},
  author={Nintsiou, Maria and Grigoriou, Elisavet and Karypidis, Paris Alexandros and Saoulidis, Theocharis and Fountoukidis, Eleftherios and Sarigiannidis, Panagiotis},
  booktitle={2023 IEEE International Conference on Cyber Security and Resilience (CSR)},
  pages={530--537},
  year={2023},
  organization={IEEE}
}

@article{adadi2018peeking,
  title={Peeking inside the black-box: a survey on explainable artificial intelligence (XAI)},
  author={Adadi, Amina and Berrada, Mohammed},
  journal={IEEE access},
  volume={6},
  pages={52138--52160},
  year={2018},
  publisher={IEEE}
}

@article{gunning2019xai,
  title={XAI—Explainable artificial intelligence},
  author={Gunning, David and Stefik, Mark and Choi, Jaesik and Miller, Timothy and Stumpf, Simone and Yang, Guang-Zhong},
  journal={Science robotics},
  volume={4},
  number={37},
  pages={eaay7120},
  year={2019},
  publisher={American Association for the Advancement of Science}
}

@article{angelov2021explainable,
  title={Explainable artificial intelligence: an analytical review},
  author={Angelov, Plamen P and Soares, Eduardo A and Jiang, Richard and Arnold, Nicholas I and Atkinson, Peter M},
  journal={Wiley Interdisciplinary Reviews: Data Mining and Knowledge Discovery},
  volume={11},
  number={5},
  pages={e1424},
  year={2021},
  publisher={Wiley Online Library}
}

@article{confalonieri2021historical,
  title={A historical perspective of explainable artificial intelligence},
  author={Confalonieri, Roberto and Coba, Ludovik and Wagner, Benedikt and Besold, Tarek R},
  journal={Wiley Interdisciplinary Reviews: Data Mining and Knowledge Discovery},
  volume={11},
  number={1},
  pages={e1391},
  year={2021},
  publisher={Wiley Online Library}
}

@inproceedings{gilpin2018explaining,
  title={Explaining explanations: An overview of interpretability of machine learning},
  author={Gilpin, Leilani H and Bau, David and Yuan, Ben Z and Bajwa, Ayesha and Specter, Michael and Kagal, Lalana},
  booktitle={2018 IEEE 5th International Conference on data science and advanced analytics (DSAA)},
  pages={80--89},
  year={2018},
  organization={IEEE}
}

@misc{vc3_ai_cybersecurity_2023,
  author       = {{VC3}},
  title        = {The Evolution of Artificial Intelligence in Cybersecurity},
  year         = {2023},
  month        = {10},
  url          = {https://www.vc3.com/blog/the-evolution-of-artificial-intelligence-in-cybersecurity},
  note         = {Accessed 27 October 2025}
}

@misc{sailpoint_ai_cybersecurity_2023,
  author       = {{SailPoint}},
  title        = {Machine learning (ML) in cybersecurity},
  year         = {2025},
  month        = {5},
  url          = {https://www.sailpoint.com/identity-library/how-ai-and-machine-learning-are-improving-cybersecurity},
  note         = {Accessed 27 October 2025}
}

@article{jiang2021industrialv2,
  title={Industrial applications of digital twins},
  author={Jiang, Yuchen and Yin, Shen and Li, Kuan and Luo, Hao and Kaynak, Okyay},
  journal={Philosophical Transactions of the Royal Society A},
  volume={379},
  number={2207},
  pages={20200360},
  year={2021},
  publisher={The Royal Society Publishing}
}

@article{charmet2022explainable,
  title={Explainable artificial intelligence for cybersecurity: a literature survey},
  author={Charmet, Fabien and Tanuwidjaja, Harry Chandra and Ayoubi, Solayman and Gimenez, Pierre-Fran{\c{c}}ois and Han, Yufei and Jmila, Houda and Blanc, Gregory and Takahashi, Takeshi and Zhang, Zonghua},
  journal={Annals of Telecommunications},
  volume={77},
  number={11},
  pages={789--812},
  year={2022},
  publisher={Springer}
}

@article{srivastava2022xai,
  title={XAI for cybersecurity: state of the art, challenges, open issues and future directions},
  author={Srivastava, Gautam and Jhaveri, Rutvij H and Bhattacharya, Sweta and Pandya, Sharnil and Maddikunta, Praveen Kumar Reddy and Yenduri, Gokul and Hall, Jon G and Alazab, Mamoun and Gadekallu, Thippa Reddy and others},
  journal={arXiv preprint arXiv:2206.03585},
  year={2022}
}

@article{alketbi2025comprehensive,
  title={A Comprehensive Survey of Explainable Artificial Intelligence Techniques for Malicious Insider Threat Detection},
  author={Alketbi, Khuloud Saeed and Mehmood, Abid},
  journal={IEEE Access},
  year={2025},
  volume={13},
  pages={121772--121798},
  publisher={IEEE}
}

@article{saqib2024comprehensive,
  title={A comprehensive analysis of explainable {AI} for malware hunting},
  author={Saqib, Mohd and Mahdavifar, Samaneh and Fung, Benjamin CM and Charland, Philippe},
  journal={ACM Computing Surveys},
  volume={56},
  number={12},
  pages={1--40},
  year={2024},
  publisher={ACM New York, NY}
}

@techreport{dynatrace2025observability,
  title        = {The State of Observability 2025},
  author       = {{Dynatrace}},
  year         = {2025},
  institution  = {Dynatrace},
  url          = {https://www.dynatrace.com/info/ebooks/the-state-of-observability/},
  note         = {Accessed October 28, 2025}
}

@book{saxe2018malware,
  title={Malware data science: attack detection and attribution},
  author={Saxe, Joshua and Sanders, Hillary},
  year={2018},
  publisher={No Starch Press}
}

@online{gartner2018ciosAI,
  author       = {{Gartner}},
  title        = {Gartner Says Nearly Half of CIOs Are Planning to Deploy Artificial Intelligence},
  year         = {2018},
  url          = {https://www.gartner.com/en/newsroom/press-releases/2018-02-13-gartner-says-nearly-half-of-cios-are-planning-to-deploy-artificial-intelligence},
  note         = {Accessed October 30, 2025},
  organization = {Gartner Newsroom}
}

@techreport{kpmg2025ai,
  title        = {Trust, attitudes and use of artificial intelligence: A global study 2025},
  author       = {{KPMG}},
  year         = {2025},
  institution  = {Klynveld, Peat, Marwick, and Goerdeler},
  url          = {https://assets.kpmg.com/content/dam/kpmgsites/xx/pdf/2025/05/trust-attitudes-and-use-of-ai-global-report.pdf},
  note         = {Accessed July 2, 2026}
}

@online{ansi2025chinaAIgovernance,
  author       = {{ANSI}},
  title        = {China Announces Action Plan for Global {AI} Governance},
  year         = {2025},
  url          = {https://www.ansi.org/standards-news/all-news/8-1-25-china-announces-action-plan-for-global-ai-governance},
  note         = {Accessed October 30, 2025},
  organization = {American National Standards Institute}
}

@inproceedings{abdul2018trends,
  title={Trends and trajectories for explainable, accountable and intelligible systems: An hci research agenda},
  author={Abdul, Ashraf and Vermeulen, Jo and Wang, Danding and Lim, Brian Y and Kankanhalli, Mohan},
  booktitle={Proceedings of the 2018 CHI conference on human factors in computing systems},
  pages={1--18},
  year={2018}
}

@inproceedings{spartalis2023balancing,
  title={Balancing xai with privacy and security considerations},
  author={Spartalis, Christoforos N and Semertzidis, Theodoros and Daras, Petros},
  booktitle={European Symposium on Research in Computer Security},
  pages={111--124},
  year={2023},
  organization={Springer}
}

@inproceedings{hara2023average,
  title={Average sensitivity of decision tree learning},
  author={Hara, Satoshi and Yoshida, Yuichi},
  booktitle={The Eleventh International Conference on Learning Representations},
  year={2023}
}

@inproceedings{pachl2025view,
  title={A View on Vulnerabilites: The Security Challenges of XAI},
  author={Pachl, Elisabeth and Langer, Fabian and Markert, Thora and Lorenz, Jeanette Miriam},
  booktitle={Symposium on Scaling {AI} Assessments},
  pages={1},
  year={2025}
}

@inproceedings{pawlicki2024explainability,
  title={Explainability versus security: The unintended consequences of xai in cybersecurity},
  author={Pawlicki, Marek and Pawlicka, Aleksandra and Kozik, Rafa{\l} and Chora{\'s}, Micha{\l}},
  booktitle={Proceedings of the 2nd ACM Workshop on Secure and Trustworthy Deep Learning Systems},
  pages={1--7},
  year={2024}
}

@inproceedings{zhang2020interpretable,
  title={Interpretable deep learning under fire},
  author={Zhang, Xinyang and Wang, Ningfei and Shen, Hua and Ji, Shouling and Luo, Xiapu and Wang, Ting},
  booktitle={29th USENIX security symposium USENIX security 20},
  year={2020}
}

@misc{prophet_howAITransforms,
	title = {How Agentic {AI} Transforms Tier 1, Tier 2, and Tier 3 SOC Analysts},
	url = {https://www.prophetsecurity.ai/blog/how-ai-transforms-tier-1-tier-2-and-tier-3-soc-analysts},
	language = {en},
	year = {2025},
	journal = {Prophet Security},
	author = {George Dimitrov},
}

@misc{eu_harmonizedstd,
	title = {The {EU} Artificial Intelligence Act. Standard Setting.},
	url = {https://artificialintelligenceact.eu/standard-setting-overview/},
	language = {en},
	year = {2025},
	author = {{EU Artificial Intelligence Act}},
    }

@article{kalakoti2025evaluating,
  title={Evaluating explainable {AI} for deep learning-based network intrusion detection system alert classification},
  author={Kalakoti, Rajesh and Vaarandi, Risto and Bahsi, Hayretdin and N{\~o}mm, Sven},
  journal={arXiv preprint arXiv:2506.07882},
  year={2025}
}

@misc{sharma_2025,
  author={Sharma, Anshika and Rani, Shalli and Shabaz, Mohammad},
  title={A comprehensive review of explainable AI in cybersecurity: Decoding the black box},
  year={2025},
  journal={Information \& Communications Technology Express},
  url={https://www.sciencedirect.com/science/article/pii/S2405959525001584}
}

@misc{siemens_whitepaperXAI,
	title = {The rise of industrial explainable artificial intelligence (XAI) – Insights across the AI life cycle},
	url = {https://assets.new.siemens.com/siemens/assets/api/uuid:3b4de373-57e2-4329-b025-2825db0172aa/WhitepaperXAI.pdf},
	language = {en},
	year = {2023},
	author = {Siemens},
    }

@article{cambria2024xai,
  title={Xai meets llms: A survey of the relation between explainable ai and large language models},
  author={Cambria, Erik and Malandri, Lorenzo and Mercorio, Fabio and Nobani, Navid and Seveso, Andrea},
  journal={arXiv preprint arXiv:2407.15248},
  year={2024}
}

@article{ayyat2025opportunities,
  title={Opportunities and challenges of foundation models in industrial manufacturing},
  author={Ayyat, Mohammed and Osman, Mohamed and Nadeem, Tamer},
  journal={IEEE Access},
  year={2025},
  publisher={IEEE}
}

@misc{s7comm,
	title = {What properties, advantages and special features does the S7 protocol offer?},
	url = {https://support.industry.siemens.com/cs/document/26483647/what-properties-advantages-and-special-features-does-the-s7-protocol-offer-?dti=0},
	language = {en},
	year = {2019},
	author = {Siemens},
    }

@misc{ukrainegrid2024,
	title = {CHRONOLOGY OF CYBER ASPECTS OF THE WAR IN UKRAINE 2022 - Present},
url = {https://nsarchive.gwu.edu/sites/default/files/2024-06/Ukraine cyber chronology - as of June 26, 2024.pdf},
	language = {en},
	year = {2024},
	author = {National Security Archive Cyber Vault},
    }

@book{lewis2013counterfactuals,
  title={Counterfactuals},
  author={Lewis, David},
  year={2013},
  publisher={John Wiley \& Sons}
}

@inproceedings{ribeiro2018anchors,
  title={Anchors: High-precision model-agnostic explanations},
  author={Ribeiro, Marco Tulio and Singh, Sameer and Guestrin, Carlos},
  booktitle={Proceedings of the AAAI conference on artificial intelligence},
  volume={32},
  number={1},
  year={2018}
}

\end{document}